\documentclass[pra,aps,showpacs,groupedaddress,superscriptaddress,twocolumn]{revtex4-1}

\usepackage[utf8x]{inputenc}
\usepackage{color}
\usepackage{bbm} 

\usepackage{amsfonts,amsmath,amssymb,stmaryrd}

\usepackage{graphicx}
\usepackage{subfigure}  % use for side-by-side figures
\usepackage{bbm} 
\usepackage{hyperref}
\usepackage{epsfig}
\usepackage{mathrsfs}
\usepackage{verbatim}
\usepackage{centernot}
\usepackage{ulem}

\renewcommand{\l}{\left(}
\renewcommand{\r}{\right)}

\newcommand{\bra}[1]{\langle#1|}
\newcommand{\ket}[1]{|#1\rangle}

\renewcommand{\H}{\hat{\mathcal{H}}}
\newcommand{\Ht}{\tilde{\mathcal{H}}}

\renewcommand{\a}{\hat{a}}

\newcommand{\ad}{\hat{a}^\dagger}
\newcommand{\bd}{\hat{b}^\dagger}
\newcommand{\dd}{\hat{d}^\dagger}
\renewcommand{\d}{\hat{d}}
\renewcommand{\b}{\hat{b}}

\newcommand{\G}{\hat{\Gamma}}

\newcommand{\hc}{\text{h.c.}}
\newcommand{\MF}{\text{MF}}

\newcommand{\f}{\text{F}}
\newcommand{\s}{\text{S}}
\renewcommand{\sf}{\text{MIX}}

\newcommand{\U}{\hat{U}}
\newcommand{\Ud}{\hat{U}^\dagger}
\newcommand{\F}{\hat{F}}
\newcommand{\Fd}{\hat{F}^\dagger}

\newcommand{\ph}{\text{ph}}
\newcommand{\IB}{\text{IB}}

\newcommand{\eff}{\text{eff}}
\newcommand{\D}{\hat{D}}
\newcommand{\Dd}{\hat{D}^\dagger~\hspace{-0.12cm}}

\usepackage[dvipsnames]{xcolor}

\usepackage{array}

\usepackage{cancel,ifthen}
\newcommand{\cmnt}[2][NoInPuT]{\ifthenelse{\equal{#1}{NoInPuT}}{}{{\color{red}\sout{#1}}} {\color{blue} #2}}

\usepackage{bm}	% bold in math mode
\renewcommand{\vec}[1]{\bm{#1}}

\begin{document}
\normalem	% changes \emph back to normal after introducing ulem package.

\title{Strong coupling Bose polarons in a BEC}

\author{F. Grusdt}
\affiliation{Department of Physics, Harvard University, Cambridge, Massachusetts 02138, USA}

\author{R. Schmidt}
\affiliation{Department of Physics, Harvard University, Cambridge, Massachusetts 02138, USA}
\affiliation{ITAMP, Harvard-Smithsonian Center for Astrophysics, Cambridge, MA 02138, USA}

\author{Y. E. Shchadilova}
\affiliation{Department of Physics, Harvard University, Cambridge, Massachusetts 02138, USA}

\author{E. Demler}
\affiliation{Department of Physics, Harvard University, Cambridge, Massachusetts 02138, USA}

\date{\today}

\begin{abstract}
We use a non-perturbative renormalization group approach to develop a unified picture of the Bose polaron problem, where a mobile impurity is strongly interacting with a surrounding Bose-Einstein condensate (BEC). A detailed theoretical analysis of the phase diagram is presented and the polaron-to-molecule transition is discussed. For attractive polarons we argue that a description in terms of an effective Fr\"ohlich Hamiltonian with renormalized parameters is possible. Its strong coupling regime is realized close to a Feshbach resonance, where we predict a sharp increase of the effective mass. Already for weaker interactions, before the polaron mass diverges, we predict a transition to a regime where states exist below the polaron energy and the attractive polaron is no longer the ground state. On the repulsive side of the Feshbach resonance we recover the repulsive polaron, which has a finite lifetime because it can decay into low-lying molecular states. We show for the entire range of couplings that the polaron energy has logarithmic corrections in comparison with predictions by the mean-field approach. We demonstrate that they are a consequence of the polaronic mass renormalization which is due to quantum fluctuations of correlated phonons in the polaron cloud.
\end{abstract}

\maketitle

%%%%%%%%%%%%%%%%%%%%%%%%%%%%%%%%%%%%%%%%%%%%%%%%%%%%%
\section{Introduction}
%%%%%%%%%%%%%%%%%%%%%%%%%%%%%%%%%%%%%%%%%%%%%%%%%%%%%

The dressing of mobile impurities with collective excitations of a surrounding many-body system -- i.e. the formation of polarons \cite{Landau1946,Landau1948} -- is a ubiquitous phenomenon in physics. Its consequences, including for example the enhancement of the effective mass \cite{PolaronsAdvMat2007,alexandrov2009advances}, have been observed in a large class of systems, ranging from electrons in a semiconductor coupled to phonon excitations in the host lattice \cite{Froehlich1954,Mahan2000} and exciton polaritons interacting with a surrounding Fermi sea \cite{Sidler2016} to mixtures of ultracold fermions \cite{Schirotzek2009,Koschorreck2012} and bosons \cite{Catani2012,Fukuhara2013,Scelle2013,Jorgensen2016PRL,Hu2016PRL} with contact interactions. 

Understanding the physics of polarons at intermediate couplings, where the phonon excitations in the polaron cloud strongly correlated, has posed a challenge for decades \cite{Feynman1955,Devreese2013,Grusdt2015Varenna}. Recently, experiments with mixtures of ultracold bosons \cite{Catani2012,Hu2016PRL,Jorgensen2016PRL} have realized this regime and they have initiated a detailed investigation of Bose polarons for a large range of couplings. Despite intense theoretical effort in understanding the physics of mobile impurities in ultracold quantum gases \cite{Mathey2004,Bruderer2007,BeiBing2009,Tempere2009,Casteels2011,Casteels2012,Rath2013,Shashi2014RF,Kain2014,Li2014,Grusdt2014BO,Yin2015PRA,Grusdt2015RG,Grusdt2016RG,Vlietinck2015,Levinsen2015,Christensen2015,Ardila2015,Grusdt2016,Shchadilova2016,Shchadilova2016PRL,Parisi2016,Kain2016}, a number of questions remains unresolved so far.

Recent calculations of the spectral function of the Bose polaron based on a time-dependent coherent state approach \cite{Shchadilova2016PRL}, closely related to the mean-field (MF) description of the polaron \cite{Lee1953,Shashi2014RF}, suggested the existence of at least two different parameter regimes. While the attractive polaron corresponds to a single sharp feature in the spectrum clearly visible at weak couplings, an entire set of lines associated with molecular bound states was predicted on the repulsive side of the Feshbach resonance at energies below the repulsive polaron peak. It remained unclear, however, whether these regimes are connected by a smooth crossover or a sharp phase transition corresponding to a non-analytical ground state energy. On the one hand the energy of the MF polaron state diverges at a critical interaction strength \cite{Shchadilova2016PRL}, suggesting a sharp transition. On the other hand, variational wavefunctions including only a few excitations \cite{Li2014,Levinsen2015,Jorgensen2016PRL} as well as closely related T-matrix calculations \cite{Rath2013} predict a smooth crossover of the ground states between the two regimes. 

Another question concerns the dependence of the polaron energy on the effective range associated with the impurity-boson scattering potential, which represents the first correction to the scattering amplitude beyond pure contact interactions. For example, the ground state energy of an interacting Bose gas has a sub-leading contribution due to quantum fluctuations which scales logarithmically with the effective range of the boson-boson scattering \cite{Wu1959,Sawada1959,Hugenholtz1959}. Similar corrections were predicted for impurities by a renormalization group (RG) analysis of the effective Fr\"ohlich polaron model \cite{Grusdt2015RG} and by analytical calculations using perturbation theory \cite{Christensen2015}.

Here we develop a unified description of Bose polarons at strong couplings, and in the limit of a weakly interacting BEC. Our analysis is based on a non-perturbative RG approach which we use to derive the polaron phase diagram. In particular, we investigate one of the theoretically most challenging regimes, where the polaron cloud contains many phonons which become correlated due to the interactions induced by the mobile impurity. An accurate description of these correlations requires either numerical quantum Monte Carlo calculations \cite{Prokofev1998,Ardila2015,Vlietinck2015,Parisi2016}, variational wavefunctions with many degrees of freedom \cite{Shchadilova2016,Kain2016} or an RG approach \cite{Grusdt2015RG,Grusdt2015Varenna,Grusdt2016RG}.

As a main result we find evidence for two sharp transitions where the nature of the polaron changes abruptly. Our work extends earlier calculations, which were either based on the effective Fr\"ohlich Hamiltonian \cite{Mathey2004,Bruderer2007,BeiBing2009,Tempere2009,Casteels2011,Casteels2012,Shashi2014RF,Kain2014,Li2014,Grusdt2014BO,Yin2015PRA,Grusdt2015RG,Grusdt2016RG,Vlietinck2015,Grusdt2016} or relied on approximate trial wavefunctions \cite{Li2014,Levinsen2015,Jorgensen2016PRL}. We include two-phonon terms going beyond the Fr\"ohlich model \cite{Rath2013} and treat them accurately up to high energies \cite{Rath2013,Shchadilova2016PRL}. Our RG analysis moreover reveals that the Fr\"ohlich terms in the Hamiltonian are the most relevant ones  for describing the Bose polaron away from the resonance where multiple bound states appear, although corrections are important for making quantitative predictions.

The paper is organized as follows. We start by summarizing our results in Sec.~\ref{sec:SummaryResults}. We proceed by introducing the model describing an impurity in a BEC in Sec.~\ref{sec:Model}. A simplified toy model is analyzed in Sec.~\ref{sec:ToyModel} to gain intuitive understanding for the polaron phase diagram. In Sec.~\ref{sec:MFreview} we summarize the mean-field theory on which the RG method is based, and discuss molecular bound states. In Sec.~\ref{sec:RGtheory} we present our RG method and the RG flow equations. Regularization schemes for divergencies of the coupling constants are also introduced. In Sec.~\ref{subsec:PolaronEnergy} we present calculations of the polaron energies which we compare to predictions by the simpler Fr\"ohlich model. Corrections to the phase diagram by the RG are discussed in Sec.~\ref{sec:phaseDiag}. We close with a summary and by giving an outlook in Sec.~\ref{sec:summary}.

%%%%%%%%%%%%%%%%%%%%%%%%%%%%%%%%%%%%%%%%%%%%%%%%%%%%%
\section{Summary of results}
\label{sec:SummaryResults}
%%%%%%%%%%%%%%%%%%%%%%%%%%%%%%%%%%%%%%%%%%%%%%%%%%%%%

\begin{figure}[t!]
\centering
\epsfig{file=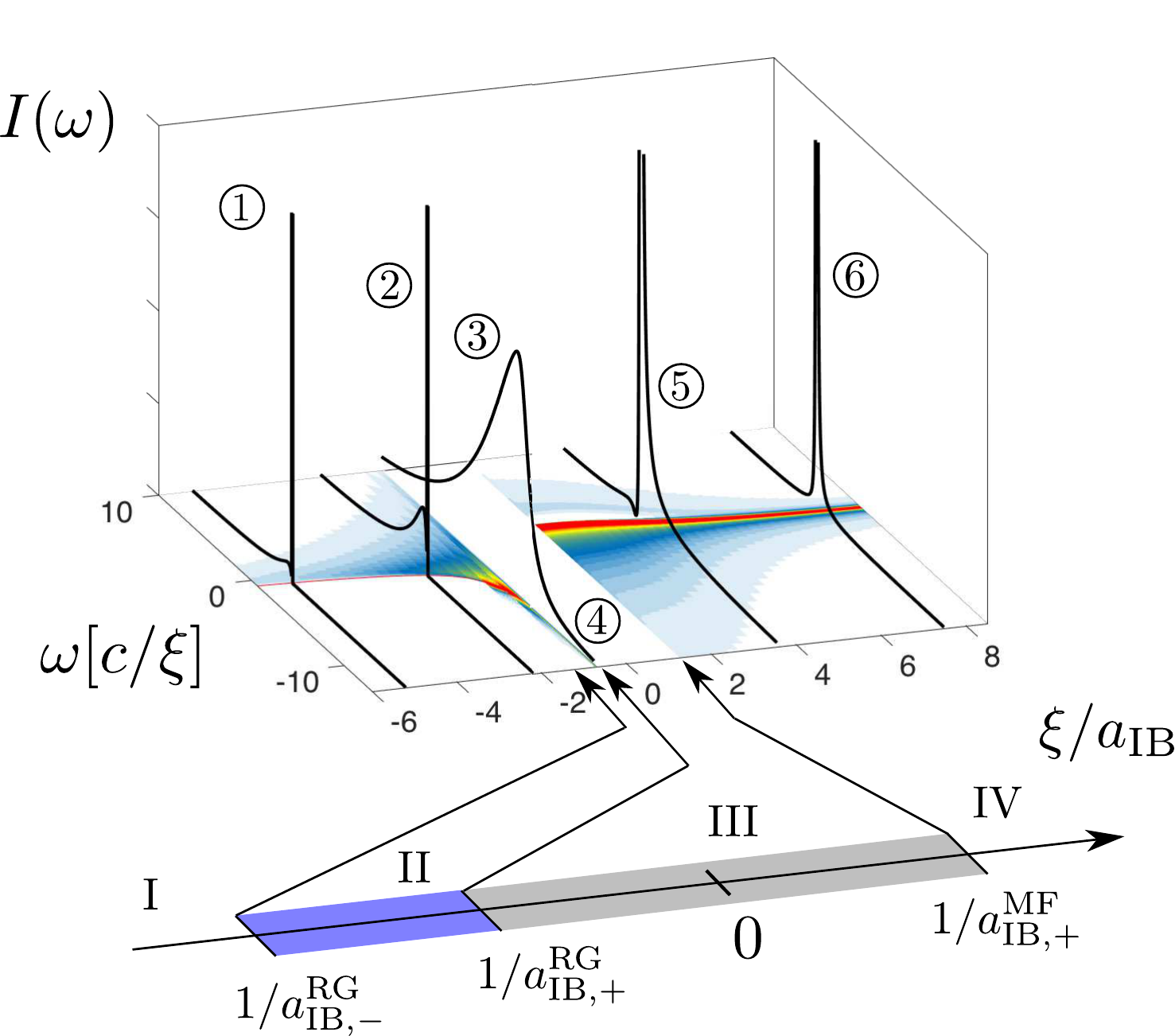, width=0.45\textwidth}
\caption{The polaron branch of the spectral function $I(\omega)$ is shown for an impurity in a BEC at different values of the impurity-boson scattering length $a_{\rm IB}$. Its qualitative features change at the critical values $a_{\rm IB,\pm}$ indicated in the phase diagram at the bottom of the figure. We neglected direct phonon-phonon interactions (Bogoliubov approximation), leading to a break-down of the RG in the central region $\textcircled{4}$. The same parameters were used as in Fig.~2 of Ref.~\cite{Rath2013},  i.e. $M/m_{\rm B}=1$ and $n_0=0.25 \xi^{-3}$, but for a UV cut-off $\Lambda_0=10^3/\xi$. Here $\xi$($c$) is the healing length (speed of sound) in the BEC, $a_{\rm IB}$ denotes the scattering length and $M$ and $m_{\rm B}$ are impurity and boson masses, respectively.}
\label{IntroSepctrum}
\end{figure}

The focus of this paper is to understand the properties of the Bose polaron, which corresponds to a sharp peak in the spectral function $I(\omega)$ at weak couplings. Changes in the nature of the Bose polaron at strong couplings are accompanied by strong modifications of the spectrum, which we discuss next. In Fig.~\ref{IntroSepctrum} we summarize our main results and sketch the shape of $I(\omega)$ for different impurity-boson scattering lengths $a_{\rm IB}$, which can be tuned by a Feshbach resonance in experiments with cold atoms.

\emph{(I) Attractive polarons.--}
On the attractive side sufficiently far from the resonance, i.e. for $1/a_\IB \ll-1/\xi$ less than the inverse of the healing length $\xi$ of the BEC, the spectrum consists of a delta function peak $Z \delta (\hbar \omega-E_0)$ located at the energy $E_0$ of the attractive polaron ($\textcircled{1}$ in Fig.~\ref{IntroSepctrum}). The corresponding quasiparticle residue $Z \approx 1$ is close to one far from the Feshbach resonance (for finite $\xi$). All the remaining spectral weight, $1-Z$, is located at energies $E>E_0$ above the polaron energy and corresponds to shake-off processes of phonons from the polaron cloud as discussed in Refs.~\cite{Rath2013,Shashi2014RF}. This regime can be described within the Fr\"ohlich model. 

For stronger interactions, but still on the attractive side, an increasing amount of spectral weight moves into the incoherent part of the spectrum. Otherwise the shape of the spectral function corresponding to the attractive polaron remains unchanged ($\textcircled{2}$ in Fig.~\ref{IntroSepctrum}). With increasing interaction strength the quantitative deviations from predictions by the Fr\"ohlich model begin to grow.

%%%%%%%%%%%%%%%%%%%%%%%%%%%%%%%%%%%%%%%%%%%%%%%%%%%%%
%%%%%%%%%%%%%%%%%%%%%%%%%%%%%%%%%%%%%%%%%%%%%%%%%%%%%

\emph{(II) Attractive polarons at strong coupling.--}
At a critical value $a_{\rm IB} = a_{\IB,-}^{\rm RG} < 0$ on the attractive side of the resonance, the shape of the spectrum changes qualitatively ($\textcircled{3}$ in Fig.~\ref{IntroSepctrum}). Here a phase transition takes place and the ground state is no longer defined. We emphasize that this conclusion relies on the Bogoliubov approximation for describing the host BEC atoms, which does not include interactions between Bogoliubov quasiparticles. Within this approximation we predict states at arbitrarily negative energies, giving rise to a finite spectral weight $I(\omega)>0$ even for $\omega < E_0$ below the polaron energy. Note however that this spectral weight quickly diminishes for $\omega$ below $E_0$. The value $a_{\IB,-}^{\rm RG} \sim \xi$ is of the order of the BEC healing length $\xi$ and depends on the impurity mass $M$; In particular $|a_{\IB,-}^{\rm RG}|=\infty$ for a localized impurity, $M=\infty$.

In this regime beyond $a_{\IB,-}^{\rm RG}$, the peak in the spectrum at energy $E_0$ corresponding to the attractive polaron acquires a finite linewidth. It can be associated with the inverse lifetime of the polaron, which can decay into the states appearing at energies below $E_0$. In the spectral function shown in Fig.~\ref{IntroSepctrum}, the broadened attractive polaron peak can not be distinguished from the even broader incoherent part of the spectrum associated with excitations of the polaron cloud at energies above $E_0$. Note, however, that these shake-off processes at high energies give rise to an asymmetric shape of $I(\omega)$ around $\omega = E_0$ which is visible in Fig.~\ref{IntroSepctrum}. 

For small but non-vanishing phonon-phonon interactions, beyond the Bogoliubov approximation considered here, we expect a continuous transition into this regime around $a_{\IB,-}^{\rm RG}$. When such phonon non-linearities are sufficiently weak, however, we think that the features in the spectral function discussed above remain valid.

\begin{figure}[t!]
\centering
\epsfig{file=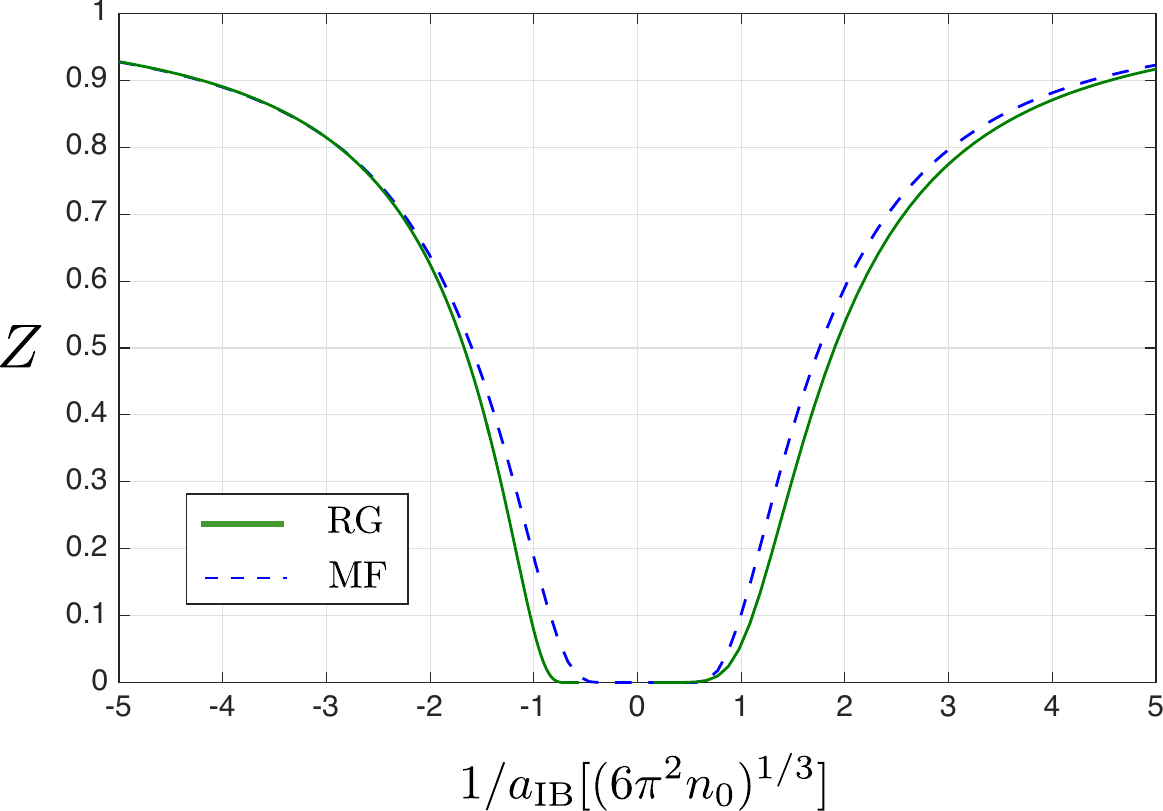, width=0.45\textwidth}
\caption{Close to the Feshbach resonance the quasiparticle weight of the polaron vanishes. Polarons in this regime can be prepared by adiabatically changing the interaction strength. For comparison mean-field (MF) calculations are shown. We used the same parameters as in the recent experimental observation of Bose polarons \cite{Jorgensen2016PRL}, $M/m_{\rm B}=1$ and $n_0=2.3 \times10^{14} {\rm cm}^{-3}$, and a UV cut-off $\Lambda_0=1/60 a_0$ corresponding to the inverse characteristic range $r^*$ of the Feshbach resonance \cite{Chin2010} estimated in Ref.~\cite{Jorgensen2016PRL} ($a_0$ is the Bohr radius).}
\label{fig:polaronZ}
\end{figure}

\begin{figure}[t!]
\centering
\epsfig{file=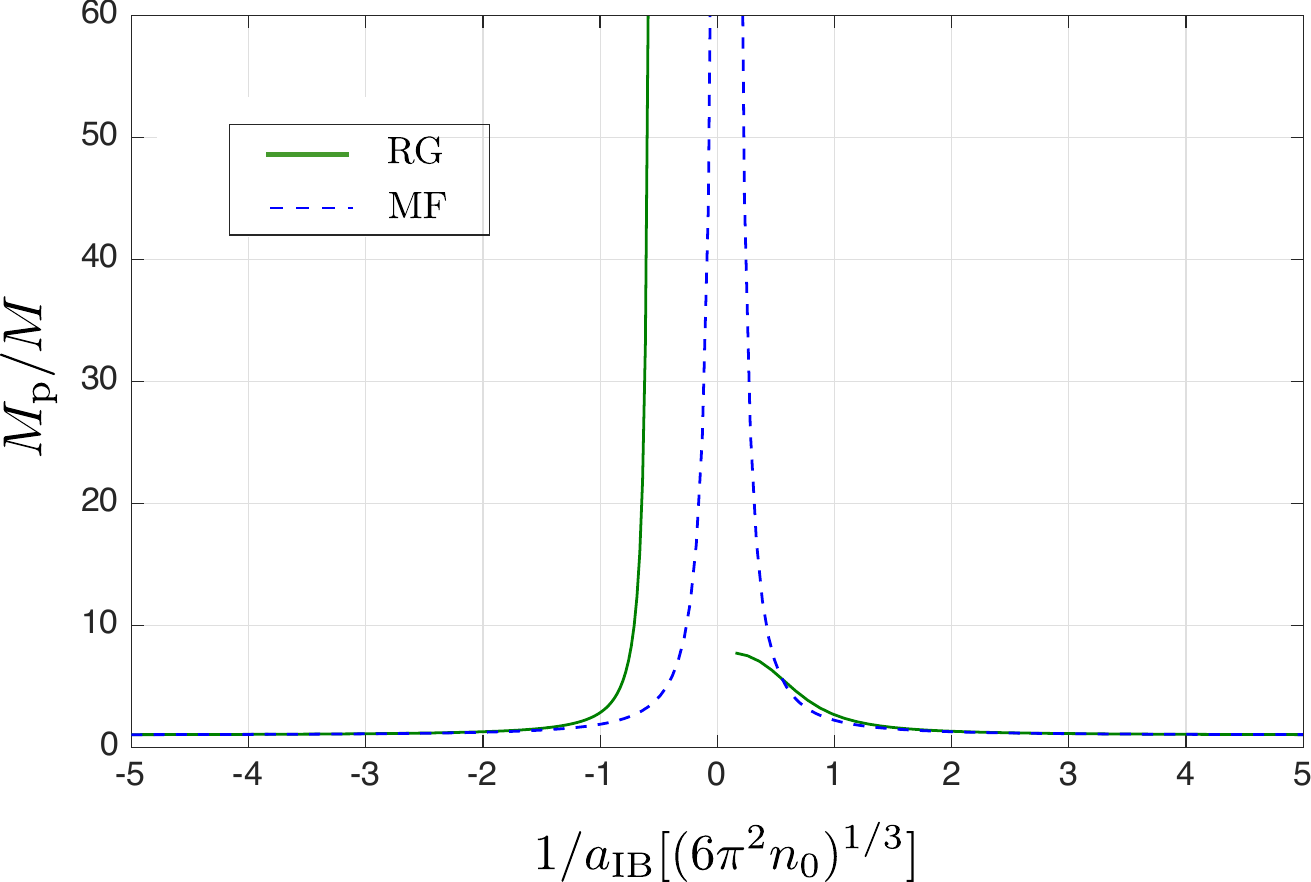, width=0.45\textwidth}
\caption{Our RG calculations predict a divergence of the effective polaron mass $M_{\rm p}$ on the attractive side of the Feshbach resonance. For the RG the renormalized mass $\mathcal{M} \approx M_{\rm p}$ was calculated and for comparison mean-field calculations are also shown. We used the same parameters as in the recent experimental observation of Bose polarons \cite{Jorgensen2016PRL}, see also Fig.~\ref{fig:polaronZ}.}
\label{fig:polaronMass}
\end{figure}

When the Feshbach resonance is approached further from the attractive side, the energy $E_0$ of the attractive polaron starts to decrease dramatically and diverges at a second critical value $a_{\rm IB} = a_{\IB,+}^{\rm RG}$. At the same time the spectral weight in the polaron peak around $E_0$, broadened by the finite lifetime, becomes strongly suppressed as a consequence of the dressing with a diverging number of phonons, see Fig.~\ref{fig:polaronZ}. Meanwhile the incoherent part of the spectrum gains weight. The polaron mass also diverges at $a_{\IB,+}^{\rm RG}$, see Fig.~\ref{fig:polaronMass}. 

These features suggest another transition located at $a_{\IB,+}^{\rm RG}$, where the nature of eigenstates at low energies changes abruptly. This transition is also predicted by MF theory \cite{Shchadilova2016PRL}, albeit at a different interaction strength $a_{\rm IB} = a_{\rm IB,+}^\MF$, where the polaron changes from attractive to repulsive. The shift of the transition to $ 1/a_{\rm IB,+}^{\rm RG} < 1/a_{\rm IB,+}^\MF$ predicted by the RG is due to additional polaronic mass renormalization which is not taken into account by the MF theory.

%%%%%%%%%%%%%%%%%%%%%%%%%%%%%%%%%%%%%%%%%%%%%%%%%%%%%
%%%%%%%%%%%%%%%%%%%%%%%%%%%%%%%%%%%%%%%%%%%%%%%%%%%%%

\emph{(III) Instability of the RG.--}
Between the two critical values predicted by the RG approach and MF theory, i.e. for $1/a_{\rm IB,+}^{\rm RG} < 1/a_{\rm IB} < 1/a_{\rm IB,+}^\MF$, the RG becomes unstable and diverges (region III in Fig.~\ref{IntroSepctrum}). It predicts an infinite number of phonons in the polaron cloud, caused by quantum fluctuations of the mobile impurity which are not included in MF theory. The effective polaron mass is also predicted to be infinite here. The range of scattering lengths within this regime depends on the impurity mass and the instability disappears for a localized impurity. To describe correctly the physics for these scattering lengths, phonon-phonon interactions neglected in our approach should be included. They are required to establish a balance with the attractive interactions between phonons induced by the mobile impurity.

The coherent state analysis of Ref.~\cite{Shchadilova2016PRL} revealed a regime in the spectrum where the eigenstates correspond to mesoscopic bound states, on the repulsive side close to the Feshbach resonance, i.e. for $1/a_{\rm IB} \gtrsim 1/a_{\rm IB,+}$. It was predicted that the weakly bound molecule state existing in this regime can be populated by multiple Bogoliubov phonons at the same time, an effect related to the formation of superpolarons observed for Rydberg atoms in a BEC \cite{Schmidt2016,Schlagmuller2016}. The instability of the RG is located precisely in this regime, $1/a_{\rm IB} \gtrsim 1/a_{\rm IB,+}^{\rm RG}$, which also indicates the presence of mesoscopic bound states. In this case ($\textcircled{4}$ in Fig.~\ref{IntroSepctrum}) we expect a broad and featureless spectrum from the coherent state analysis \cite{Shchadilova2016PRL}, difficult to distinguish qualitatively from the spectrum in the region $\textcircled{3}$ of Fig.~\ref{IntroSepctrum}.

%%%%%%%%%%%%%%%%%%%%%%%%%%%%%%%%%%%%%%%%%%%%%%%%%%%%%
%%%%%%%%%%%%%%%%%%%%%%%%%%%%%%%%%%%%%%%%%%%%%%%%%%%%%

\emph{(IV) Repulsive polarons.--}
On the repulsive side of the Feshbach resonance, for $1/a_{\rm IB} > 1/a_{\rm IB,+}^\MF>0$, we predict a repulsive polaron peak at positive energies $E_0 > 0$ ($\textcircled{5}$ and $\textcircled{6}$ in Fig.~\ref{IntroSepctrum}). Because molecular bound states exist at lower energies (not shown in the spectrum), the polaron peak broadens and acquires a finite lifetime \cite{Rath2013,Li2014, Shchadilova2016PRL}. This effect is particularly pronounced close to the resonance ($\textcircled{5}$ in Fig.~\ref{IntroSepctrum}). 

For weaker interactions, $1/a_{\rm IB} \gg 1/\xi$, the quasiparticle weight $Z$ of the repulsive polaron approaches unity, associated with a corresponding loss of weight in the incoherent part of the spectrum. For sufficiently weak couplings ($\textcircled{6}$ in Fig.~\ref{IntroSepctrum}) the Fr\"ohlich model can be used to describe the repulsive polaron.

\emph{(V) Efimov states.--}
In Refs.~\cite{Levinsen2015,Sun2017} the possibility of Efimov bound states existing in the presence of the BEC has been discussed. Such states can not be captured by our description of the Bose polaron. In the spectral function shown in Fig.~\ref{IntroSepctrum}, they were predicted to appear first on the attractive side \cite{Levinsen2015,Sun2017} and extend all the way to the repulsive side. Although we do not expect dramatic changes of the polaron branch in the spectrum due to Efimov states, we note that they could give rise to additional broadening of the polaron peak. They, too, should become dressed by phonons from the surrounding BEC, but understanding the effect of strong dressing close to the Feshbach resonance remains an open problem.

\emph{Relation to experiments.--} 
In experiments with ultracold atoms \cite{Catani2012,Spethmann2012,Fukuhara2013,Scelle2013,Hohmann2015,Jorgensen2016PRL,Hu2016PRL,Rentrop2016} both polaron branches (attractive and repulsive) can be explored. Recently the first observations of strong coupling Bose polarons in radio-frequency absorption spectra have been reported in two independent experiments \cite{Jorgensen2016PRL,Hu2016PRL}. In Appendix \ref{sec:compExp} we compare the results of our RG calculations to the measured absorption spectra \cite{Jorgensen2016PRL,Hu2016PRL}. We find that our results are consistent with the experimental observations.

To distinguish different theoretical predictions for the polaron properties measurements are required which go beyond the spectral function. For strong couplings we expect the quasiparticle weight $Z$ of the polaron to be small, see Fig.~\ref{fig:polaronZ}, and therefore most of the spectral weight corresponds to excited states. Ultracold atoms also offer the possibility to study dynamics of polarons and explore states which are inaccessible in the spectrum because of their negligibly small quasiparticle residue by preparing them adiabatically. 

For example, by studying the dynamics of polarons oscillating in a harmonic trap, their effective mass can be accurately measured \cite{Catani2012}. In Fig.~\ref{fig:polaronMass} we show our prediction of the effective polaron mass for an impurity in a BEC, for parameters as in the experiment by J\o rgensen et al. \cite{Jorgensen2016PRL}. A similar behavior is obtained for parameters as in the experiment of Hu et al. \cite{Hu2016PRL}. We expect that the large increase of the polaron mass close to the resonance can be observed experimentally, enabling a more detailed comparison of different theoretical descriptions of strongly coupled Bose polarons in a BEC. The effective mass can also be used to study polaronic self-trapping, for a discussion see Appendix \ref{sec:polSelfTrapping}.

\emph{Relation to other theoretical approaches.--}
We emphasize that the goal of our theory is to provide an accurate description of the polaron branch in the spectrum. However the RG formulated for the polaron also provides indirect indications for the existence of bound states, in the form of divergencies of the coupling constants and imaginary parts of the polaron energy \cite{Schmidt2013}. Similar RG flows in the complex plane were introduced for the description of Efimov four-body bound states \cite{Schmidt2010,Moroz2010}. This connects our work to the T-matrix theory of Ref.~\cite{Rath2013} and to variational treatments of the impurity-BEC problem \cite{Li2014,Levinsen2015} where molecular bound states were explicitly included in the analysis. In contrast to these approaches we start from a MF theory of the polaron \cite{Lee1953,BeiBing2009,Shashi2014RF,Shchadilova2016PRL} which allows for an infinite number of phonon excitations (see also Ref.~\cite{Rath2013}). In addition we include correlations between phonons non-perturbatively and solve the resulting RG flow equations fully self-consistently. As a consequence the predictions of our RG method in the strong coupling regime are vastly different from those of truncated basis methods including only a few quasiparticle excitations at a time.

Like in Refs.~\cite{Rath2013,Li2014,Levinsen2015,Jorgensen2016PRL,Shchadilova2016PRL}, our discussion is based on the use of Bogoliubov theory, assuming non-interacting phonons to describe the weakly interacting BEC. We expect that the Bogoliubov approximation is justified for the description of the polaron branch, as long as only virtual phonons are loosely attached to the impurity. On the other hand, the instability of the RG at strong couplings is expected to disappear when phonon-phonon interactions are included as in Refs.~\cite{Ardila2015,blinova2013single}. The properties of the bound states, which are not the main subject of this paper however, may depend more sensitively on residual phonon-phonon interactions. 

An important difference between various theoretical approaches to polaron problems is the predicted dependence of the polaron energy on the large-momentum (UV) cut-off. In Ref.~\cite{Grusdt2015RG} a logarithmic divergence of the polaron energy with the UV cut-off was identified in the Fr\"ohlich model, for which indications were found in diagrammatic quantum Monte Carlo calculations \cite{Vlietinck2015}. Using Gaussian variational wavefunctions, this divergence of the Fr\"ohlich polaron energy was also confirmed \cite{Shchadilova2016}. Here we show analytically that the RG predicts the same logarithmic UV divergence as in the Fr\"ohlich model for the full Bose polaron problem. It can be regularized by introducing a finite range interaction which effectively introduces a UV cut-off in the microscopic Hamiltonian, see also Ref.~\cite{Tempere2009}. This log-divergence is a direct manifestation of quantum fluctuations in the polaron cloud which are not included by other approaches \cite{Rath2013,Li2014,Levinsen2015,Shchadilova2016PRL}. Note that a closely related log-divergence was confirmed in a rigorous higher-order perturbative analysis \cite{Christensen2015}.

%%%%%%%%%%%%%%%%%%%%%%%%%%%%%%%%%%%%%%%%%%%%%%%%%%%%%
\section{Model}
\label{sec:Model}
%%%%%%%%%%%%%%%%%%%%%%%%%%%%%%%%%%%%%%%%%%%%%%%%%%%%%
We consider an ultracold impurity atom of mass $M$ inside a weakly interacting, homogeneous BEC in three-dimensions. The condensate density will be denoted by $n_0$. The interaction of the impurity with the bosons (mass $m_{\rm B}$) in the BEC is modeled by a contact interaction characterized by the scattering length $a_\IB$. The interaction between the bosons, characterized by the scattering length $a_{\rm BB}$, is assumed to be sufficiently weak so that Bogoliubov theory is valid for the description of the BEC and its excitations \cite{Pitaevskii2003,Pethick2008}.

Following Refs.~\cite{Tempere2009,Grusdt2015Varenna,Shchadilova2016PRL} we obtain the following Hamiltonian ($\hbar = 1$):
\begin{widetext}
\begin{multline}
\H_{\rm B} =  g_\IB^{(0)} n_0 + \int^{\Lambda_0} d^3 k ~ \Biggl[  \omega_{k} \ad_{\vec{k}} \a_{\vec{k}}  + V_{k} e^{i \vec{k} \cdot \hat{\vec{x}}} \l \a_{\vec{k}} + \ad_{-\vec{k}} \r  \Biggr]  +  \frac{\hat{\vec{p}}^2}{2 M}  +\\
+ \frac{g_\IB^{(0)}}{(2 \pi)^3} \int^{\Lambda_0} d^3 k ~ d^3 k'~ e^{i (\vec{k}-\vec{k}') \cdot \hat{\vec{x}}} ~ \l \cosh \theta_k \ad_{\vec{k}} - \sinh \theta_k \a_{-\vec{k}} \r \l \cosh \theta_{k'} \a_{\vec{k}'} - \sinh \theta_{k'} \ad_{-\vec{k}'} \r.
\label{eq:BogoPolaron}
\end{multline}
\end{widetext}
While the first line correspond to the Bogoliubov-Fr\"ohlich Hamiltonian \cite{Froehlich1954}, the second line describes processes involving two phonons, where bosons of finite momentum scatter off the impurity. Here we introduced Bogoliubov phonons $\a_{\vec{k}}$ by the relation
\begin{equation}
\d_{\vec{k}} = \cosh \theta_k \a_{\vec{k}} - \sinh \theta_k \ad_{-\vec{k}},
\end{equation}
where $\d_{\vec{k}}$ annihilates a boson at momentum $\vec{k}$. Their dispersion relation is given by $\omega_k = c k \sqrt{1 + k^2 \xi^2/2}$, with $\xi = 1/\sqrt{2 m_{\rm B} g_{\rm BB} n_0}$ and $c = \sqrt{g_{\rm BB} n_0/m_{\rm B}}$ being the healing length and speed of sound in the BEC, respectively. $g_{\rm BB}$ denotes the strength of the delta interactions between the bosons. In this paper we will mostly use $c/\xi$ and $\xi$ as units of energy and length, making the following relation useful: $m_{\rm B} = 1 /( \sqrt{2} c \xi )$. Moreover, the momentum and position operators of the impurity are defined by $\hat{\vec{p}}$ and $\hat{\vec{x}}$, and the Fr\"ohlich type impurity-phonon interaction is described by $V_k = g_\IB^{(0)} \sqrt{n_0} (2 \pi)^{-3/2} W_k$, where
\begin{equation}
W_k = \cosh \theta_k - \sinh \theta_k =\l \frac{k^2 \xi^2}{2 + k^2 \xi^2} \r^{1/4}.
\end{equation}
The definition of $\theta_k$ can be found for example in \cite{Pitaevskii2003,Pethick2008}. Note that $W_k^{-1}= \cosh \theta_k + \sinh \theta_k$.

The microscopic interaction strength $g_\IB^{(0)}$ is related to the scattering length $a_\IB$ through the Lippmann-Schwinger equation. After introducing the UV cut-off $\Lambda_0$ for the regularization of the contact interaction one obtains
\begin{equation}
\frac{1}{g_\IB^{(0)}} = \frac{m_{\rm red}}{2 \pi a_\IB} - \int^{\Lambda_0} \frac{d^3 k}{(2 \pi)^3} \frac{2 m_{\rm red}}{k^2},
\label{eq:gIBaIB}
\end{equation}
where the reduced mass is defined by $m_{\rm red}^{-1} = M^{-1} + m_{\rm B}^{-1}$. Note that the right-hand side depends explicitly on the UV cut-off, and, as will become clear later, for $\Lambda_0 \to \infty$ we need to keep the following terms in the expansion, 
\begin{equation}
g_\IB^{(0)} = - \frac{\pi^2}{m_{\rm red}} \Lambda_0^{-1} - \frac{\pi^3}{2 m_{\rm red} a_\IB} \Lambda_0^{-2}.
\end{equation}
~

The Hamiltonian \eqref{eq:BogoPolaron} can be simplified by eliminating the impurity coordinate $\hat{\vec{x}}$. This is achieved by the Lee-Low-Pines (LLP) canonical transformation \cite{Lee1953}, which leads to the following Hamiltonian in the polaron frame depending explicitly on the conserved momentum $\vec{p}$, 
\begin{widetext}
\begin{multline}
\H_{\rm LLP} = g_\IB^{(0)} n_0 +  \int^{\Lambda_0} d^3 k ~ \Biggl[ \omega_{k} \ad_{\vec{k}} \a_{\vec{k}}  +  V_{k}  \l \a_{\vec{k}} + \ad_{\vec{k}} \r  \Biggr] + \frac{1}{2 M} \l \vec{p} - \int^{\Lambda_0} d^3 k~ \vec{k} \ad_{\vec{k}} \a_{\vec{k}} \r^2 +\\
+ \frac{g_\IB^{(0)}}{(2 \pi)^3} \int^{\Lambda_0} d^3 k ~ d^3 k'~ \ \l \cosh \theta_k \ad_{\vec{k}} - \sinh \theta_k \a_{-\vec{k}} \r \l \cosh \theta_{k'} \a_{\vec{k}'} - \sinh \theta_{k'} \ad_{-\vec{k}'} \r.
\label{eq:HLLPfull}
\end{multline}
\end{widetext}
For more details on the LLP transformation see e.g. the reviews \cite{Devreese2013,Grusdt2015Varenna}. In the following we will mostly be concerned with the spherically symmetric case $\vec{p}=0$, but a generalization of our RG to $\vec{p} \neq 0$ can be found in Appendix \ref{sec:FiniteMomentumRG}.

%%%%%%%%%%%%%%%%%%%%%%%%%%%%%%%%%%%%%%%%%%%%%%%%%%%%%
\section{Toy model: 0d polaron}
\label{sec:ToyModel}
%%%%%%%%%%%%%%%%%%%%%%%%%%%%%%%%%%%%%%%%%%%%%%%%%%%%%

Before we discuss the full model Eq.~\eqref{eq:BogoPolaron}, let us gain some intuitive understanding of the polaron phase diagram. To this end we assume that the impurity is localized at $\vec{x}=0$ and has an infinite mass $M \to \infty$. Moreover we simplify the Hamiltonian by considering only a single $\vec{k}$ mode of Bogoliubov phonons, effectively making the system zero-dimensional. This leads us to the following toy model,
\begin{multline}
\H_{\rm toy} =  g_\IB n_0 + \omega \ad \a + g_\IB \sqrt{n_0} W \l \ad + \a \r + \\
 + \frac{g_\IB}{4} \left[ (W^2 + W^{-2}) \ad \a + (W^2 - W^{-2}) \a \a + \hc \right].
 \label{eq:Htoy}
\end{multline}
We omitted factors of $2 \pi$ and the $\vec{k}$ indices, set $V=g_\IB \sqrt{n_0} W$ and simplified the two-phonon terms from the second line of Eq.~\eqref{eq:HLLPfull}. Moreover we dropped a constant energy shift $g_\IB (W - W^{-1})^2/4$. We are not interested in a physical realization of this model, but rather it serves as a mathematical analog for understanding the different regimes in the spectral function, see Fig.~\ref{IntroSepctrum}.

\emph{MF theory.--} We start by eliminating linear terms in $\a$ from the toy model. This is equivalent to a MF treatment of the problem and can be achieved by finding the saddle point of the classical Hamiltonian corresponding to Eq.~\eqref{eq:Htoy}. Formally we perform a unitary transformation
\begin{equation}
\U_\MF = \exp \l \alpha^\MF \ad - \hc \r,
\end{equation}
and make the choice $\alpha^\MF = - \beta_\MF g_\IB \sqrt{n_0} W / \omega$ with $\beta_\MF = 1 / (1 + g_\IB W^2/\omega)$. As a result we obtain for the Hamiltonian in the new frame:
\begin{multline}
\tilde{\mathcal{H}}_{\rm toy} = \U_\MF^\dagger \H_{\rm toy} \U_\MF =  \beta_\MF g_\IB n_0 + \omega \ad \a +\\
+ \frac{g_\IB}{4} \left[ (W^2 + W^{-2}) \ad \a + (W^2 - W^{-2}) \a \a + \hc \right].
 \label{eq:HtoyTilde}
\end{multline}

Physically the saddle point corresponds to a MF polaron state, with the energy $E_0^\MF = \beta_\MF g_\IB n_0$. This is not necessarily the lowest energy state however. The properties of the polaron, and in particular the spectral function $I(\omega)$, also depend on the states in the vicinity of the MF saddle point, which we study next.

\emph{Spectrum.--} 
To calculate the excitation spectrum of $\tilde{\mathcal{H}}_{\rm toy}$, we introduce the following conjugate variables,
\begin{flalign}
\delta \hat{n} &= \sqrt{n_0} W \l \a + \ad \r, \\  
\hat{\vartheta} &=  \frac{1}{2 i \sqrt{n_0}} W^{-1}  \l \a - \ad \r,
\end{flalign}
with $[\delta \hat{n},  \hat{\vartheta}]=i$, which are the analogues of the position and momentum operator for a harmonic oscillator. In our case of a BEC, they correspond to particle number and phase fluctuations. In terms of these new variables, the Hamiltonian becomes
\begin{equation}
\tilde{\mathcal{H}}_{\rm toy} =  n_0 W^2 A_-  \hat{\vartheta}^2  + \frac{A_+}{4 n_0 W^2} \delta \hat{n}^2
% \beta_\MF g_\IB n_0 + \delta E_0
% $\delta E_0 = \omega/2 + g_\IB (W^2 + W^{-2}) /4$
\label{eq:HtoyQuad}
\end{equation}
up to an overall energy offset, where
\begin{equation}
A_\pm =  \omega + g_\IB W^{\pm 2}.
\end{equation}
Note that particle-number and phase fluctuations are decoupled in Eq.~\eqref{eq:HtoyQuad}.

Now we would like to understand how the eigenenergies of our zero-dimensional toy model Hamiltonian change as the interaction strength $g_\IB$ is varied. As in the original model we assume that $ \omega>0$ is positive and $0 < W \leq 1$. From Eq.~\eqref{eq:HtoyQuad} we recognize three different regimes with different signs of $A_\pm$, which we label by (I), (II) and (IV) in analogy with Fig.~\ref{IntroSepctrum}:
\begin{itemize}
\item[(I)] When $A_\pm > 0 $ are both positive, the polaron corresponds to the true ground state in our toy model. The spectrum is bounded from below and there exists a discrete set of phonon excitations above the polaron energy. 
\item[(II)] When $A_+>0$ and $A_-<0$, the polaron corresponds to a meta-stable saddle point. The spectrum of the toy model is dense and unbounded both from below and from above. The polaron is dynamically unstable in this case, with exponentially many phonons accumulating over time. 
\item[(IV)] When $A_\pm <0$ are both negative, the polaron is the highest-energy state in our toy model. The spectrum is bounded from above and there exists a discrete set of excited states below the polaron energy. They can be understood as analogs of the molecular bound states found in Refs.~\cite{Schmidt2016,Shchadilova2016PRL}. Without additional terms in the Hamiltonian, the polaron is stable because it can not decay into the low-lying eigenstates.
\end{itemize}

%%%%%%%%%%%%%%%%%%%%%%%%%%%%%%%%%%%%%%%%%%%%%%%%%%%%%
\begin{figure}[t!]
\centering
\epsfig{file=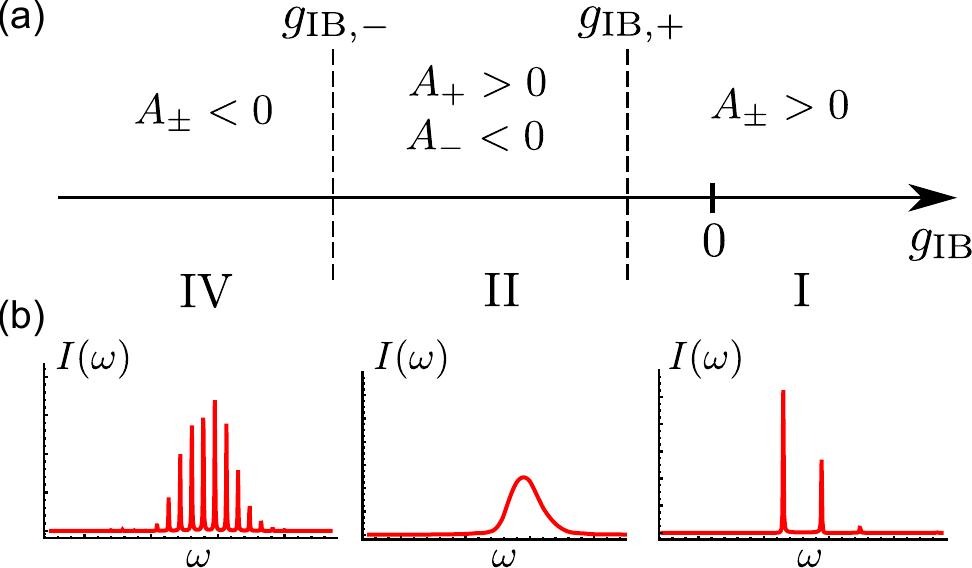, width=0.47\textwidth}
\caption{(a) The phase diagram of the zero-dimensional toy model Eq.~\eqref{eq:Htoy} is shown as a function of the interaction strength $g_{\rm IB}$. Different regimes are distinguished by the properties of the polaron saddle point. (b) The spectral function is calculated for the toy model in the three different regions discussed in the text.}
\label{fig:0dtoyPhaseDiag}
\end{figure}
%%%%%%%%%%%%%%%%%%%%%%%%%%%%%%%%%%%%%%%%%%%%%%%%%%%%%

The transitions between the three different regimes are determined by the conditions $A_\pm = 0$. Therefore the critical interaction strengths are given by
\begin{equation}
g_{\IB,\pm} = - \omega / W^{\pm 2},
\end{equation}
as sketched in Fig.~\ref{fig:0dtoyPhaseDiag} (a). While the polaron is dynamically stable for strong attraction (IV) where discrete bound states can form at lower energies, it is unstable for moderate attraction (II). In that case we expect phase separation because the condensate is depleted in time. This situation is analogous to a quantum particle with negative mass in a harmonic trapping potential, which only has a bounded trajectory if also the curvature of the potential is negative.

In Fig.~\ref{fig:0dtoyPhaseDiag} (b) we calculate the spectral function of the impurity for the toy model. It shows the same characteristic features as expected for the full three-dimensional polaron model, see Fig.~\ref{IntroSepctrum} and Ref.~\cite{Shchadilova2016PRL}. In regions (I) and (IV) we observe a series of equidistant spectral lines, which correspond to multi-particle bound states with different numbers $n$ of phonons \cite{Schmidt2016,Schlagmuller2016,Shchadilova2016PRL}. Their energies $E_n = n \Omega + \beta_\MF g_\IB n_0$ can be determined by diagonalization of the Hamiltonian \eqref{eq:HtoyTilde} using a Bogoliubov transformation; $\tilde{\mathcal{H}}_{\rm toy} =  \Omega ~ \bd \b$ up to an energy shift, and the binding energy is given by $\Omega  = \sqrt{A_+ A_-}$. Note that $\Omega$ is imaginary in the dynamically unstable regime (II).

In the following section we repeat our analysis of the MF polaron saddle point for the full three-dimensional problem in Eq.~\eqref{eq:HLLPfull}. We will show that the same three regimes as identified for the toy model exist in that case. Although we include an infinite number of phonon modes in the higher dimensional case, we will argue that still only a single mode becomes dynamically unstable and acquires negative energies for sufficiently strong attraction. It corresponds to a bound state of phonons to the impurity.

%%%%%%%%%%%%%%%%%%%%%%%%%%%%%%%%%%%%%%%%%%%%%%%%%%%%%
\section{MF theory of impurities in a BEC}
\label{sec:MFreview}
%%%%%%%%%%%%%%%%%%%%%%%%%%%%%%%%%%%%%%%%%%%%%%%%%%%%%
In the polaron frame, i.e. after applying the LLP transformation, a variational MF theory of the polaron can be constructed by assuming a product wavefunction of coherent states  
\begin{equation}
\ket{\psi_{\rm MF}} = \U_\MF \ket{0},
\label{eq:defMFstate}
\end{equation}
where the coherent displacement operator is defined by
\begin{equation}
\U_\MF[\alpha^\MF_{\vec{k}} ] = \exp \l \int^{\Lambda_0} d^3k ~ \alpha^\MF_{\vec{k}} \ad_{\vec{k}} -\hc \r.
\end{equation}
This ansatz wavefunction was originally suggested by Lee, Low and Pines \cite{Lee1953} and has also been applied to the Fr\"ohlich model in a BEC \cite{BeiBing2009,Shashi2014RF}.

The MF theory for the Bose polaron in a BEC, as described by the Hamiltonian \eqref{eq:BogoPolaron}, has been formulated in Ref.~\cite{Shchadilova2016PRL}. Because it represents  the starting point for our RG approach as well as for our analysis of the MF saddle point, we now provide a brief review of the key results relevant to our discussion. Afterwards we study the polaron phase diagram by extending the MF analysis to include molecular bound states around the saddle point.

%%%%%%%%%%%%%%%%%%%%%%%%%%%%%%%%%%%
\subsection{Review of the MF polaron state}
\label{subsec:reviewMFsaddlePoint}
%%%%%%%%%%%%%%%%%%%%%%%%%%%%%%%%%%%
Starting from the MF variational energy $\mathscr{H}[\alpha_{\vec{k}}] = \bra{0} \hat{U}_\MF^\dagger[\alpha_{\vec{k}}]  \H_{\rm LLP} \hat{U}_\MF[\alpha_{\vec{k}}] \ket{0}$, the saddle point equations
\begin{equation}
\frac{\delta \mathscr{H}[\alpha_{\vec{k}}^\MF]}{\delta \alpha_{\vec{k}}^\MF}=0
\label{eq:SaddlePoint}
\end{equation}
can be used to find a local minimum of the polaron energy. Their unique solution yields the coherent amplitudes corresponding to the MF Bose polaron \cite{Shchadilova2016PRL,Shashi2014RF},
\begin{equation}
\alpha_k = - \beta_\MF \frac{V_k}{\Omega_{k}^\MF}, \qquad  \Omega_{k}^\MF=\omega_k+\frac{k^2}{2M},
\end{equation}
where we used spherical symmetry, $\alpha_k\equiv \alpha_{\vec k}$. The scattering amplitude $V_k$ is renormalized by a factor
\begin{equation}
\beta_\MF = \left[ 1 + \frac{g_\IB^{(0)}}{(2 \pi)^3} \int^{\Lambda_0} d^3k ~ \frac{W_{k}^2}{\Omega_{k}^\MF} \right]^{-1},
\label{eq:selfConsEquations}
\end{equation}
which is absent in the Fr\"ohlich model \cite{Shashi2014RF} where $\beta_\MF=1$. The MF variational energy $E_0^\MF=\mathscr{H}[\alpha_{\vec{k}}^\MF] $ is given by
\begin{equation}
E_0^\MF = n_0 \beta_\MF g_\IB^{(0)} + \frac{g_\IB^{(0)}}{(2 \pi)^3} \int^{\Lambda_0} d^3k ~ \sinh^2 \theta_k.
\label{eq:EoMF}
\end{equation}

Note that $g_\IB^{(0)}$ needs to be expressed in terms of the scattering length, see Eq.~\eqref{eq:gIBaIB}, and the second term in Eq.~\eqref{eq:EoMF} vanishes in the limit $\Lambda_0\to\infty$. Therefore we will discard it in our calculations. While $\beta_\MF \to \infty$ is UV divergent when $\Lambda_0 \to \infty$, the product $\beta_\MF g_\IB^{(0)}$ becomes UV convergent after introducing the scattering length $a_\IB$,
\begin{equation}
E_0^\MF = n_0 \beta_\MF g_\IB^{(0)} =n_0 \frac{2 \pi}{m_{\rm red}} \l 1/a_\IB + 1/a_{\IB,+}^\MF \r^{-1},
\label{eq:E0MFres}
\end{equation}
where
\begin{equation}
a_{\IB,+}^\MF  = \pi \left[  2 \Lambda_0 - m_{\rm red}^{-1} \int_0^{\Lambda_0} dk ~ k^2 \frac{W_k^2}{\Omega^\MF_k}  \right]^{-1} \geq 0.
\label{eq:aIBcMF}
\end{equation}
This result is remarkable because it predicts a mere shift of the resonance, $1/a_\IB \to 1/a_\IB + 1/a_{\IB,+}^\MF$, relative to the weak coupling result $n_0 2 \pi a_\IB / m_{\rm red}$.

Sometimes in the calculations below we find it convenient to retain the combination $\beta_\MF g_\IB^{(0)}$ in formal expressions. It should be kept in mind that $\beta_\MF g_\IB^{(0)}$ is UV convergent and can be expressed solely in terms of the scattering length $a_\IB$, see Eq.~\eqref{eq:E0MFres}.

%%%%%%%%%%%%%%%%%%%%%%%%%%%%%%%%%%%
\subsection{Polaron phase diagram and MF theory of molecule formation}
\label{subsec:MFpolaronPhaseDiag}
%%%%%%%%%%%%%%%%%%%%%%%%%%%%%%%%%%%
Now we extend our MF theory and analyze the nature of the saddle point \eqref{eq:SaddlePoint} corresponding to the Bose polaron. To this end we study low-energy excited states around the MF polaron state. In the vicinity of the MF saddle point, $\alpha_{\vec{k}}=\alpha_{\vec{k}}^\MF + \delta \alpha_{\vec{k}}$, the variational energy can be written as a quadratic form 
\begin{equation}
\mathscr{H}[\alpha_{\vec{k}}] =E_0^\MF + \frac{1}{2} \int d^3k d^3k' ~  \delta \uline{\alpha}_{\vec{k}}^\dagger ~ \uuline{H}(\vec{k},\vec{k}')  ~\delta \uline{\alpha}_{\vec{k}'},
\label{eq:QuadForm}
\end{equation}
where $\delta \uline{\alpha}_{\vec{k}} = ( \delta \alpha_{\vec{k}}, \delta \alpha^*_{\vec{k}} )^T$ and the Hessian matrix $\uuline{H}$ has entries $\uuline{H}_{1,1}(\vec{k},\vec{k}') = \delta^2 \mathscr{H}[\alpha_{\vec{k}}] / \delta \alpha_{\vec{k}}\delta \alpha_{\vec{k}'}$, etc.; note that the dimension of $H$ is determined by the system size.

\begin{figure}[t!]
\centering
\epsfig{file=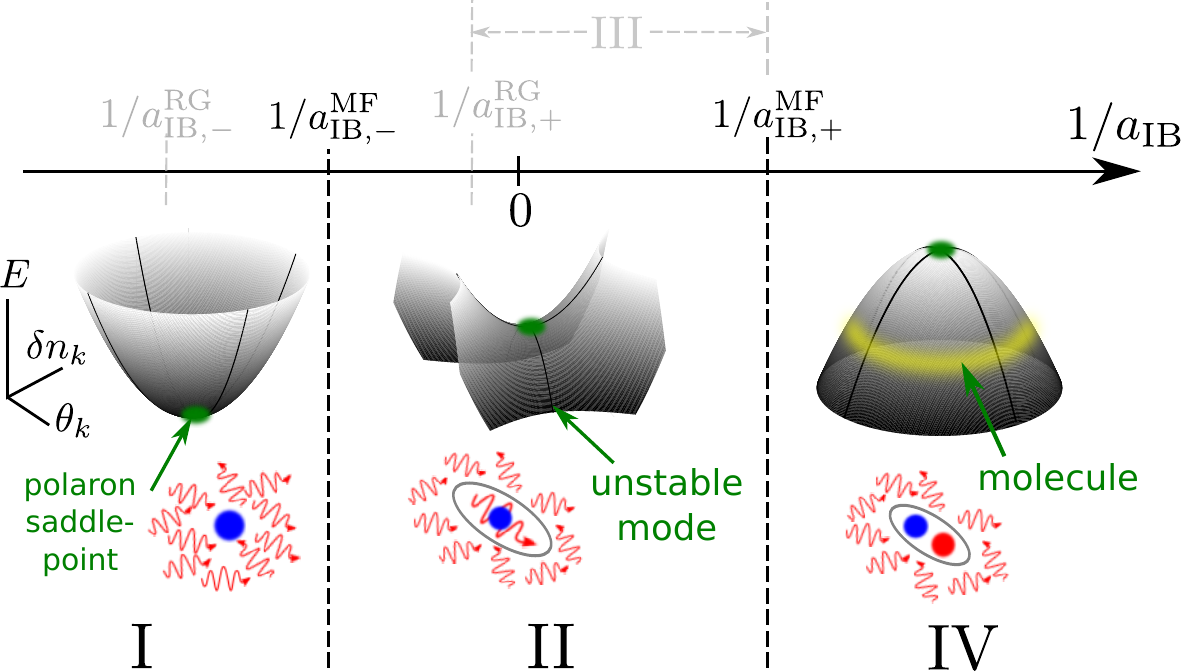, width=0.5\textwidth}
\caption{The phase diagram of an impurity in a BEC obtained from MF theory: On the attractive side the saddle point solution describes a stable polaron where both phase ($\vartheta_k$) and particle-number ($\delta n_k$) fluctuations cost a positive energy $E$ (region I). Beyond $1/a_{\IB,-}^\MF$ an unstable mode appears which is bound to the impurity, where phase fluctuations provide an unstable direction (region II). For $1/a_{\IB} > 1/a_{\IB,+}^\MF$ particle-number fluctuations provide another unstable direction. Together with the phase-fluctuations they form a molecular bound state at energies below the polaron energy $E_0^\MF$ (region IV). For comparison we show how the phase diagram changes when quantum fluctuations are taken into account in the RG (gray).}
\label{fig:polaronMFPhaseDiag}
\end{figure}

When the Hessian is positive definite (I in Fig.~\ref{fig:polaronMFPhaseDiag}), the unique saddle point corresponds to the global minimum of $\mathscr{H}[\alpha_{\vec{k}}]$ and the MF polaron state is a good candidate for the ground state. If only one eigenvalue of $H$ becomes negative (II in Fig.~\ref{fig:polaronMFPhaseDiag}), there exists a dynamically unstable mode \cite{Arnold1989}. In this case the MF polaron state becomes metastable and it is no longer a candidate for the ground state. Finally, if a pair of conjugate variables with negative eigenvalues exists (IV in Fig.~\ref{fig:polaronMFPhaseDiag}), they describe dynamically stable modes at energies below the MF polaron energy $E_0^\MF$. In this case, too, the MF polaron state is metastable and not a ground state candidate. These features are in direct analogy with our toy model from Sec.~\ref{sec:ToyModel}.

%%%%%%%%%%%%%%%%%%%%%%%%%%%%%%%%%%%%%%%%%%%%%%%%%%%%%
%%%%%%%%%%%%%%%%%%%%%%%%%%%%%%%%%%%%%%%%%%%%%%%%%%%%%

%%%%%%%%%%%%%%%%%%%%
\subsubsection{Quadratic Hamiltonian}
We start by expanding the Hamiltonian \eqref{eq:HLLPfull} around the MF saddle point solution. To this end we express the two-phonon scattering processes from the second line of Eq.~\eqref{eq:HLLPfull} in terms of
\begin{flalign}
\delta \hat{n}_{\vec{k}} &= \sqrt{n_0} W_k \l \a_{\vec{k}} + \ad_{-\vec{k}}  \r, \\  
\hat{\vartheta}_{\vec{k}} &=  \frac{1}{2 i \sqrt{n_0}} W_k^{-1}  \l \a_{\vec{k}} - \ad_{-\vec{k}} \r,
\end{flalign}
and obtain an exact expression for $\tilde{\mathcal{H}} =  \U^\dagger_\MF \H_{\rm LLP} \U_\MF$,
\begin{multline}
\tilde{\mathcal{H}} = E_0^\MF+\int^{\Lambda_0} d^3k  d^3k'  ~ \frac{\vec{k} \cdot \vec{k}'}{2 M}  : \G_{\vec{k}} \G_{\vec{k}'} :+\\
+\int^{\Lambda_0} d^3k ~ \Omega^\MF_{k} \left[ \frac{: \delta \hat{n}_{-\vec{k}} \delta \hat{n}_{\vec{k}} :}{4 n_0 W_k^2}  + n_0 W_k^2  :  \delta \hat{\vartheta}_{\vec{k}} \delta \hat{\vartheta}_{\vec{k}'} :\right] + \\
+ \int^{\Lambda_0} d^3k  d^3k' \left[ \frac{G_+}{2n_0}  :\delta \hat{n}_{\vec{k}} \delta \hat{n}_{\vec{k}'}: + 2 n_0 G_- : \hat{\vartheta}_{\vec{k}}  \hat{\vartheta}_{\vec{k}'} : \right].
\label{eq:HuniversalBoundStatesUnified}
\end{multline} 
Here $:...:$ denotes normal ordering of the Bogoliubov phonons, and to simplify notations we introduced the operator
\begin{equation}
\G_{\vec{k}} = \ad_{\vec{k}} \a_{\vec{k}} + \alpha^\MF_{k} \l \ad_{\vec{k}} + \a_{\vec{k}} \r.
\label{eq:defGdefOmega}
\end{equation}
The coupling constants $G_\pm =  \frac{1}{2} \frac{g_\IB^{(0)}}{(2 \pi)^3}$ are equal, although later we find that they acquire different RG flows.

To understand the physical meaning of the new operators $\delta \hat{n}_{\vec{k}}$ and $\hat{\vartheta}_{\vec{k}}$ we note that, to leading order, the boson field operator $\hat{\psi}(\vec{r})$ can be expressed as
\begin{equation}
\hat{\psi}(\vec{r}) \approx \sqrt{n_0 + \delta \hat{n}(\vec{r})} e^{i \hat{\vartheta}(\vec{r})},
\end{equation}
where $\delta \hat{n}(\vec{r})$ ($\hat{\vartheta}(\vec{r})$) is the Fourier transform of $\delta \hat{n}_{\vec{k}}$ ($\hat{\vartheta}_{\vec{k}}$). Therefore they correspond to particle number and phase fluctuations of the BEC, respectively. Note that $\delta \hat{n}(\vec{r})$ and $\hat{\vartheta}(\vec{r})$ represent a pair of conjugate fields, 
\begin{equation}
[\delta \hat{n}_{\vec{k}},\delta \hat{\vartheta}_{-\vec{k}'}] = i \delta(\vec{k} - \vec{k}').
\end{equation}

When we consider only spherically symmetric perturbations $\delta \alpha_{k}$ (we assume $\vec{p}=0$), the non-linear term $\sim 1/M$ in the first line of Eq.~\eqref{eq:HuniversalBoundStatesUnified} vanishes in the expression \eqref{eq:QuadForm} for the variational energy. In this case we end up with a purely quadratic Hamiltonian, from which we can directly read off the Hessian matrix required for our analysis of the MF saddle point. Note that particle-number and phase fluctuations decouple, allowing us to treat these two conjugate variables separately now.

%%%%%%%%%%%%%%%%%%%%
\subsubsection{Phase fluctuations}
First we consider perturbations $\delta \alpha_k = i \beta_k^- \in i \mathbb{R}$ describing only phase fluctuations, $\langle \hat{\vartheta}_{\vec{k}} \rangle =  \beta_k^- W_k^{-1} / \sqrt{n_0} \equiv \vartheta_k$ and $\langle \delta \hat{n}_{\vec{k}} \rangle = 0$. To find out whether the resulting quadratic energy functional $\mathscr{H}[\vartheta_k] = E_0^\MF + \frac{1}{2} \int d^3k d^3k' ~ \vartheta_k H_\vartheta(k,k') \vartheta_{k'}$ is positive definite, we search for a minimum $E_0^\vartheta$ of $\mathscr{H}[\vartheta_k]$ on a hyper-sphere around the MF saddle point defined by the condition 
\begin{equation}
\int d^3k ~ |\delta \alpha_k|^2 \stackrel{!}{=}1.
\label{eq:defHyperSphere}
\end{equation}
$H_\vartheta$ is positive definite only if the minimum energy $E_0^\vartheta > E_0^\MF$ is larger than the MF polaron energy; Otherwise the MF polaron state is metastable. In this case the same analysis needs to be repeated for the conjugate number fluctuations to check of which type the instability is. 

Condition \eqref{eq:defHyperSphere} can be taken into account in the minimization of $\mathscr{H}[\vartheta_k]$ by introducing a Lagrange multiplier $\mu_-$. Because the hyper-sphere is bounded, its global minimum is also a local minimum and we have to solve
\begin{equation}
\frac{\delta}{\delta \vartheta_k} \l \mathscr{H}[\vartheta_k] - \mu_-  n_0 W_k^2 \int d^3k ~ \vartheta_k^2 \r = 0.
\end{equation}
This equation has the solution
\begin{equation}
\beta_{k}^-=-2 G_- \frac{W_k^{-1}}{\Omega_{k}^\MF - \mu_-}  I_- 
\label{eq:solMinus}
\end{equation}
where $I_- = \left[ \int d^3k ~ 4 G_-^2 W_k^{-2} (\Omega_k^\MF - \mu_-)^{-2} \right]^{-1/2}$ is a real-valued normalization constant; The Lagrange multiplier is determined by the following equation,
\begin{equation}
G_-^{-1} = -2 \int^{\Lambda_0} d^3k~ \frac{W_k^{-2}}{\Omega_{k}^\MF - \mu_-},
\label{eq:GminMuTheta}
\end{equation}
and the variational energy at the local extremum is
\begin{equation}
E_0^\vartheta = E_0^\MF + \mu_-.
\label{eq:E0theta}
\end{equation}

Eq.~\eqref{eq:GminMuTheta} only has solutions for $\mu_-<0$, because otherwise the integral on the right hand side diverges. According to Eq.~\eqref{eq:E0theta} all such solutions describe modes with energies below the polaron saddle point energy. Using Eq.~\eqref{eq:GminMuTheta} again we find that solutions $\mu_-<0$ only exist if 
\begin{equation}
G_-^{-1}  > -2 \int^{\Lambda_0} d^3k~ \frac{W_k^{-2}}{\Omega_{k}^\MF},
\end{equation}
which is equivalent to the condition that
\begin{equation}
\frac{1}{a_{\IB}}> \frac{1}{a_{\IB,-}^{\MF}} = \pi \left[  2 \Lambda_0 - m_{\rm red}^{-1} \int_0^{\Lambda_0} dk ~ k^2 \frac{W_k^{-2}}{\Omega_k^\MF}  \right]^{-1}.
\label{eq:defaIBminusMF}
\end{equation}

Therefore at $a_{\IB,-}^{\MF}$ the MF polaron state becomes unstable, giving rise to broadening of the polaron peak in the spectrum as discussed in connection with Fig.~\ref{IntroSepctrum}. In the regime where $1/a_\IB > 1/a_{\IB,-}^{\MF}$ there always exist states with energies arbitrarily deep below that of the MF polaron state, $E_0^\MF$. Because $0 < W_k < 1$, and by comparison with Eq.~\eqref{eq:aIBcMF}, we find that 
\begin{equation}
1/a_{\IB,-}^{\rm MF} < 0 < 1 / a_{\IB,+}^{\MF},
\end{equation}
i.e. the instability is located on the attractive side of the Feshbach resonance.

%%%%%%%%%%%%%%%%%%%%
\subsubsection{Particle-number fluctuations}
Now we repeat our analysis and study the stability of the MF polaron state with respect to particle-number fluctuations. We consider real-valued variations $\delta \alpha_k = \beta_k^+ \in \mathbb{R}$ for which $\langle \delta \hat{n}_{\vec{k}} \rangle =  2 \beta_k^+ W_k \sqrt{n_0} \equiv \delta n_k$ and $\langle \hat{\vartheta}_{\vec{k}} \rangle = 0$. Similar to the case of phase fluctuations we minimize the variational energy $\mathscr{H}[\delta n_k]$ on the hyper-sphere defined by Eq.~\eqref{eq:defHyperSphere}.

After introducing the Lagrange multiplier $\mu_+$ we obtain the solution
\begin{equation}
\beta_{k}^+=-2 G_+ \frac{W_k}{\Omega_{k}^\MF - \mu_+}  I_+,
\label{eq:solPlus}
\end{equation}
where $I_+ = \left[ \int d^3k ~ 4 G_+^2 W_k^2 (\Omega_k^\MF - \mu_+)^{-2} \right]^{-1/2}$ is a normalization constant. The Lagrange multiplier $\mu_+$ is determined by
\begin{equation}
G_+^{-1} = -2 \int^{\Lambda_0} d^3k~ \frac{W_k^2}{\Omega_{k}^\MF - \mu_+},
\label{eq:GpinMuN}
\end{equation}
which only has a solution for $\mu_+ < 0$ and requires
\begin{equation}
1/a_{\IB} > 1 / a_{\IB,+}^{\MF}.
\end{equation}

As in the case of phase fluctuations the variational energy at the local extremum is given by
\begin{equation}
E_0^n = E_0^\MF + \mu_+.
\label{eq:E0n}
\end{equation}
Thus we conclude that the variational energy functional $\mathscr{H}[\delta n_k]$ is positive definite for $1/a_{\IB} < 1 / a_{\IB,+}^{\MF}$, see Fig.~\ref{fig:polaronMFPhaseDiag}. The MF polaron state becomes unstable with respect to particle-number fluctuations only on the repulsive side of the Feshbach resonance.

%%%%%%%%%%%%%%%%%%%%%
\subsubsection{Molecule formation in a BEC}
\label{subsubsec:moleculeFormationInBEC}
Combining our results for particle-number and phase fluctuations around the MF polaron state, we identify three characteristic regimes in the polaron phase diagram, see Fig.~\ref{fig:polaronMFPhaseDiag}. On the attractive side of the Feshbach resonance, I in Fig.~\ref{fig:polaronMFPhaseDiag}, the MF polaron state is stable and represents a good ground state candidate. At $a_{\IB,-}^\MF$ the polaron becomes unstable with regard to phase fluctuations, while the conjugate particle-number fluctuations remain stable. In this regime, II in Fig.~\ref{fig:polaronMFPhaseDiag}, there exists a dynamically unstable mode at energies below the MF polaron energy. 

At $a_{\IB,+}^\MF$, in addition, the polaron becomes unstable with respect to particle-number fluctuations. In this regime, IV in Fig.~\ref{fig:polaronMFPhaseDiag}, the polaron is unstable but there exists a mode at an energy \emph{below} that of the MF polaron state. This state corresponds to a molecular bound state, as can be seen by considering the two-particle case where $n_0 \to  0$. The healing length $\xi \to \infty$ diverges in this limit and we find that 
\begin{equation}
a_{\IB,+}^\MF = a_{\IB,-}^\MF=\infty,
\end{equation}
corresponding to unitarity. Indeed, here a weakly bound Feshbach molecule appears at the continuum threshold \cite{Chin2010} whose wavefunction in momentum space is correctly described by Eqs.\eqref{eq:solMinus}, \eqref{eq:solPlus}.

So far we only discussed the properties of the classical variational energy landscape $\mathscr{H}[\alpha_{\vec{k}}]$ but we ignored that $\hat{\vartheta}_{\vec{k}}$ and $\delta \hat{n}_{\vec{k}}$ are actually non-commuting conjugate variables. In Appendix \ref{subsubsec:MFmoleculeTheory} we take into account quantum fluctuations and construct a quantum theory of the negative energy modes which we have identified above.

%%%%%%%%%%%%%%%%%%%%%%%%%%%%%%%%%%%%%%%%%%%%%%%%%%%%%
\section{Renormalization group theory}
\label{sec:RGtheory}
%%%%%%%%%%%%%%%%%%%%%%%%%%%%%%%%%%%%%%%%%%%%%%%%%%%%%
Within the MF ansatz \eqref{eq:defMFstate} the phonons in the polaron cloud are treated as uncorrelated excitations. The goal of this paper is to remedy this shortcoming. To this end we will go beyond the MF description summarized in the last section and include correlations between phonons in the polaron cloud. Our starting point is the Hamiltonian 
\begin{equation}
\Ht = \U^\dagger_\MF \H_{\rm LLP} \U_\MF
\end{equation}
which describes quantum fluctuations around the MF state. Our method is a generalization of the RG approach developed for the Fr\"ohlich Hamiltonian in Refs.~\cite{Grusdt2015RG,Grusdt2015Varenna,Grusdt2016RG}. Unlike more traditional formulations of the RG \cite{Altland2010}, we use only wavefunctions and unitary transformations and there is no need to employ path integrals. 

Our approach is very similar in spirit to Wegner's flow equation method \cite{Wegner1994}. In both approaches the basic idea is to subsequently diagonalize the Hamiltonian by applying a series of infinitesimal unitary transformations. We make use of the separation of energy scales of phonons at different momenta, i.e. $\omega_{k_1} \gg \omega_{k_2}$ for $k_1 \gg k_2$, to formulate a momentum-shell RG for an effective Hamiltonian $\Ht(\Lambda)$. Here $\Lambda<\Lambda_0$ defines a high momentum cut-off beyond which fluctuations have already been accounted for. While Wegner's approach is more general because it does not rely on the existence of a separation of energy scales, it produces significantly more complicated flow equations (at least for the polaron problem at hand). 

The derivation of the RG flow equations is closely related to the treatments presented in Refs.~\cite{Grusdt2015RG,Grusdt2015Varenna,Grusdt2016RG}. Therefore we will not provide a description of our method in the main text but we delegated an overview as well as detailed calculations to Appendix \ref{apdx:RGderivation}. The remainder of this section is organized as follows. In subsection \ref{subsec:UniversalHamiltonian} we introduce the Hamiltonian used in the RG and explain the meaning of the coupling constants. Their flow equations are presented in part \ref{subsec:RGflowEq}. In subsection \ref{subsec:summaryPhaseDiag} we summarize how the flow of the coupling constants gives rise to the polaron phase diagram. We proceed by a dimensional analysis in subsection \ref{subsec:DimAna}. When resonant three-body collisions are possible the RG has divergencies in the polaron energy which we regularize in subsection \ref{subsec:EnergyReg}. We also discuss physical implications of these divergencies and draw analogies with the well-understood case of two-particle scattering.

%%%%%%%%%%%%%%%%%%%%%%%%%%%%%%%%%%%
\subsection{Universal Hamiltonian and coupling constants}
\label{subsec:UniversalHamiltonian}
%%%%%%%%%%%%%%%%%%%%%%%%%%%%%%%%%%%

The universal form of the Hamiltonian in the spherically symmetric case (i.e. for zero total polaron momentum, $\vec{p}=0$) is given by
\begin{multline}
\tilde{\mathcal{H}}(\Lambda) = E_0 + \int^\Lambda d^3k \biggl\{  \Omega_{k} \ad_{\vec{k}} \a_{\vec{k}} +  \int^\Lambda  d^3k' \frac{\vec{k} \cdot \vec{k}'}{2 \mathcal{M}} : \G_{\vec{k}} \G_{\vec{k}'} :  \\
+ \frac{G_+}{2 n_0} : \delta \hat{n}_{\vec{k}} \delta \hat{n}_{\vec{k}'}:  +  2 n_0 G_-  :\hat{\vartheta}_{\vec{k}} \hat{\vartheta}_{\vec{k}'}:   \biggr\}.
\label{eq:Huniversal}
\end{multline}
The first line corresponds to the Fr\"ohlich polaron case \cite{Grusdt2015RG}. Here $\Omega_{k}$ can be interpreted as the renormalized dispersion of the phonons in the polaron frame. The second term describes phonon-phonon interactions induced by the impurity, where $\mathcal{M}(\Lambda) \geq M$ is the renormalized impurity mass flowing in the RG. We defined the operator $\G_{\vec{k}}$ as in Eq.~\eqref{eq:defGdefOmega}, 
\begin{equation}
\G_{\vec{k}}(\Lambda) = \ad_{\vec{k}} \a_{\vec{k}} + \alpha_{k}(\Lambda) \l \ad_{\vec{k}} + \a_{\vec{k}} \r,
\label{eq:defGdefOmegaRG}
\end{equation}
but replaced the coherent MF amplitude $\alpha_{\vec{k}}^\MF$ by $\alpha_{k}(\Lambda)$ which is also flowing in the RG \cite{Grusdt2016RG}, 
\begin{equation}
\alpha_{k}(\Lambda) = - \beta(\Lambda) \frac{V_k}{\Omega_{k}(\Lambda)}, \qquad \Omega_{k}(\Lambda) = \omega_k+\frac{k^2}{2 \mathcal{M}(\Lambda)}.
\label{eq:defAlphaMFflowing}
\end{equation}
The second line of Eq.~\eqref{eq:Huniversal} describes two-phonon terms beyond the Fr\"ohlich model, see also Eq.~\eqref{eq:HuniversalBoundStatesUnified}.

Summarizing, the coupling constants of the RG which depend on the cut-off $\Lambda$ are the renormalized impurity mass $\mathcal{M}$, the two interaction strengths $G_\pm$ of particle number and phase fluctuations, as well as the renormalized MF amplitude $\beta$. The energy $E_0(\Lambda)$ also flows in the RG procedure and approaches the polaron energy when $\Lambda \to 0$. The associated initial conditions are
\begin{align}
&G_\pm(\Lambda_0) =  \frac{1}{2} \frac{g_\IB^{(0)}}{(2 \pi)^3},  &\mathcal{M}(\Lambda_0) = M  \\
&E_0(\Lambda_0) = E_0^\MF, &\beta(\Lambda_0)=\beta_\MF.
\end{align}

Note that the Hamiltonian $\tilde{\mathcal{H}}(\Lambda)$ in Eq.~\eqref{eq:Huniversal} is exact for $\Lambda=\Lambda_0$, while for smaller cut-offs $\Lambda < \Lambda_0$ it keeps its form within our approximate RG procedure, see Appendix \ref{apdx:RGderivation}. In particular, the coupling constants $G_\pm(\Lambda)$ only depend on $\Lambda$, related to an energy scale, but not on the momenta $\vec{k}$, $\vec{k}'$ of particle-number and phase fluctuations. A more accurate description including also the momentum dependence was provided for the Fr\"ohlich terms of the Hamiltonian in Ref.~\cite{Enss2009}.

%%%%%%%%%%%%%%%%%%%%%%%%%%%%%%%%%%%
\subsection{RG flow equations}
\label{subsec:RGflowEq}
%%%%%%%%%%%%%%%%%%%%%%%%%%%%%%%%%%%
The RG flow equations for the coupling constants are derived in Appendix \ref{apdx:RGderivation} by applying infinitesimal unitary transformations $\hat{U}_\Lambda$ to decouple fast from slow phonons. This is achieved by explicitly constructing $\hat{U}_\Lambda^{-1}$ perturbatively in $1/\Omega_k$. In the spherically symmetric case considered here (i.e. assuming $\vec{p}=0$) we obtain
\begin{flalign}
\frac{\partial \mathcal{M}}{\partial \Lambda} &= - \frac{2}{3} \int_\f d^2k~ k^2 \frac{\alpha_k^2}{\Omega_k} \label{eq:RGflowMSymm},\\
\frac{\partial G_+^{-1}}{\partial \Lambda} &= - 2 \int_\f d^{2}k~ \frac{W_k^2}{\Omega_k},\\
\frac{\partial G_-^{-1}}{\partial \Lambda} &= - 2 \int_\f d^{2}k~ \frac{W_k^{-2}}{\Omega_k}.
\label{eq:RGflowg1Symm}
\end{flalign}
Here $\int_\f d^{2}k = 4 \pi \Lambda^2$ denotes an integral over a fast phonon shell (F) evaluated at the momentum cut-off $k=\Lambda$.

The RG flow equation \eqref{eq:RGflowMSymm} of the mass $\mathcal{M}$ is identical to its counterpart in the Fr\"ohlich model \cite{Grusdt2015RG}. Note however that the MF amplitude $\alpha_k$ is renormalized as a consequence of the two-phonon terms, which modifies the resulting mass as compared to the Fr\"ohlich case.

The flow of the groundstate energy is given by
\begin{multline}
\frac{\partial E_0}{\partial \Lambda} = \frac{1}{2} \frac{ \partial \mathcal{M}^{-1}}{\partial \Lambda} \l \int_\s d^3p ~ \alpha_{p}^2 p^2 \r +\\
 + \int_\s d^3p  \int_\f d^{2}k ~ \frac{1}{\Omega_{k}}  \l W_k W_p G_+ - W_p^{-1} W_k^{-1} G_- \r^2,
 \label{eq:RGflowEnergySymm}
\end{multline}
where $\int_\s d^3p = \int_0^\Lambda dp ~ 4 \pi p^2$ denotes an integral over the slow phonons. The first term on the right hand side is identical to the expression derived for the Fr\"ohlich model and describes the effect of quantum fluctuations of the impurity, see Ref.~\cite{Grusdt2015RG}. As in the case of the effective mass $\mathcal{M}$ the result is renormalized because the MF amplitude $\alpha_k$ is modified by two-phonon terms. The last term in Eq.~\eqref{eq:RGflowEnergySymm} is new and entirely due to two-phonon scattering. Note that the RG flow of $E_0(\Lambda)$ depends on the MF amplitude $\alpha_{p}$ at lower energies, which is determined fully self-consistently in every RG step as described in Ref.~\cite{Grusdt2016RG}.

The flow equation for the coupling constant $\beta$ introduced in Eq.~\eqref{eq:defAlphaMFflowing} above,
\begin{align}
\frac{\partial \beta}{\partial \Lambda} &= - \frac{(2 \pi)^{3/2}}{g_\IB^{(0)} \sqrt{n_0}} \frac{ \partial \mathcal{M}^{-1}}{\partial \Lambda} \frac{3 I^x }{G_+^{-1} + 2 J_1 }, \label{eq:RGflowBeta}\\
J_1 &= \int_\s d^3 p ~ \frac{W_p^2}{\Omega_p}, \\
I^x &= \int_\s d^3 p ~ p_x^2 \alpha_p \frac{W_p}{\Omega_p},
\label{eq:defIxintegral}
\end{align}
can be solved exactly and we obtain the solution
\begin{equation}
 \beta(\Lambda) =\frac{G_+^{-1}(\Lambda_0)}{G_+^{-1}(\Lambda) + 2 J_1(\Lambda)}.
\label{eq:g1FormalSolutionSymm}
\end{equation}

%%%%%%%%%%%%%%%%%%%%%%%%%%%%%%%%%%%
\subsection{Main features of the RG flows}
\label{subsec:summaryPhaseDiag}
\begin{figure}[t!]
\centering
\epsfig{file=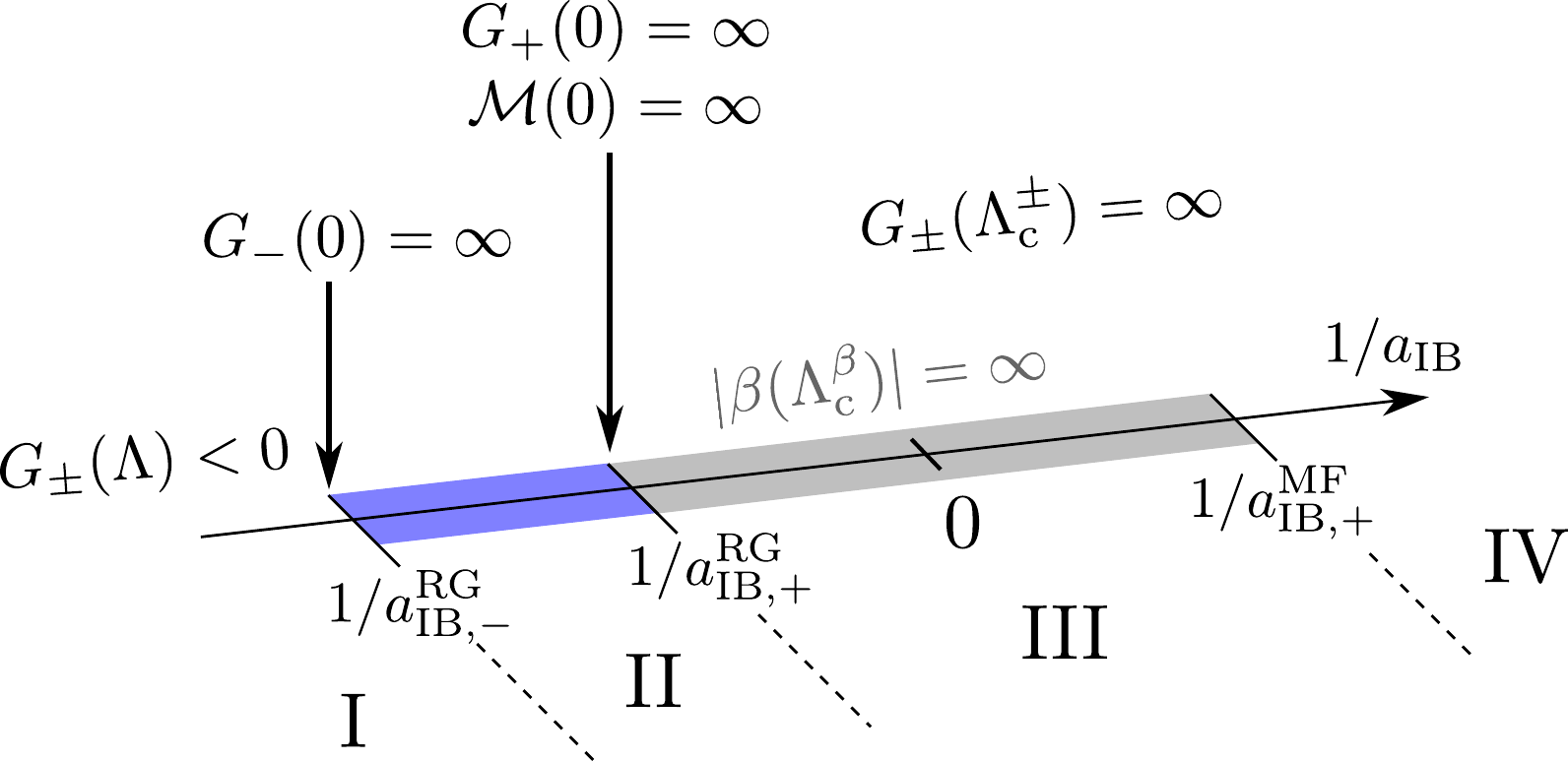, width=0.49\textwidth}
\caption{We summarize the main features of the RG flow which is obtained from analyzing divergencies of the coupling constants $G_\pm(\Lambda)$, $\mathcal{M}(\Lambda)$ and $\beta(\Lambda)$.}
\label{fig:summaryPolaronPhaseDiag}
\end{figure}

%%%%%%%%%%%%%%%%%%%%%%%%%%%%%%%%%%%
In Fig.~\ref{fig:summaryPolaronPhaseDiag} we briefly summarize the main features of the RG flow equations \eqref{eq:RGflowMSymm} - \eqref{eq:RGflowBeta} to simplify the following analysis. To this end we analyze for which cut-offs the coupling constants diverge. The connection between the bound states derived from a MF analysis and these divergencies is explained in Sec.~\ref{subsec:EnergyReg}, and a detailed discussion of the polaron phase diagram is provided in Sec.~\ref{sec:phaseDiag}.

For $1/a_{\rm IB} < 1/a_{\rm IB,-}^{\rm RG}$ (region I in Fig.~\ref{fig:summaryPolaronPhaseDiag}) there are no divergencies. At $a_{\rm IB,-}^{\rm RG}$ the coupling constant $G_-$ diverges in the low energy (IR) limit for $\Lambda \to 0$. Beyond this point, for $1/a_{\rm IB} > 1/a_{\rm IB,-}^{\rm RG}$ (regions II-IV in Fig.~\ref{fig:summaryPolaronPhaseDiag}), this divergence is shifted to finite cut-offs $\Lambda_{\rm c}^->0$. We will argue in subsection \ref{subsec:EnergyReg} that this divergence can be regularized and gives rise to an imaginary contribution to the polaron energy. 

Similarly, at $a_{\rm IB,+}^{\rm RG}$ the coupling constant $G_+$ diverges at $\Lambda =0$. At the same point the effective mass $\mathcal{M}$ diverges for $\Lambda \to 0$. For all $1/a_{\rm IB} > 1/a_{\rm IB,+}^{\rm RG}$ (regions III, IV in Fig.~\ref{fig:summaryPolaronPhaseDiag}) the coupling constants $G_\pm$ diverge at finite cut-offs $\Lambda_{\rm c}^\pm>0$ and both divergencies can be regularized, see \ref{subsec:EnergyReg}.

In the regime between $1/a_{\rm IB,+}^{\rm RG}<1/a_{\rm IB} < 1/a_{\rm IB,+}^{\rm MF}$ (region III in Fig.~\ref{fig:summaryPolaronPhaseDiag}) the coupling constant $\beta(\Lambda)$ diverges at a finite cut-off $\Lambda_{\rm c}^\beta > 0$. This divergence can not be regularized and the RG breaks down in this regime.

%%%%%%%%%%%%%%%%%%%%%%%%%%%%%%%%%%%%%%%%%%%%%%%%%%%%%
%%%%%%%%%%%%%%%%%%%%%%%%%%%%%%%%%%%%%%%%%%%%%%%%%%%%%

%%%%%%%%%%%%%%%%%%%%%%%%%%%%%%%%%%%
\subsection{Dimensional analysis}
\label{subsec:DimAna}
%%%%%%%%%%%%%%%%%%%%%%%%%%%%%%%%%%%
Now we perform a dimensional analysis to understand when the RG is reliable and in order to identify the most relevant terms in the Hamiltonian. We will show that the Fr\"ohlich terms in the Hamiltonian provide a universal description of the polaron branch in a regime not too close to the critical points $a_{\IB,\pm}^{\rm RG}$ where $G_\pm$ diverge in the IR limit, respectively. We keep the discussion in this section general and consider arbitrary dimensionality $d$. For the Hamiltonian \eqref{eq:Huniversal} studied in this paper, $d=3$.

The basic idea is to rescale phonon operators $\a_{\vec{k}} \sim \Lambda^\gamma$ in such a way that the free part $\int^\Lambda d^d k ~ \Omega_{k} \ad_{\vec{k}} \a_{\vec{k}} \simeq \Lambda^0 =1$ of the universal Hamiltonian \eqref{eq:Huniversal} becomes of order unity, for discussion see e.g.\cite{Altland2010}. Then we compare the scaling of the remaining interaction terms with the cut-off $\Lambda$. We distinguish between two regimes of our theory which we treat separately: The high energy regime where $k \gg 1/\xi$ and the low energy regime where $k \ll 1/\xi$.

%%%%%%%%%%%%%%%%%%%%%
\begin{table}[b]
\begin{tabular}{  p{4cm} | p{1.2cm} | p{1.2cm} | p{1.2cm} }
 scaling for $k \ll 1/\xi$  & $d=1$ & $d=2$  & $d=3$ \\[1.2ex]
 \hline \hline
 $\a_{\vec{k}}$  & $\Lambda^{-1}$   & $\Lambda^{-3/2}$   & $\Lambda^{-2}$  \\ [1.2ex]
 \hline
  $\frac{1}{M} \int^\Lambda d^d k  d^d k'  ~ \vec{k} \cdot \vec{k}' :\G_{\vec{k}} \G_{\vec{k}'}: $  & $\bf \Lambda^0$  & $\bf \Lambda^0$ & $\bf \Lambda^0$ \\[1.2ex]
   \hline
 $G_+  \int^\Lambda d^d k  d^d k' ~ : \delta \hat{n}_{\vec{k}} \delta \hat{n}_{\vec{k}'} : $  &$ \Lambda^1$  & $ \Lambda^2$ & $ \Lambda^3$ \\[1.2ex]
  \hline
 $G_-  \int^\Lambda d^d k  d^d k'  ~ : \hat{\vartheta}_{\vec{k}} \hat{\vartheta}_{\vec{k}'} : $  &$\bf \Lambda^{-1}$   & $\bf \Lambda^{0}$  & $ \Lambda^{1}$
\end{tabular}
\caption{Dimensional analysis of quantum fluctuations around the MF polaron state at low energies, $k\ll 1/\xi$. Engineering dimensions were used for deriving this table. Relevant and marginal terms are highlighted.}
\label{tab:dimAnNoDivergenciesIRengin}
\end{table}
%%%%%%%%%%%%%%%%%%%%%

%%%%%%%%%%%%%%%%%%%%%
\begin{table}[t]
\begin{tabular}{  p{4cm} | p{1.2cm} | p{1.2cm} | p{1.2cm} }
 scaling for $k \gg 1/\xi$  & $d=1$ & $d=2$  & $d=3$ \\[1.2ex]
 \hline \hline
 $\a_{\vec{k}}$  & $\Lambda^{-3/2}$ & $\Lambda^{-2}$  &  $\Lambda^{-5/2}$ \\ [1.2ex]
 \hline
  $\frac{1}{M}  \int^\Lambda d^d k  d^d k' ~ \vec{k} \cdot \vec{k}' :\G_{\vec{k}} \G_{\vec{k}'}: $  & $\bf \Lambda^{-3}$ & $\bf \Lambda^{-2}$ & $\bf \Lambda^{-2}$ \\[1.2ex]
   \hline
 $G_+  \int^\Lambda d^d k  d^d k' ~ : \delta \hat{n}_{\vec{k}} \delta \hat{n}_{\vec{k}'} : $  & $\bf \Lambda^{-1}$ &  $\bf \Lambda^{0}$ &  $\Lambda^{1}$  \\[1.2ex]
  \hline
 $G_-  \int^\Lambda d^d k  d^d k' ~ : \hat{\vartheta}_{\vec{k}} \hat{\vartheta}_{\vec{k}'} : $  & $\bf \Lambda^{-1}$ &  $\bf \Lambda^{0}$  &  $\Lambda^{1}$
\end{tabular}
\caption{The same analysis as in Tab.\ref{tab:dimAnNoDivergenciesIR} is shown, but for the high energy regime where $k \gg 1/\xi$.}
\label{tab:dimAnNoDivergenciesUVengin}
\end{table}
%%%%%%%%%%%%%%%%%%%%%

Our results are summarized in Tabs. \ref{tab:dimAnNoDivergenciesIRengin} and \ref{tab:dimAnNoDivergenciesUVengin}, where we used engineering dimensions of the operators. We provide the derivation of these results in Appendix \ref{apdx:dimAna}. Our analysis shows that in $d=3$ dimensions the two-phonon terms $\sim G_\pm$ extending the Fr\"ohlich Hamiltonian are irrelevant. This suggests that the Fr\"ohlich Hamiltonian provides a universal description of the Bose polaron for $d=3$. Quantum fluctuations $\sim \mathcal{M}^{-1} :\G_{\vec{k}} \G_{\vec{k}'}:$ around the Fr\"ohlich MF polaron state become relevant in the regime where $k \gg 1/\xi$. For $k \ll 1/\xi$ they are marginal, and when the polaron mass $\mathcal{M}$ is sufficiently large we expect that MF polaron theory is valid. Therefore, as in the Fr\"ohlich model, the most relevant phonons are those with momenta $k \simeq 1/\xi$.

Our simple analysis above suggests that in $d=3$ dimensions the two-phonon terms can not modify the fixed point of the Fr\"ohlich Hamiltonian where $\mathcal{M} \to \infty$ diverges. Note however that $\G_{\vec{k}}$ depends explicitly on $G_+$, see Eqs.\eqref{eq:defGdefOmegaRG}, \eqref{eq:defAlphaMFflowing} and \eqref{eq:g1FormalSolutionSymm}. Therefore the RG flow of $G_+$ determines the position of the strong coupling fixed point of the Fr\"ohlich Hamiltonian. Indeed, as shown in Fig.~\ref{fig:summaryPolaronPhaseDiag}, the divergencies of $G_+$ and $\mathcal{M}$ are located at the same scattering length $a_{\rm IB,+}^{\rm RG}$. This is not in contradiction with the fact that the $G_+$ term appears to be irrelevant when engineering dimensions are used as in Tabs. \ref{tab:dimAnNoDivergenciesIRengin} and \ref{tab:dimAnNoDivergenciesUVengin}.

In Appendix \ref{apdx:dimAna} we also include the scalings of the coupling constants with the cut-off $\Lambda$ in the analysis, i.e. we consider the anomalous scaling dimensions of the different terms. For $d=3$ dimensions the same conclusions can be drawn as from engineering dimensions, provided that $a_{\rm IB}$ is sufficiently far from $a_{\rm IB,+}^{\rm RG}$ where $G_+$ diverges. However, we also show in Appendix \ref{apdx:dimAna} that the anomalous scaling dimension of $G_+$ changes close to $a_{\rm IB,+}^{\rm RG}$. As a result, two-phonon scattering can become marginal close to $a_{\rm IB,+}^{\rm RG}$. In this regime a competition between two-phonon and Fr\"ohlich type terms in the Hamiltonian is expected.

The second coupling constant $G_-$ does not enter the effective Fr\"ohlich Hamiltonian. Because it is irrelevant according to Tabs. \ref{tab:dimAnNoDivergenciesIRengin} and \ref{tab:dimAnNoDivergenciesUVengin}, we expect that it has only a small effect on the polaron wavefunction. As will be shown later however, the divergence of $G_-$ leads to a change of the ground state of the system. As a result, terms which are irrelevant in the RG can lead to couplings of the polaron to the new ground state, giving rise to a finite lifetime of the polaron and thus to broadening of the spectral function as discussed in the introduction. This strong influence of irrelevant terms in the Hamiltonian on qualitative properties of the polaron state is analogous to the more general concept of dangerously irrelevant operators \cite{Amit1982}.

As in the case of $G_+$, the two-phonon term with strength $G_-$ becomes marginal close to a divergence of $G_-$ when its anomalous scaling dimension is taken into account. For details see Appendix \ref{apdx:dimAna}.

%%%%%%%%%%%%%%%%%%%%%%%%%%%%%%%%%%%
\subsection{Regularization of $G_\pm$ divergencies}
\label{subsec:EnergyReg}
%%%%%%%%%%%%%%%%%%%%%%%%%%%%%%%%%%%
For some scattering lengths the coupling constants $G_\pm \to \infty$ diverge during the RG flow, as summarized in Fig.~\ref{fig:summaryPolaronPhaseDiag}. When $G_\pm(\Lambda)$ diverge in the IR limit for $\Lambda \to 0$, this indicates a transition where the nature of the ground state changes abruptly. We will now examine these divergencies more closely and describe their physical meaning. To this end we relate them to the appearance of modes below the polaron energy which we discussed in Secs.~\ref{sec:ToyModel}, \ref{subsec:MFpolaronPhaseDiag} using MF theory. We will start from the well-understood case of two-particle scattering where $n_0 = 0$.

While the polaron wavefunction flowing in the RG remains well-defined even when $G_\pm$ diverge at some point during the RG, the corresponding polaron energy diverges. We will now show that this divergence of the polaron energy can be regularized by introducing an infinitesimal imaginary part to $G_\pm$. This allows us to make predictions for the polaron properties despite the fact that $G_\pm$ may be divergent during the RG flow.

%%%%%%%%%%%%%%%%%%%%%
\subsubsection{The two-particle case}
\label{sec:GpmDivergencies2particle}
We start by revisiting the two-particle scattering problem where $n_0=0$ and the Fr\"ohlich part of the Hamiltonian becomes trivial. In this case $W_k=1$, $\Omega_k = k^2/2 m_{\rm red}$ and the RG flows of $G_+(\Lambda) = G_-(\Lambda) = g_\IB(\Lambda) / 2 (2 \pi)^3$ coincide. They reproduce the familiar result of the Lippmann-Schwinger equation:
\begin{equation}
g^{-1}_\IB(\Lambda) = \frac{m_{\rm red}}{2 \pi a_\IB} - \int^\Lambda \frac{d^3k}{(2 \pi)^{3}} ~ \frac{2 m_{\rm red}}{k^2}.
\end{equation}

Let us consider $a_\IB > 0$ corresponding to the repulsive side of the resonance, such that $g_\IB(0)>0$. Because $g_\IB^{(0)} < 0$ (for sufficiently large $\Lambda_0$, as considered here), the interaction strength $g_\IB(\Lambda)$ diverges at some cut-off $\Lambda_c \geq 0$ during the RG flow (i.e. $g_\IB^{-1}(\Lambda_c)=0$). Nevertheless the Lippmann-Schwinger equation is exactly reproduced and the divergence of the coupling constants $G_\pm \to \infty$ does not invalidate the RG procedure, despite the fact that the two-particle scattering terms are only treated to lowest order in $G_\pm$ in every RG step.

The divergence of $g_\IB(0) \to \infty$ when $a_\IB \to \infty$ is related to the appearance of a two-particle bound state on the repulsive side of the Feshbach resonance \cite{Chin2010}. Its binding energy can be calculated e.g. from MF theory, see Eq.~\eqref{eq:MFbindingEnergy}, or from a calculation of the full T-matrix using a two-body RG formalism \cite{Mitra1998}. Instead of following these exact approaches, we will now use a simple argument to estimate the bound state energy $E_b$ and connect it to the divergence of the interaction strength $g_\IB(\Lambda)$ in the RG.

When $a_\IB=\infty$ the divergence of $g_\IB(\Lambda)$ is located in the deep IR limit at $\Lambda_c=0$. For $0< a_\IB < \infty$ on the other hand, the divergence of $g_\IB(\Lambda_c) \to \infty$ appears at a finite momentum cut-off $\Lambda_c$, which can be associated with a characteristic energy of the bound state, $\Omega_{\Lambda_c} \approx E_b$. From this equation, and using $g_\IB^{-1}(\Lambda_c)=0$, we obtain the following estimate for the binding energy
\begin{equation}
E_b \approx \frac{\pi^2}{8 m_{\rm red} a_\IB^2}, \qquad \Lambda_c=\frac{\pi}{2} a_\IB^{-1}.
\end{equation}

Up to a numerical factor $\pi^2/4$ this result yields the correct expression for the universal binding energy of the dimer $E_{\rm dim} = 1/2m_{\rm red} a_\IB^2$. We can understand this factor by recalling the simplifications used in the RG and using the exact equation for the dimer energy,
\begin{equation}
1 / g_\IB^{(0)} = - \int^{\Lambda_0} \frac{d^3k}{(2 \pi)^3}~ \frac{1}{\Omega_k + E_{\rm dim}}.
\end{equation}
The RG procedure is perturbative in $1/\Omega_k = 2 m_{\rm red}/k^2$ and at the cut-off $\Lambda_c$ only bosons at momenta $k> \Lambda_c$ are taken into account. If we evaluate the last expression in this regime and use perturbation theory in $\Omega_k^{-1}$ we arrive at the following approximation:
\begin{equation}
m_{\rm red} a_\IB^{-1} \approx - (2 \pi)^{-2} 4 \pi (2 m_{\rm red})^2 \int_{\Lambda_c}^{\Lambda_0} dk ~ \frac{E_{\rm dim}}{k^2}.
\end{equation}
For $\Lambda_0 \to \infty$ this yields the same result as the RG, namely $E_{\rm dim} \approx \pi^2 / 8 m_{\rm red} a_\IB^2$. This indicates that the RG gives the correct physics. 

In the two-particle case, the ground state energy $E_0(\Lambda)$ is not flowing at all. The Fr\"ohlich term in Eq.~\eqref{eq:RGflowEnergySymm} vanishes because $n_0=0$ and the two-particle term vanishes because $G_+ = G_-$ and $W_k=1$. 

%%%%%%%%%%%%%%%%%%%%%
\subsubsection{The many-body case}
\label{sec:GpmDivergenciesManyParticle}
In the many-body case we can also interpret the divergencies of $G_\pm$ as indicators for bound state formation of phonons with the impurity. In contrast to the two-particle case we obtain two separate divergencies associated with particle number ($G_+ \to \infty$) and phase fluctuation scattering ($G_- \to \infty$) respectively. These two divergencies correspond to the instabilities of the MF polaron state with respect to particle-number and phase fluctuations, see our discussion in Sec.~\ref{subsec:MFpolaronPhaseDiag}.

Unlike in the two-particle case, the divergencies of $G_+ \neq G_-$ affect the polaron energy $E_0$. From Eq.~\eqref{eq:RGflowEnergySymm} we can see that corrections due to two-phonon terms diverge when $G_\pm^{-1}(\Lambda_c^\pm) =0$, i.e. $E_0 \to \infty$. Now we introduce a regularization scheme to deal with these divergencies. Their origin are resonant three-particle processes, where bosons recombine with the impurity to form a bound state. Unlike in the two-particle case such processes are explicitly allowed because the total phonon number is no longer conserved in the presence of the condensate. 

To regularize the polaron energy we include an imaginary part in $G_\pm$ as suggested in Refs.~\cite{Schmidt2010,Moroz2010},
\begin{equation}
G_\pm^{-1}(\Lambda) \to G_\pm^{-1}(\Lambda) + i \epsilon, \qquad \epsilon \to 0.
\end{equation}
We start by discussing the case when only $G_-$ diverges but $G_+$ remains finite. 

The contribution to the ground state energy from phase fluctuation scattering on the impurity becomes
\begin{equation}
\Delta E_0 |_{\vartheta} = - (4 \pi)^2 \int_0^{\Lambda_0}  d\Lambda \Lambda^2 \int^\Lambda dp p^2 \frac{W_p^{-2}}{(G_-^{-1} + i \epsilon)^2} \frac{W_\Lambda^{-2}}{\Omega_\Lambda}.
\end{equation}
Now we use $-\partial_\Lambda ( G_-^{-1} + i \epsilon )^{-1} = ( G_-^{-1} + i \epsilon )^{-2} (\partial_\Lambda G_-^{-1})$ for partial integration. In combination with the RG flow equation \eqref{eq:RGflowg1Symm} and the identity $( G_-^{-1} + i \epsilon )^{-1} = \mathcal{P} \frac{1}{G_-^{-1}} - i \pi \delta (G_-^{-1})$ valid for $\epsilon \to 0^+$, we obtain for $G_-^{-1}(\Lambda_c^-)=0$:
\begin{multline}
\Delta E_0 |_{\vartheta} = - i \frac{\pi}{4} \Omega_{\Lambda_c^-} + 2 \pi \int_0^{\Lambda_0} d \Lambda~ \Lambda^2 ~ \mathcal{P} \frac{W_\Lambda^{-2}}{G_-^{-1}} \\
- \frac{G_-(\Lambda_0)}{2} \int^{\Lambda_0} d^3 p ~ W_p^{-2}.
\end{multline}
The principle-value integral in the first line is finite when $G_-$ diverges at $\Lambda_c^-$, and we obtain a well-defined energy. It has an imaginary part which can be interpret as a decay rate into the bound state. In this case we expect a broadened polaron peak in the spectral function, and additional weight appearing below.

When both couplings $G_\pm$ diverge at cut-offs $\Lambda_c^\pm$ during the RG, we can use a similar regularization scheme to obtain a finite energy with an imaginary part.

%%%%%%%%%%%%%%%%%%%%%%%%%%%%%%%%%%%
\subsection{UV log-divergence of the energy}
\label{sec:Divergencies}
%%%%%%%%%%%%%%%%%%%%%%%%%%%%%%%%%%%
Now we study how the polaron energy $E_0$ depends on the UV cut-off $\Lambda_0$. For Fr\"ohlich polarons in $d=3$ dimensions we have shown that $E_0 \simeq - \log(\Lambda_0 \xi)$ diverges logarithmically with the UV cut-off $\Lambda_0$ \cite{Grusdt2015RG}. Because the Fr\"ohlich Hamiltonian provides a universal description of the polaron branch (see Sec.~\ref{subsec:DimAna}) we expect the same log-divergence for the Bose polaron with zero range contact interactions. Indeed we will show by an explicit calculation that the ground state energy diverges logarithmically with the UV cut-off,
\begin{equation}
E_0 ~ \stackrel{\rm UV}{\simeq} ~ - 2^7 \frac{m_{\rm red} n_0^2}{M^2}  \l \frac{1}{a_\IB} + \frac{1}{a_{\IB,+}^\MF} \r^{-4} \log \l \frac{\Lambda_0}{\Lambda_1} \r.
\label{eq:UVlogDivergence}
\end{equation} 
Here $\Lambda_1$ denotes a momentum scale at low energies which is given by $\Lambda_1 \approx \xi^{-1}$ if the total mass renormalization is small, $\mathcal{M}-M \ll M$. If the total mass renormalization is large, $\mathcal{M} \gg M$, it is given by $\Lambda_1 \approx n_0 a_\IB^2 m_{\rm red} / M$.

The UV log-divergence is a direct consequence of the polaron mass renormalization caused by quantum fluctuations of the mobile impurity, for a discussion see Refs.~\cite{Grusdt2015RG,Grusdt2015Varenna}. It was confirmed by diagrammatic Monte Carlo calculations for the Fr\"ohlich model \cite{Vlietinck2015}. To regularize the polaron energy, we suggest taking into account the finite range $r_e \neq 0$ of the impurity-boson interactions. Effectively this leads to a UV cut-off $\Lambda_0 \approx 1/r_e$ which is of the order of $\Lambda_0 \approx 200 / \xi$ (in units of $\xi$) for typical experimental parameters, see Refs. \cite{Tempere2009,Jorgensen2016PRL}. 

We believe that the log-divergence provides an important test-case for theories of the Bose polaron. Simpler approaches like MF theory yield UV convergent polaron energies and thus do not fully include the quantum fluctuations leading to the log-divergence. In the diffusive Monte Carlo calculations of Ref.~\cite{Ardila2015} finite range interactions have been used, but the dependence of the energy on the effective range has not been studied in detail. In the self-consistent T-matrix calculations of Ref.~\cite{Rath2013} zero-range interactions were considered, but the energy was UV convergent. This implies that the quantum fluctuations included in the self-consistent T-matrix approach are different from the ones captured by the RG.

Now we derive the log-divergence of the polaron energy from the Fr\"ohlich term. We show in Appendix \ref{apdx:2PhUVlogDiv} that two-phonon terms only lead to a UV-convergent correction to the polaron energy. In Appendix \ref{apdx:relationLogDivergence} we discuss how the log-divergence is connected to similar logarithmic corrections predicted by different theoretical methods in closely related physical systems.

%%%%%%%%%%%%
\subsubsection{Derivation of the Fr\"ohlich type log-divergence}
We start by solving the RG flow of the mass $\mathcal{M}$ in the UV limit. There, to leading order in $\Lambda^{-1}$, it holds $\Omega_\Lambda \simeq \Lambda^2 / 2 m_{\rm red}$ with the bare reduced mass $m_{\rm red} = 1 / \l m_{\rm B}^{-1} + M^{-1} \r$, and $g_\IB^{(0)} \beta(\Lambda) \simeq  g_\IB^{(0)} \beta_\MF$. Hence
\begin{equation}
\frac{\partial \mathcal{M}}{\partial \Lambda} = -\frac{8}{\pi^2} n_0 m^3_{\rm red} \l \beta_\MF g_\IB^{(0)} \r^2 \Lambda^{-2}.
\end{equation}
Integrating this separable differential equation yields the UV behavior,
\begin{equation}
\mathcal{M} = M +  \frac{8}{\pi^2} n_0 m^3_{\rm red} \l \beta_\MF g_\IB^{(0)} \r^2  \l  \frac{1}{\Lambda} - \frac{1}{\Lambda_0} \r,
\label{eq:MUVsol}
\end{equation}
showing that deviations of $\mathcal{M}$ from $M$ correspond to higher order terms in $\Lambda^{-1}$.

In the UV limit we further obtain,
\begin{equation}
 \int_\s^\Lambda d^3p ~  \alpha_{\vec{p}}^2 p^2 ~ \stackrel{\rm UV}{\simeq} ~ \frac{2}{\pi^2} m_{\rm red}^2 n_0 \l \beta_\MF g_\IB^{(0)} \r^2 \Lambda,
\end{equation} 
and upon integration we find that the first term in Eq.~\eqref{eq:RGflowEnergySymm} leads to the UV divergence \eqref{eq:UVlogDivergence}. The only difference to the result previously encountered in the Fr\"ohlich model is the renormalization of the pre-factor $g_\IB^{(0)} \to \beta_\MF g_\IB^{(0)}$ because we expand around the MF solution of the full model.

To understand how the log-divergence is cut-off in the IR limit at $\Lambda_1$, we consider two separate cases. When impurity-boson interactions are sufficiently weak the mass renormalization is small for all cut-offs, $\mathcal{M}(\Lambda)-M \ll M$. Therefore $\partial_\Lambda \mathcal{M}^{-1} \approx -M^{-2} \partial_\Lambda \mathcal{M} \simeq \Lambda^{-2}$ until the RG flow of $\mathcal{M}^{-1}$ stops around $\Lambda_1=1/\xi$. This cuts off the log-divergence at 
\begin{equation}
\Lambda_1 = 1/\xi \qquad  \text{for} \qquad \mathcal{M} \approx M.
\end{equation}
In the second case impurity-boson interactions are so strong that the mass renormalization becomes dramatic, $\mathcal{M} \gg M$. This happens when $\mathcal{M}(\Lambda_1) - M \approx M$, which yields
\begin{equation}
\Lambda_1 = \frac{m_{\rm red}}{M} \frac{32 n_0}{ \l \frac{1}{a_\IB} + \frac{1}{a_{\IB,+}^\MF} \r^{2}} \qquad  \text{for} \quad \mathcal{M} \gg M,
\end{equation}
i.e. $\Lambda_1 \sim  n_0 a_\IB^2 m_{\rm red}/M$, using Eq.~\eqref{eq:MUVsol} for large $\Lambda_0 \to \infty$. For smaller cut-offs $\Lambda < \Lambda_1$ we find $\partial_\Lambda \mathcal{M}^{-1} = - \mathcal{M}^{-2} \partial_\Lambda \mathcal{M} \simeq \Lambda^2 \Lambda^{-2} = 1$. Hence the logarithmic flow of the ground state energy ends, $\partial_\Lambda E_0 \simeq \Lambda$ for $1/\xi \leq \Lambda \lesssim \Lambda_1$, see Eq.~\eqref{eq:RGflowEnergySymm}.

%%%%%%%%%%%%%%%%%%%%%%%%%%%%%%%%%%%
\section{Results: Polaron energies}
\label{subsec:PolaronEnergy}
%%%%%%%%%%%%%%%%%%%%%%%%%%%%%%%%%%%
In this section we use the RG approach for the calculation of the polaron energy $E_0$ using Eq.~\eqref{eq:RGflowEnergySymm}. In the regime I, where $1/a_\IB < 1/a_{\IB,-}^{\rm RG}$ and before $G_\pm \to \infty$ diverge, we expect no low lying states in the spectrum below $E_0$. Within our theory the polaron energy corresponds to the ground state energy in this regime. Note however that we did not include Efimov states \cite{Levinsen2015,Sun2017} which may have even lower energies in the case of short-range interactions \cite{Braaten2006}. On the other hand, when $G_\pm(\Lambda) \to \infty$ diverge during the RG, we expect additional states at lower energies (see also Ref.~\cite{Shchadilova2016PRL}) and $E_0$ corresponds to the energy of the polaron branch in the spectrum. In this regime we make use of the regularization scheme for the polaron energy discussed in Sec.~\ref{subsec:EnergyReg}.

\begin{figure}[t!]
\centering
\epsfig{file=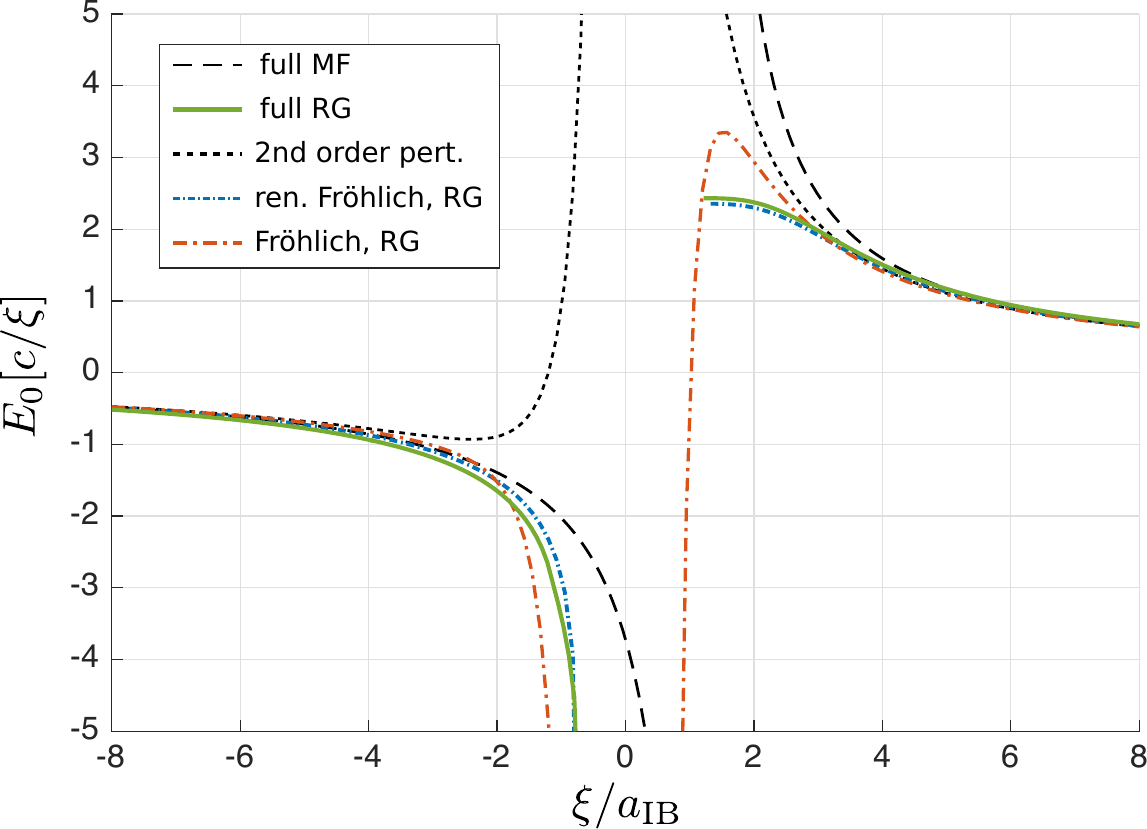, width=0.48\textwidth}
\caption{The polaron energy $E_0$ is calculated using different approaches. It is dominated by the Fr\"ohlich part of the Hamiltonian. We used the same parameters as in Fig.~2 of Ref.~\cite{Rath2013}, i.e. $M/m_{\rm B}=1$ and $n_0=0.25 \xi^{-3}$, but for a UV cut-off $\Lambda_0=10^3/\xi$.}
\label{fig:polaronEnergy}
\end{figure}

In Fig.~\ref{fig:polaronEnergy} we show the polaron energy as a function of $a_\IB^{-1}$ (solid line) and find an attractive and a repulsive polaron branch. Close to the two-particle Feshbach resonance (at $a_\IB^{-1}=0$) we predict large deviations from MF theory due to quantum fluctuations. 

%%%%%%%%%%%%%%%%%%%%%%%%%%%%%%%%%%%%%%%%%%%%%%%%%%%%%
%%%%%%%%%%%%%%%%%%%%%%%%%%%%%%%%%%%%%%%%%%%%%%%%%%%%%

On the repulsive side of the resonance, the polaron energy saturates before the strong-coupling phase is reached where the RG breaks down. Similar saturation of the polaron properties in this regime has been observed in Figs.~\ref{fig:polaronZ}, \ref{fig:polaronMass}. Such behavior would be expected for bubble polarons which have been predicted in this regime by Gross-Pitaevskii MF calculations \cite{blinova2013single}. 

We have argued by a dimensional analysis in \ref{subsec:DimAna} that the Fr\"ohlich terms in the Hamiltonian are most relevant in the RG. Sufficiently far from divergencies of $G_\pm$ we expect that the RG flows towards an effective Fr\"ohlich Hamiltonian with renormalized parameters. To demonstrate this, we considered a renormalized Fr\"ohlich model where we ignored the additional contributions of two-phonon terms in the energy, i.e. the second line of Eq.~\eqref{eq:RGflowEnergySymm}. As can be seen in Fig.~\ref{fig:polaronEnergy} the polaron energy predicted by this renormalized Fr\"ohlich model is in excellent agreement with the full calculation based on Eq.~\eqref{eq:RGflowEnergySymm} for the extended Fr\"ohlich model.

To understand how much the Fr\"ohlich model is renormalized by two-phonon scattering, we also compare our results to the original Fr\"ohlich model introduced in Ref.~\cite{Tempere2009}, for which we developed the RG in Ref.~\cite{Grusdt2015RG}. In this case two-phonon terms are completely neglected. On the attractive side we find reasonable quantiative agreement with the full calculation, and qualitatively the correct behavior is predicted. On the repulsive side the original Fr\"ohlich model does not break down at strong couplings, while the curvature of $E_0(a_\IB^{-1})$ is correctly captured. For sufficiently weak repulsion the agreement is reasonable.

In Fig.~\ref{fig:polaronEnergy} we also compare our results to second order perturbation theory in $a_\IB$ (dotted line). To this order the original Fr\"ohlich model is exact, but it was shown in Ref.~\cite{Christensen2015} that it misses higher order terms. This has caused serious concerns that the Fr\"ohlich Hamiltonian can not be used to make predictions beyond those of second order perturbation theory. In Ref.~\cite{Bruderer2007} a condition was derived when two-phonon terms can be safely neglected, see also Ref.~\cite{Grusdt2015Varenna}. For the parameters used in Fig.~\ref{fig:polaronEnergy} it becomes
\begin{equation}
|a_\IB|^{-1} \gg \pi \sqrt{2} \xi^{-1} = 4.44 \xi^{-1}.
\end{equation}
In this regime all theories basically coincide, and beyond it the perturbative result deviates substantially, in particular on the attractive side. Remarkably, the original Fr\"ohlich model provides a reasonable description of the polaron energy derived from the full model even beyond the careful estimate above.

%%%%%%%%%%%%%%%%%%%%%%%%%%%%%%%%%%%%%%%%%%%%%%%%%%%%%
\section{Polaron phase diagram}
\label{sec:phaseDiag}
%%%%%%%%%%%%%%%%%%%%%%%%%%%%%%%%%%%%%%%%%%%%%%%%%%%%%
Now we discuss the phase diagram predicted by the RG, including quantum fluctuations around the MF saddle point solution. Their most dramatic effect is to cause divergencies of the interaction strengths $G_\pm$ which we related to the appearance of new states below the polaron energy in Secs. \ref{subsec:MFpolaronPhaseDiag}, \ref{subsec:EnergyReg}. This gives rise to the two critical values $a_{\IB,\pm}$ where the features in the spectral function change as discussed in Sec.~\ref{sec:SummaryResults}.

\begin{figure}[t!]
\centering
\epsfig{file=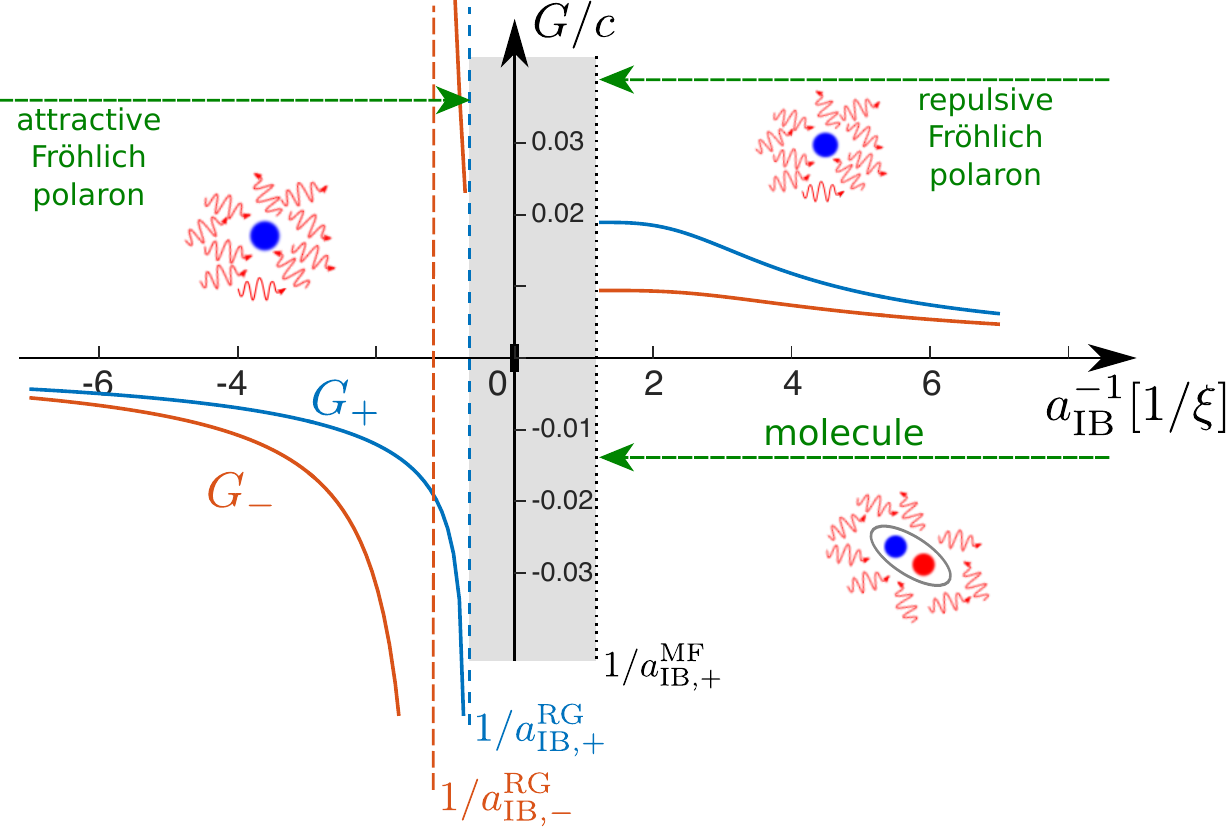, width=0.5\textwidth}
\caption{The polaron phase diagram can be obtained from the coupling constants $G_\pm$, as described in the text. We plot their fully converged values in the IR limit as a function of the inverse two-particle scattering length $a_\IB$. Sufficiently far on the attractive side neither of the two coupling constants diverges in the RG and the ground state is an attractive Fr\"ohlich type polaron. At $a_{\IB,-}^{\rm RG}$ the coupling $G_-$ associated with phase-fluctuations diverges and an unstable mode appears at energies below the polaron energy. At $a_{\IB,+}^{\rm RG}$ the coupling $G_+$ associated with particle-number fluctuations also diverges and molecular bound states appears below the polaron energy. In the shaded region the RG breaks down. On the repulsive side for $1/a_{\rm IB}>1/a_{\IB,+}^{\rm MF}$ we obtain a repulsive Fr\"ohlich polaron which can decay into the molecular states. We used the same parameters as in Fig.~\ref{fig:polaronEnergy}, cf. Ref.~\cite{Rath2013}.}
\label{fig:polaronPhaseDiag}
\end{figure}

In Fig.~\ref{fig:polaronPhaseDiag} we calculate the IR values of the coupling constants $G_\pm(\Lambda=0)$ as a function of the inverse two-particle scattering length $1/a_\IB$. Let us start our discussion far on the attractive side, $1/a_\IB \ll -1/\xi$, where the ground state is an attractive Fr\"ohlich polaron \cite{Tempere2009,Rath2013,Shashi2014RF}. Here $G_\pm(\Lambda) < 0$ during the entire RG flow (note that $G_\pm(\Lambda_0)<0$ is always attractive in the UV).

%%%%%%%%%%%%%%%%%%%%%%%%%%%%%%%%%%%%%%%%%%%%%%%%%%%%%
%%%%%%%%%%%%%%%%%%%%%%%%%%%%%%%%%%%%%%%%%%%%%%%%%%%%%

%%%%%%%%%%%%%%%%%%%%%%%%%%%%%%%%%%%
\subsection{Phase fluctuations}
\label{subsubsec:ThetaRes}
%%%%%%%%%%%%%%%%%%%%%%%%%%%%%%%%%%%
When we approach the Feshbach resonance the interaction strength $G_-$ associated with phase-fluctuation diverges first, on the attractive side of the two-particle resonance. Note that because $W_k<1$ it always holds $G_-<G_+$. From the exact solution of the RG flow equation,
\begin{equation}
G_-^{-1}(\Lambda) = - 8 \pi  \l  m_{\rm red} (2 \Lambda_0 - \pi a_\IB^{-1}) -  \int_\Lambda^{\Lambda_0}d \tilde{\Lambda}~ \tilde{\Lambda}^2 \frac{W_{\tilde{\Lambda}}^{-2}}{\Omega_{\tilde{\Lambda}}} \r,
\end{equation}
we obtain the divergence $G_-^{-1}(0)=0$ for $a_\IB=a_{\IB,-}^{\rm RG}$, with
\begin{equation}
a_{\IB,-}^{\rm RG} = \pi \left[  2 \Lambda_0 - m_{\rm red}^{-1} \int_0^{\Lambda_0} d\Lambda ~ \Lambda^2 \frac{W_\Lambda^{-2}}{\Omega_\Lambda}  \right]^{-1} \leq 0.
\label{eq:defAIBminusRG}
\end{equation}
The last inequality follows from $W_\Lambda \leq 1$.

Eq.~\eqref{eq:defAIBminusRG} is the analogue of $a_{\IB,-}^\MF$, see Eq.~\eqref{eq:defaIBminusMF}, corrected for the effects of quantum fluctuations included in the RG. Because $\mathcal{M} \geq M$ it follows that
\begin{equation}
1/a_{\rm IB,-}^{\rm RG}  \leq 1/a_{\IB,-}^\MF.
\end{equation}

For $a_\IB^{-1} > 1/a_{\IB,-}^{\rm RG}$ there exists a divergence of $G_-(\Lambda_c^-) \to \infty$ in the RG at some cut-off $\Lambda_c^-$, because $G_-(0)>0$ and $\partial_\Lambda G_-^{-1}(\Lambda)<0$. A divergence of $G_-$ indicates the existence of new states below the polaron energy for $a_\IB^{-1} \geq 1/a_{\IB,-}^{\rm RG}$, see Sec.~\ref{sec:GpmDivergenciesManyParticle}. Because $G_-$ describes scattering of a phase fluctuation on the impurity, they contain predominantly phase fluctuations. This is confirmed by the mean-field theory presented in Sec.~\ref{subsec:MFpolaronPhaseDiag}. Because the impurity atom can only couple to particle number fluctuations due to the underlying density-density interactions, the polaron state is not much affected by the divergence of $G_-$.

%%%%%%%%%%%%%%%%%%%%%%%%%%%%%%%%%%%
\subsection{Number fluctuations}
\label{subsubsec:NRes}
%%%%%%%%%%%%%%%%%%%%%%%%%%%%%%%%%%%
For larger $a_\IB^{-1}$ eventually also the interaction strength $G_+ \to \infty$ diverges at $a_\IB = a_{\IB,+}^{\rm RG}$. From the exact solution of the RG flow of $G_+^{-1}(\Lambda)$ we obtain
\begin{equation}
a_{\IB,+}^{\rm RG} = \pi \left[  2 \Lambda_0 - m_{\rm red}^{-1} \int_0^{\Lambda_0} d\Lambda ~ \Lambda^2 \frac{W_\Lambda^2}{\Omega_\Lambda}  \right]^{-1}.
\end{equation}
This expression is closely related to the critical scattering length $a_{\IB,+}^\MF$ where the MF polaron energy diverges, see Eq.~\eqref{eq:aIBcMF}. The only difference is the appearance of the renormalized mass $\mathcal{M}$ (instead of $M$) in the dispersion relation $\Omega_\Lambda$, because of quantum fluctuations included in the RG. 

In this case we find that $a_{\IB,+}^{\rm RG}$ can become either positive or negative, but obeys the following inequality:
\begin{equation}
\frac{1}{a_{\IB,-}^{\rm RG}} \leq \frac{1}{a_{\IB,+}^{\rm RG}} \leq \frac{1}{a_{\IB,+}^{\MF}}. 
\end{equation}
$a_{\IB,+}^{\rm RG}<0$ can become negative only when the mass renormalization is large, $\mathcal{M} \gg M$.

As in the case of $G_-$, the divergence of $G_+$ indicates the existence of states below the polaron energy for $a_\IB^{-1} \geq 1/a_{\IB,+}^{\rm RG}$. In this regime $G_+(\Lambda)$ diverges at some cut-off $\Lambda_c^+$ during the RG. Because $G_+$ describes the interaction of a particle number fluctuation with the impurity, we expect states bound to the impurity which contain number fluctuations. As discussed in Sec.~\ref{subsec:MFpolaronPhaseDiag}, in this regime molecular states form together with the phase fluctuations bound to the impurity.

The shift of the particle-number scattering resonance from $a_\IB^{-1}=0$ in the two-particle case to $a_{\IB,+}^{\rm RG}$ in the many-body case has a simple physical interpretation. Let us first consider an infinitely heavy impurity $M=\infty$, where $a_{\IB,+}^{\rm RG}=a_{\IB,+}^{\MF}$. In the two-particle case the coupling constants $G_+ = G_-$ diverge in the IR when the energy of the two-particle bound state, which is universally given by $1/2 m_{\rm red} a_\IB^2$, reaches the continuum threshold. This happens for $a_\IB \to \infty$. In the many-body case the molecular bound state acquires a density-induced shift to higher energies by an amount $\mu_{\rm BEC} = g_{\rm BB} n_0$ due to repulsive boson-boson interactions. Therefore, to first order, it reaches the two-particle scattering continuum when $g_{\rm BB} n_0 - 1/2 m_{\rm red} a_\IB^2 = 0$. From this simple argument we expect a shift of the resonance to $a_{\IB,c}^{-1} = \xi^{-1} > 0$. On the other hand, the MF result for $M = \infty$ is $1/a_{\IB,+}^{\MF} = \sqrt{2} \xi^{-1}$, which has the same scaling with the coherence length $\xi$.

When the impurity has a finite mass, it becomes increasingly heavy during the RG flow and its effective mass should be replaced by $\mathcal{M} > M$. Therefore the universal binding energy of the dimer $\sim 1/\mathcal{M} a_\IB^2$ decreases in comparison to the two-particle case. This shifts the energy of the bound state upwards and therefore the bound state reaches the scattering continuum at smaller values of $1/a_\IB$, explaining why $1/a_{\IB,\pm}^{\rm RG}<1/a_{\IB,\pm}^{\MF}$.

%%%%%%%%%%%%%%%%%%%%%%%%%%%%%%%%%%%
\subsection{Instability of the RG}
\label{subsubsec:InstabilityRG}
%%%%%%%%%%%%%%%%%%%%%%%%%%%%%%%%%%%
When $|g_\IB^{(0)} \beta(\Lambda)| \to \infty$ in the RG, the MF amplitude $|\alpha_{\vec{k}}| \to \infty$ diverges and the RG breaks down. In this regime quantum fluctuations lead to a collapse of all bosons onto the impurity, and MF theory is no longer justified. We expect that this strong-coupling regime is an artifact of the Bogoliubov approximation because an infinite number of phonons can only be sustained in the polaron cloud if their mutual interactions can be neglected.

We find that the RG breaks down in the extended regime (III in Fig.~\ref{IntroSepctrum})
\begin{equation}
1/a_{\IB,+}^{\rm RG} < 1/a_\IB < 1/a_{\IB,+}^\MF
\label{eq:SCregimeRG}
\end{equation}
where it runs into the strong coupling phase described above. To understand this, we first note that $\partial_\Lambda g_\IB^{(0)} \beta(\Lambda) > 0$, i.e. the impurity-boson coupling becomes more attractive during the RG. Next recall the initial condition for the RG, $g_\IB^{(0)} \beta(\Lambda_0) = E_0^\MF(a_\IB) / n_0$. Therefore on the attractive side of the MF resonance, for $1/a_\IB < 1/a_{\IB,+}^\MF$ where $E_0^\MF<0$, we see that $g_\IB^{(0)} \beta(\Lambda)$ becomes increasingly more negative during the RG. In fact we can show using Eq.~\eqref{eq:g1FormalSolutionSymm} that already for $a_\IB = a_{\IB,+}^{\rm RG}$ the strong-coupling fixed point is reached where $g_\IB^{(0)} \beta(\Lambda=0) = -\infty$ diverges in the IR. In the entire regime \eqref{eq:SCregimeRG} $g_\IB^{(0)} \beta(\Lambda)$ always diverges during the RG flow. Only for $1/a_\IB > 1/a_{\IB,+}^\MF$, where $g_\IB^{(0)} \beta(\Lambda_0)$ starts out positive, a finite value can be reached in the IR limit.

%%%%%%%%%%%%%%%%%%%%%%%%%%%%%%%%%%%%%%%%%%%%%%%%%%%%%
\section{Discussion}
\label{sec:summary}
%%%%%%%%%%%%%%%%%%%%%%%%%%%%%%%%%%%%%%%%%%%%%%%%%%%%%
We have investigated the problem of a mobile impurity interacting with a surrounding BEC in the presence of an inter-species Feshbach resonance. Our theoretical analysis is based on a renormalization group approach to go beyond mean-field calculations and describe quantum fluctuations around the polaron branch non-perturbatively. This gives rise to a logarithmic correction of the polaron energy and we predict new qualitative features appearing in the spectral function close to the Feshbach resonance.

Already on the attractive side, beyond a critical interaction strength $a_{\IB,-}^{\rm RG}$, the polaron peak in the spectrum broadens and the quasiparticle weight $Z=0$ vanishes. In this regime there exists a dynamically unstable mode bound to the impurity at energies below that of the polaron. At a second critical interaction strength $a_{\IB,+}^{\rm RG}$ the number of phonons in the polaron cloud diverges and beyond this point stable molecular states exist at low energies. We predict an extended strong coupling regime where the number of bosons bound to the impurity diverges as a consequence of quantum fluctuations. In reality this divergency is expected to be prevented by direct phonon-phonon interactions not included in our approach.

At the critical point $a_{\IB,+}^{\rm RG}$ we also predict a divergence of the effective polaron mass. It can be extracted experimentally from dipole oscillations of the impurity \cite{Catani2012}. Different theoretical approaches predict vastly different polaron masses. In Ref.~\cite{Rath2013} it was suggested that the polaron hybridizes with the molecule, leading to an effective quasiparticle mass which is only slightly larger than the bare molecular mass. The RG presented in this paper, in contrast, suggests a diverging polaron mass at $a_{\IB,+}^{\rm RG}$. Therefore we expect that a measurement of the effective mass will provide valuable physical insights into the Bose polaron problem. More generally we expect that far-from-equilibrium experiments may provide additional insights, beyond the physics that can be probed by a measurement of the spectral function.

We have argued that the attractive polaron can be accurately described by a renormalized Fr\"ohlich Hamiltonian, and the corresponding strong coupling regime can be realized. We suggest to prepare this polaronic state by adiabatically increasing the interaction strength. This scenario can be realized in experiments with multi-component atomic mixtures which have been experimentally realized in laboratories around the world \cite{Hadzibabic2002,Roati2002,Schreck2001,Catani2008,Spethmann2012,Scelle2013,Ferrier-Barbut2014,Hohmann2015,Moses2015,Park2015,Wacker2015,Hu2016PRL,Jorgensen2016PRL}. The mass ratio can be tuned by coupling the impurity atom to a light field \cite{Grusdt2016}, which can also be used to realize light polarons in an exciton-polariton condensate \cite{Sidler2016}. Such systems should be less vunerable to three-body losses because there are no additional low-lying molecular bound states into which the polaron can decay. This should furthermore allow for an accurate measurement of the three-body losses into the dynamically unstable mode and the molecular state bound to the impurity which we predicted in this paper. 

In this work we employed the Bogoliubov approximation and assumed sufficiently weak boson-boson interactions. To describe molecular bound states, or polarons with a large phonon cloud, we expect that residual phonon-phonon interactions need to be considered. In the diffusive Monte Carlo calculations by Ardila and Giorgini \cite{Ardila2015} boson-boson interactions have been fully taken into account. The predictions of our RG and the MF method yield considerably lower energies than those predicted in \cite{Ardila2015}, and a detailed analysis will be required to understand whether this has to do with a shortcoming of the Bogoliubov approximation. In one dimension a detailed comparison of our RG method with diffusive Monte Carlo calculations has been performed \cite{Grusdt2016RG1D} and the results are in agreement with recent calculations using flow-equation techniques \cite{Volosniev2017}.

\section*{Acknowledgements}
We would like to thank Christoph Gohle, Lars Wacker, Artur Widera, Atac Imamoglu, Frauke See\ss elberg, Nikolaus Buchheim, Immanuel Bloch, Anatoli Polkovnikov, Michael Fleischhauer, Gregory Astrakharchik, Annabelle Bohrdt, Lode Pollet, Valentin Kasper, Tobias Lausch, Vladimir Stojanovic and Dries Sels for fruitful discussions. We acknowledge support from Harvard-MIT CUA, NSF Grant No. DMR-1308435, AFOSR Quantum Simulation MURI, AFOSR MURI Photonic Quantum Matter. F.G. gratefully acknowledges support from the Gordon and Betty Moore foundation. R.S. acknowledges support by the NSF through a grant for the Institute for Theoretical Atomic, Molecular, and Optical Physics at Harvard University and the Smithsonian Astrophysical Observatory.

\appendix

%%%%%%%%%%%%%%%%%%%%%%%%%%%%%%%%%%%%%%%%%%%%%%%%%%%%%
\section{Comparison to experiments}
\label{sec:compExp}
%%%%%%%%%%%%%%%%%%%%%%%%%%%%%%%%%%%%%%%%%%%%%%%%%%%%%

%%%%%%%%%%%%%%%%%%%%%%%%%%%%%%%%%%%%%%%%%%%%%%%%%%%%%
\begin{figure}[b!]
\centering
\epsfig{file=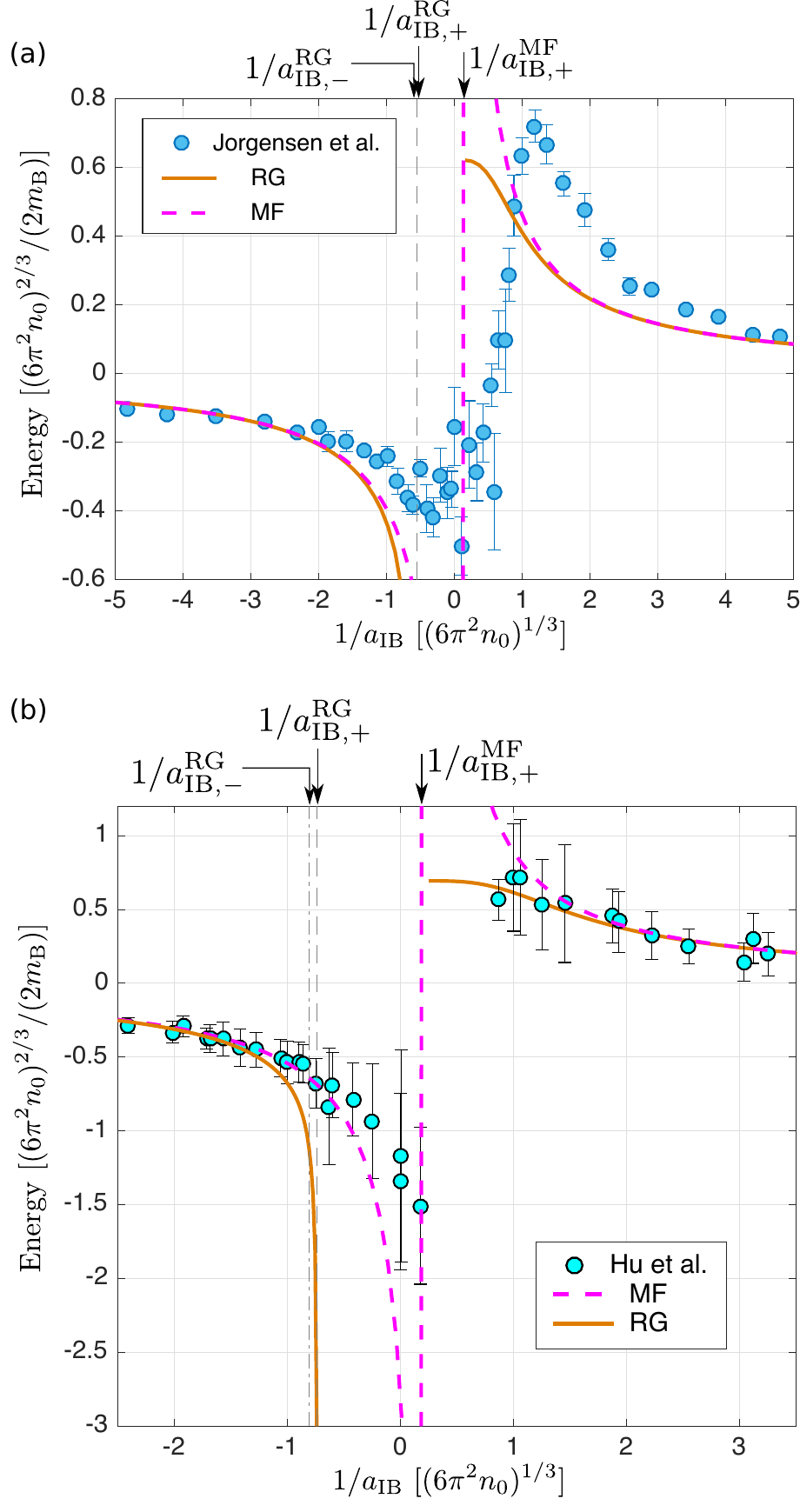, width=0.45\textwidth}
\caption{We compare predictions for the polaron energy, obtained by the RG approach presented here, with recent measurements from radio-frequency absorption spectra. The experimental data points correspond to the center of the peak in the absorption spectrum whereas our theoretical curves correspond to the polaron energy and do not take into account incoherent excitations. In (a) we compare to the experiment by J\o rgensen et al. \cite{Jorgensen2016PRL} with a mixture of different hyperfine states of $~^{39}{\rm K}$. The parameters used for our theoretical analysis are the same as in Figs.~\ref{fig:polaronMass} and \ref{fig:polaronZ}. In (b) we compare to the experiment by Hu et al. \cite{Hu2016PRL} with a mixture of $~^{87}{\rm Rb}$ and $~^{40}{\rm K}$ impurities. In this case we used $n_0=1.8 \times 10^{14} {\rm cm}^{-3}$, $a_{\rm BB}=100 a_0$ and we employed a UV cut-off $\Lambda_0=10^3/\xi$.}
\label{fig:polaronExperiments}
\end{figure}
%%%%%%%%%%%%%%%%%%%%%%%%%%%%%%%%%%%%%%%%%%%%%%%%%%%%%

In Fig.~\ref{fig:polaronExperiments} we compare the polaron energies predicted by our RG approach to the peak positions measured recently in radio-frequency absorption spectra \cite{Jorgensen2016PRL,Hu2016PRL}. 

In the first experiment \cite{Jorgensen2016PRL} the BEC is very weakly interacting and the model which we use for the RG, based on the Bogoliubov approximation, should be accurate. The agreement of our data with the experiment is reasonable, although we would have expected smaller deviations for weak interactions. We notice however that the agreement with the experimental data is excellent when we shift the resonance position horizontally, but the reason for this is unclear. Our prediction from the RG is in excellent agreement with the theoretical calculations presented in Ref.~\cite{Jorgensen2016PRL}, except at very strong couplings where the spectrum is extremely broad and the comparison in Fig.~\ref{fig:polaronExperiments} is not meaningful.

For the second experiment \cite{Hu2016PRL} we obtain excellent agreement of the repulsive polaron energies. In this case shifting the resonance position does not improve the comparison. For weak attractive interactions the agreement is also excellent. For strong attractive interactions the RG predicts a divergence of the polaron energy which is not seen in the experiment. We attribute this to the fact that, in the experiment, the position of a broad feature with a large weight in the spectrum is recorded. As has been shown in Ref.~\cite{Shchadilova2016PRL} by a calculation of the spectral function, this position does not coincide with the polaron energy in the strong coupling regime. Hence a comparison of our results with the data is meaningless in this regime.

%%%%%%%%%%%%%%%%%%%%%%%%%%%%%%%%%%%%%%%%%%%%%%%%%%%%%
\section{Polaronic self-trapping}
\label{sec:polSelfTrapping}
%%%%%%%%%%%%%%%%%%%%%%%%%%%%%%%%%%%%%%%%%%%%%%%%%%%%%
Investigating the polaron mass allows to study the self-trapping of the impurity predicted in the strong-coupling regime by Landau and Pekar \cite{Landau1946,Landau1948}. It can be characterized by a dramatic increase of the polaron mass, as observed in Fig.~\ref{fig:polaronMass}, which effectively localizes the impurity \cite{Casteels2011}. There has been considerable debate in the literature whether this self-trapping is possible with ultracold atoms, and if it is, whether there exists is a smooth cross-over or a true polaronic phase transition. 

So far the discussion in the literature was restricted to the Fr\"ohlich model, where Feynman's variational path-integral method suggests a sharp self-trapping transition \cite{Tempere2009} for sufficiently light impurities \cite{Casteels2012}. The applicability of Feynman's method in the intermediate coupling regime has been questioned, however \cite{Feranchuk2005,Vlietinck2015,Grusdt2015RG}. Using more sophisticated methods to describe quantum fluctuations in this regime \cite{Tulub1962,Shchadilova2016,Grusdt2015RG,Grusdt2016RG,Kain2016} it has has been shown that self-trapping takes place by a cross-over through an extended intermediate coupling regime. This effect becomes particularly pronounced for light impurities \cite{Grusdt2016}. 

In Ref.~\cite{Rath2013} the self-consistent T-matrix approximation has been applied to go beyond the Fr\"ohlich model, but no indications for self-trapping were found. Instead the T-matrix calculations suggest a hybridization of the attractive polaron with a molecule, with an effective mass only slightly exceeding the molecule mass. This is in stark contrast to the predictions of the RG method presented in Fig.~\ref{fig:polaronMass}, where we find a smooth \cite{Gerlach1991} self-localization in the spirit of Landau and Pekar on the attractive side of the Feshbach resonance. We expect the scenario described by T-matrix calculations to become relevant for sufficiently large boson-boson interactions however.

%%%%%%%%%%%%%%%%%%%%%%%%%%%%%%%%%%%%%%%%%%%%%%%%%%%%%
\section{RG at finite momentum}
\label{sec:FiniteMomentumRG}
%%%%%%%%%%%%%%%%%%%%%%%%%%%%%%%%%%%%%%%%%%%%%%%%%%%%%
In this appendix we summarize the RG flow equations for finite polaron momentum $\vec{p} \neq 0$. We keep our discussion general and consider arbitrary dimensionality $d$. In this case the universal Hamiltonian reads
\begin{widetext}
\begin{multline}
\tilde{\mathcal{H}}(\Lambda) = E_0 + \int^\Lambda d^dk ~ \Omega_{\vec{k}} \ad_{\vec{k}} \a_{\vec{k}} +  \int^\Lambda d^dk  d^dk' \Biggl\{ \frac{1}{2} k_\mu \mathcal{M}_{\mu \nu}^{-1} k_\nu'  : \G_{\vec{k}} \G_{\vec{k}'} :
+ \frac{G_+}{2 n_0} : \delta \hat{n}_{\vec{k}} \delta \hat{n}_{\vec{k}'}:  +\\
+  2 n_0 G_-  :\hat{\vartheta}_{\vec{k}} \hat{\vartheta}_{\vec{k}'}: + k_\mu b_\mu W_{k'}  :\G_{\vec{k}} \l \ad_{\vec{k}'} + \a_{\vec{k}'} \r : \Biggr\}.
\label{eq:HuniversalAppdx}
\end{multline}
\end{widetext}
Here we introduced the new coupling constant $b_\mu$, where $\mu,\nu=x,y,...$ are summed over cartesian coordinates using Einstein convention, and we consider a tensor-valued effective impurity mass $\mathcal{M}_{\mu \nu}$. The effective phonon dispersion in the polaron frame reads 
\begin{equation}
\Omega_{\vec{k}} = \omega_k+\frac{1}{2} k_\mu \mathcal{M}_{\mu \nu}^{-1} k_\nu - M^{-1} k_\mu Q_\mu,
\label{eq:defGdefOmegaApdx}
\end{equation}
where $Q_\mu$ is an additional coupling constant. The universal form of the MF amplitude $\alpha_{\vec{k}}(\Lambda)$, appearing in $\G_{\vec{k}}$, is
\begin{align}
\alpha_{\vec{k}} &= - \frac{\eta_{\vec{k}}}{\Omega_{\vec{k}}}, \\
 \eta_{\vec{k}}(\Lambda) &= g_1(\Lambda) W_k + \frac{1}{2} b_\mu(\Lambda) k_\mu W_k + g_2^\mu(\Lambda) k_\mu W_k^{-1},
\label{eq:AlphakUniversal}
\end{align}
where two additional coupling constants $g_1(\Lambda)$ and $g_2^\mu(\Lambda)$ appear.

The initial conditions for the coupling constants are
\begin{align}
E_0(\Lambda_0) &= E_0^\MF,  & \mathcal{M}_{\mu \nu}^{-1}(\Lambda_0) &= M^{-1} \delta_{\mu \nu},  \\
 \vec{Q}(\Lambda_0) &= \vec{p} - \vec{P}_\ph^\MF,  &G_{\pm}(\Lambda_0) &=  \frac{1}{2} \frac{g_\IB^{(0)}}{(2 \pi)^d}  \\
  \vec{b}(\Lambda_0) &= 0  & \alpha_{\vec{k}}(\Lambda_0) &= \alpha_{\vec{k}}^\MF.
\end{align}
The MF polaron state at finite momentum is determined by
\begin{equation}
\alpha_{\vec{k}}^\MF = - \beta_\MF \frac{V_k}{\Omega^\MF_{\vec{k}}},
\end{equation}
where
\begin{align}
\Omega_{\vec{k}}^\MF&=\omega_k+\frac{k^2}{2M} - \frac{\vec{k}}{M} \cdot \l \vec{p} - \vec{P}_\ph^\MF \r, \\ 
\beta_\MF &= \left[ 1 + \frac{g_\IB^{(0)}}{(2 \pi)^d} \int^{\Lambda_0} d^dk ~ \frac{W_{k}^2}{\Omega_{\vec{k}}^\MF} \right]^{-1}, \\
 \vec{P}_\ph^\MF &=  \beta_\MF^2 \int^{\Lambda_0} d^dk ~ \vec{k} \frac{V_k^2}{\l \Omega_{\vec{k}}^\MF \r^2}.
\end{align}
The MF polaron energy is given by
\begin{multline}
E_0^\MF = \beta_\MF g_\IB^{(0)} n_0 + \frac{g_\IB^{(0)}}{(2 \pi)^d} \int^{\Lambda_0} d^dk ~ \sinh^2 \theta_k \\
 + \frac{\vec{p}^2}{2M} - \frac{\l \vec{P}_\ph^\MF \r^2}{2M}.
\label{eq:EoMFfinitep}
\end{multline}

%%%%%%%%%%%%%%%%%%%%%%%%%%%%
\emph{RG flow equations.--}
The RG flow equation for the effective impurity mass reads
\begin{multline}
\frac{\partial \mathcal{M}_{\mu \nu}^{-1}}{\partial \Lambda} = 2 \int_\f d^{d-1}k ~ \frac{1}{\Omega_{\vec{k}}} \Bigl\{ \alpha_{\vec{k}}^2 \mathcal{M}_{\mu \lambda}^{-1} k_\lambda k_\tau \mathcal{M}_{\tau \nu}^{-1} \\
+ \frac{1}{2} \alpha_{\vec{k}} W_k k_\lambda \l b_\mu \mathcal{M}_{\lambda \nu}^{-1} + \mathcal{M}_{\mu \lambda}^{-1} b_\nu \r \Bigr\}.
\end{multline}
For $G_\pm$ we obtain
\begin{flalign}
\frac{\partial G_+}{\partial \Lambda} &= 2 \int_\f d^{d-1}k ~ \frac{1}{\Omega_{\vec{k}}} \l \alpha_{\vec{k}} k_\mu b_\mu +W_k G_+ \r^2 , \label{eq:RGflowEqGp} \\
\frac{\partial G_-}{\partial \Lambda} &= 2 \int_\f d^{d-1}k ~ \frac{1}{\Omega_{\vec{k}}} W_k^{-2} G_-^2. \label{eq:RGflowEqGm}
\end{flalign}

The flow of $b_\mu$, which is only non-vanishing for finite polaron momentum $\vec{p} \neq 0$, is given by
\begin{widetext}
\begin{equation}
\frac{\partial b_\mu}{\partial \Lambda} = \int_\f d^{d-1}k ~ \frac{1}{\Omega_{\vec{k}}} \l 2 \mathcal{M}_{\mu \nu}^{-1} k_\nu \alpha_{\vec{k}} + b_\mu W_k \r \l W_k G_+ + \alpha_{\vec{k}} k_\lambda b_\lambda \r,
\end{equation}
and for $g_2^\mu$ we derive
\begin{equation}
\frac{\partial g_2^\mu}{\partial \Lambda} = - G_- \int_\f d^{d-1}k ~\frac{W_k^{-1}}{\Omega_{\vec{k}}} \l \mathcal{M}_{\mu \nu}^{-1} k_\nu \alpha_{\vec{k}} + b_\mu W_k \r.
\end{equation}
The flow of $g_1$ is determined from the flows of $b_\mu$, $g_2^\mu$ and $\mathcal{M}_{\mu \mu}^{-1}$ (sum over the index $\mu$ is implied),
\begin{equation}
\frac{\partial g_1}{\partial \Lambda}  = - \l G_+ , b_x \r  \l
\begin{array}{cc}
1+2 G_+ J_1  + 2  b_x J_2  &  2 b_x J_1 + 2 \mathcal{M}_{xx}^{-1} J_2 \\
2 G_+ J_2   + 2 b_x J_7  &    1+2 b_x J_2 + 2 \mathcal{M}_{xx}^{-1} J_7
\end{array} \r^{-1}
 \l
\begin{array}{ccc}
J_3 & 2J_4  & J_5^\mu \\
J_5^x & 2 J_6 & J_8^\mu 
\end{array} \r
\l
\begin{array}{c}
\partial b_x / \partial \Lambda \\
\partial g_2^x / \partial \Lambda \\
\partial \mathcal{M}_{\mu \mu}^{-1} / \partial \Lambda
\end{array} \r.
\end{equation}
Here, and in the following, we assume $\vec{p} = p \vec{e}_x$, such that $b_\mu = b_x \delta_{\mu x}$, $g_2^\mu = g_2^x \delta_{\mu x}$, $Q_\mu = Q_x \delta_{\mu  x}$ and $\mathcal{M}_{\mu \nu}^{-1} = {\rm diag} (\mathcal{M}_{xx}^{-1}, \mathcal{M}_{yy}^{-1}, \mathcal{M}_{yy}^{-1},...)$. Moreover we introduced the following integrals (where $J_5^\mu = I^x$ in the spherically symmetric case, see Eq.~\eqref{eq:defIxintegral})
\begin{align}
 J_1 &= \int_\s d^d p~ \frac{W_p^2}{\Omega_{\vec{p}}},  & J_2 &= \int_\s d^d p~ \alpha_{\vec{p}} p_x \frac{W_p}{\Omega_{\vec{p}}}, \label{eq:IntDef1} \\
  J_3 &= \int_\s d^d p~ p_x \frac{W_p^2}{\Omega_{\vec{p}}},  & J_4 &= \int_\s d^d p~  \frac{p_x}{\Omega_{\vec{p}}}, \label{eq:IntDef2}\\
  J_5^\mu &= \int_\s d^d p~ \alpha_{\vec{p}} p_\mu^2 \frac{W_p}{\Omega_{\vec{p}}}, & J_6 &= \int_\s d^d p~ \alpha_{\vec{p}} p_x^2 \frac{W_p^{-1}}{\Omega_{\vec{p}}},\label{eq:IntDef3}\\
 J_7 &= \int_\s d^d p~ \alpha_{\vec{p}}^2 p_x^2 \frac{1}{\Omega_{\vec{p}}},  & J_8^{\mu} &= \int_\s d^d p~ p_x \alpha_{\vec{p}}^2 p_\mu^2 \frac{1}{\Omega_{\vec{p}}} \label{eq:IntDef4}.
\end{align}

The coupling constant $Q_\nu$ has an RG flow given by
\begin{equation}
\frac{\partial Q_x}{\partial \Lambda}  = M \l b_x , \mathcal{M}_{xx}^{-1} \r  \l
\begin{array}{cc}
1+2 G_+ J_1  + 2  b_x J_2  &  2 b_x J_1 + 2 \mathcal{M}_{xx}^{-1} J_2 \\
2 G_+ J_2   + 2 b_x J_7  &    1+2 b_x J_2 + 2 \mathcal{M}_{xx}^{-1} J_7
\end{array} \r^{-1}
 \l
\begin{array}{ccc}
J_3 & 2J_4  & J_5^\mu \\
J_5^x & 2 J_6 & J_8^\mu 
\end{array} \r
\l
\begin{array}{c}
\partial b_x / \partial \Lambda \\
\partial g_2^x / \partial \Lambda \\
\partial \mathcal{M}_{\mu \mu}^{-1} / \partial \Lambda
\end{array} \r.
\end{equation}
The RG flow of the ground state energy is given by
\begin{multline}
\frac{\partial E_0}{\partial \Lambda} = \frac{1}{2} \frac{ \partial \mathcal{M}_{\mu \nu}^{-1}}{\partial \Lambda} \l \int_\s d^dp ~ \alpha_{\vec{p}}^2 p_\mu p_\nu \r + 
\frac{\partial b_\mu}{\partial \Lambda} \l \int_\s d^dp ~ \alpha_{\vec{p}} p_\mu W_p \r + 
2 \frac{\partial g_2^\mu}{\partial \Lambda} \l \int_\s d^dp ~ \alpha_{\vec{p}} p_\mu W_p^{-1} \r -\\
- \int_\s d^dp  \int_\f d^{d-1}k ~ \Biggl\{ \frac{2}{\Omega_{\vec{k}}} \l G_+ + k_\mu b_\mu \alpha_{\vec{k}} W_k^{-1} \r G_- - W_p^2  \frac{1}{\Omega_{\vec{k}}} \left[ \l k_\mu b_\mu \alpha_{\vec{k}} \r^2 + W_k^2 G_+^2 \right] - W_p^{-2} \frac{W_k^{-2}}{\Omega_{\vec{k}}}  G_-^2 \Biggr\}.
\label{eq:RGflowEnergy}
\end{multline}
\end{widetext}

%%%%%%%%%%%%%%%%%%%%%
\section{Quantum theory of molecule formation}
\label{subsubsec:MFmoleculeTheory}
In this appendix we introduce a quantum theory of molecular bound states on the repulsive side of the Feshbach resonance, for $1/a_\IB > 1/a_{\IB,+}^\MF$. One approach would be to diagonalize the full Hamiltonian \eqref{eq:HuniversalBoundStatesUnified}, which can be done in the limit $M \to \infty$ using Gaussian states \cite{Shchadilova2016PRL}. Here, instead, we introduce an effective Hamiltonian describing only particle-number and phase fluctuation modes bound to the impurity. To this end we introduce the following two operators,
\begin{equation}
\hat{\vartheta}_b = \int d^3k~ \frac{\beta_k^-}{i \sqrt{2}} (\a_{\vec{k}} - \ad_{\vec{k}}), \quad \hat{n}_b = \int d^3k~ \frac{\beta_k^+}{\sqrt{2}} (\a_{\vec{k}} + \ad_{\vec{k}}),
\end{equation}
which represent two conjugate variables,
\begin{equation}
[\hat{n}_b, \hat{\vartheta}_b] = i \lambda, \qquad \lambda = \int d^3k~ \beta_k^+ \beta_k^-.
\label{eq:conjVar}
\end{equation}
Because particle-number and phase fluctuation modes $\beta_k^\pm$ are both normalized, $\int d^3k~ (\beta_k^\pm)^2=1$, we obtain $\lambda \leq 1$. For sufficiently small boson-boson interactions $\mu_+\approx \mu_-$ and we thus obtain $\lambda \approx 1$.

In the regime $1/a_{\IB,-}^\MF < 1/a_\IB < 1/a_{\IB,+}^\MF$ (II in Fig.~\ref{fig:polaronMFPhaseDiag}) only $\mu_-<0$ exists. The unstable mode of phase fluctuations can be described by the following effective Hamiltonian,
\begin{equation}
\H_\eff = E_0^\MF + \frac{1}{2} \mu_- : \hat{\vartheta}_b^2 :.
\end{equation}
This corresponds to a dynamically unstable harmonic oscillator where the potential has negative curvature. 

In the regime $1/a_\IB > 1/a_{\IB,+}^\MF$ (IV in Fig.~\ref{fig:polaronMFPhaseDiag}) both $\mu_\pm<0$ exist. To describe both particle-number and phase fluctuations, the following effective Hamiltonian can be used,
\begin{equation}
\H_\eff = E_0^\MF + \frac{1}{2} \mu_- : \hat{\vartheta}_b^2 : + \frac{1}{2} \mu_+ : \hat{n}_b^2 :.
\end{equation}
Its variational energy landscape has the same qualitative properties as derived in the proceeding section and summarized in Fig.~\ref{fig:polaronMFPhaseDiag}. Using Eq.~\eqref{eq:conjVar} we can easily diagonalize this  Hamiltonian and obtain
\begin{equation}
\H_\eff = E_0^\MF - E_b  \bd \b,
\label{eq:boundStates}
\end{equation}
where $[\b,\bd]=1$ are ladder operators constructed from $\hat{\vartheta}_b/\sqrt{\lambda}$ and $\hat{n}_b/\sqrt{\lambda}$. The binding energy is given by
\begin{equation}
E_b = \lambda \sqrt{\mu_+ \mu_-}.
\label{eq:MFbindingEnergy}
\end{equation}

Eq.~\eqref{eq:boundStates} describes a series of molecular bound states at integer multiples of the binding energy $E_b$ below the polaron energy. In Ref.~\cite{Shchadilova2016PRL} we have performed dynamical simulations of the polaron spectrum, where we found a series of equally spaced peaks corresponding to these bound states. Our expression for their energies matches their positions very accurately. Note that this series of molecular bound states has been observed experimentally \cite{Schmidt2016,Schlagmuller2016} for finite range potentials.

%%%%%%%%%%%%%%%%%%%%%%%%%%%%%%%%%%%%%%%%%%%%%%%%%%%%%
\section{Derivation of the RG equations}
\label{apdx:RGderivation}
%%%%%%%%%%%%%%%%%%%%%%%%%%%%%%%%%%%%%%%%%%%%%%%%%%%%%

For the derivation of the RG flow equations presented in Appendix \ref{sec:FiniteMomentumRG} we choose slightly different representation of the two phonon terms in the universal Hamiltonian. We define
\begin{equation}
\D_{\vec{k}}^\alpha = \l \d_{\vec{k}}, \dd_{\vec{k}} \r^T, \qquad G_{\alpha \beta}(\Lambda_0) = \frac{1}{2} \frac{g_\IB^{(0)}}{(2\pi)^d} \delta_{\alpha,\beta},
\end{equation}
for $\alpha,\beta = 0,1$. Recall that $\d_{\vec{k}}$ denote the bare boson operators before applying the Bogoliubov transformation. The two-phonon terms can be written as
\begin{equation}
 \frac{G_+}{2 n_0} : \delta \hat{n}_{\vec{k}} \delta \hat{n}_{\vec{k}'}:  +  2 n_0 G_-  :\hat{\vartheta}_{\vec{k}} \hat{\vartheta}_{\vec{k}'}: ~ =~ : \Dd^\alpha_{\vec{k}} G_{\alpha \beta} \D^\beta_{\vec{k}'}:
\end{equation}
and the coupling constants are related by
\begin{equation}
G_\pm = G_{00} \pm G_{01}.
\end{equation}

%%%%%%%%%%%%%%%%%%%%%
\subsection{Sketch of the RG procedure}
\label{sec:RGsketch}
%%%%%%%%%%%%%%%%%%%%%
Now we will explain how the RG procedure works, and derive the universal form of the Hamiltonian in Eq.~\eqref{eq:HuniversalAppdx}. The basic idea is to consider separately slow phonons ($\s$) at momenta $p \leq \Lambda-\delta \Lambda$ and fast phonons ($\f$) from a thin shell $\Lambda-\delta \Lambda < k \leq \Lambda$. Interactions between fast phonons can be ignored, because they are $\mathcal{O}(\delta \Lambda)$, and the fast-phonon Hamiltonian reads
\begin{equation}
\Ht_\f = \int_\f d^dk~ \Omega_{\vec{k}} \ad_{\vec{k}} \a_{\vec{k}} + \mathcal{O}(\delta \Lambda^2).
\end{equation}
Slow phonons, on the other hand, are described by the Hamiltonian in Eq.~\eqref{eq:HuniversalAppdx}, but with all integrals restricted to momenta $|\vec{p}|\leq \Lambda - \delta \Lambda$. We denote the slow-phonon Hamiltonian by $\Ht_\s$. Finally, the coupling between slow and fast phonons is described by
\begin{widetext}
\begin{equation}
\H_\sf = \int_\s d^dp \int_\f d^dk \left\{ k_\mu \mathcal{M}_{\mu \nu}^{-1} p_\nu  \G_{\vec{k}} \G_{\vec{p}} +  \l \Dd^\alpha_{\vec{k}} G_{\alpha \beta} \D^\beta_{\vec{p}}  + \hc \r + k_\mu b_\mu W_p  \G_{\vec{k}} \l \ad_{\vec{p}} + \a_{\vec{p}} \r + p_\mu b_\mu W_k  \G_{\vec{p}} \l \ad_{\vec{k}} + \a_{\vec{k}} \r \right\}.
\label{eq:Hmix}
\end{equation}
\end{widetext}
We made use of the symmetries $\mathcal{M}_{\mu \nu}^{-1} = \mathcal{M}_{\nu \mu}^{-1}$ and $G_{\alpha \beta} = G_{\beta \alpha}$, which can always be enforced by proper ordering of operators. 

Next we decouple fast and slow degrees of freedom, by applying a unitary transformation 
\begin{align}
\U_\Lambda &= \exp \l  \int_\f d^dk~ \Fd_{\vec{k}} \a_{\vec{p}} - \hc \r, \\
 \Ud_\Lambda &\a_{\vec{k}} \U_\Lambda = \a_{\vec{k}} - \F_{\vec{k}}, \qquad \vec{k} \in \f.
\label{eq:defUlambda}
\end{align}
The operator $\F_{\vec{k}}$ depends only on slow variables. To calculate the transformation properties of slow-phonon operators $\hat{O}_\s$, we make use of a perturbative expansion in the inverse of the fast-phonon frequency $\Omega_{\vec{k}}^{-1}$, which provides the largest energy scale. It holds
\begin{equation}
\hat{U}_\Lambda^\dagger \hat{O}_\s \hat{U}_\Lambda = \hat{O}_\s + \int_\f d^dk  ~ \left\{  \a_{\vec{k}} [\hat{O}_\s , \F^\dagger_{\vec{k}}]  - \ad_{\vec{k}} [\hat{O}_\s , \F_{\vec{k}}] \right\}
\label{eq:ULambdaOnOs}
\end{equation}
up to terms of order $\mathcal{O}(\Omega_{\vec{k}}^{-2})$.

The resulting Hamiltonian becomes 
\begin{widetext}
\begin{equation}
\Ud_\Lambda \Ht_\Lambda \U_\Lambda = \Ht_\f + \Ht_\s + \int_\f d^dk ~ \Omega_{\vec{k}} \Fd_{\vec{k}} \F_{\vec{k}} + \int_\f d^dk ~\left\{ - \ad_{\vec{k}}  \l \Omega_{\vec{k}} \F_{\vec{k}} + [\Ht_\s,\F_{\vec{k}}] \r + \hc \right\} + \Ud_\Lambda \Ht_\sf \U_\Lambda + \mathcal{O}(\Omega_{\vec{k}}^{-2}).
\end{equation}
\end{widetext}
Our basic strategy is to choose $\F_{\vec{k}}$ such that all terms which linear in $\a_{\vec{k}}$ (in $\ad_{\vec{k}}$) vanish, to order $\mathcal{O}(\Omega_{\vec{k}}^{-1},\delta\Lambda)$. In $\Ud_\Lambda \Ht_\sf \U_\Lambda$ the following terms appear: From $\G_{\vec{k}}\G_{\vec{p}}$ we get $\ad_{\vec{k}} p_\mu \G_{\vec{p}}$ because $\G_{\vec{k}}$ contains $\alpha_{\vec{k}} \ad_{\vec{k}}$; Moreover there is a term $\ad_{\vec{k}} [p_\mu \G_{\vec{p}},\F_{\vec{k}}]$ originating from the $\ad_{\vec{k}} \a_{\vec{k}}$-term in $\G_{\vec{k}}$ and using
\begin{multline}
\int_\f d^dk~ \ad_{\vec{k}} \a_{\vec{k}} \int_\f d^d k'~ \ad_{\vec{k}'} [\G_{\vec{p}},\F_{\vec{k}'}] =\\
= \int_\f d^dk~ \ad_{\vec{k}} [\G_{\vec{p}},\F_{\vec{k}'}] + \mathcal{O}(\delta \Lambda^2).
\end{multline}
Note the importance of normal-ordering, $\int_\f d^dk d^d k'  :\ad_{\vec{k}} \a_{\vec{k}} \ad_{\vec{k}'}: = \mathcal{O}(\delta \Lambda^2)$ whereas $\int_\f d^dk d^d k' ~ \ad_{\vec{k}} \a_{\vec{k}} \ad_{\vec{k}'} = \int_\f d^dk ~ \ad_{\vec{k}} + \mathcal{O}(\delta \Lambda^2)$. Next, from $\Dd_{\vec{k}} \D_{\vec{p}}$ we derive $\ad_{\vec{k}} \D_{\vec{p}}$; The hermitian conjugate $\Dd_{\vec{k}} \D_{\vec{p}}$ yields $\ad_{\vec{k}} \Dd_{\vec{p}}$. From the last two terms in Eq.~\eqref{eq:Hmix} we obtain $\ad_{\vec{k}} p_\mu \G_{\vec{p}}$ and $\ad_{\vec{k}} W_p \l \ad_{\vec{p}} + \a_{\vec{p}} \r$. This final term can be simplified by noting that $  W_p \l \ad_{\vec{p}} + \a_{\vec{p}} \r = \D^\alpha_{\vec{p}} + \Dd_{\vec{p}}^\alpha$ for arbitrary $\alpha=0,1$. 

Therefore, we obtain the following list of terms linear in $\ad_{\vec{k}}$, ordered in powers of $\Omega_{\vec{k}}^{-1}$,
\begin{multline}
\ad_{\vec{k}}  \Omega_{\vec{k}} \F_{\vec{k}}, \qquad \ad_{\vec{k}} p_\mu \G_{\vec{p}}, \qquad \ad_{\vec{k}} \D^\alpha_{\vec{p}}, \\
\qquad \ad_{\vec{k}} \Dd^\alpha_{\vec{p}}, \qquad \ad_{\vec{k}} [p_\mu \G_{\vec{p}},\F_{\vec{k}}], \qquad \ad_{\vec{k}} [\Ht_\s,\F_{\vec{k}}],
\label{eq:adkList}
\end{multline}
where we included the full $\vec{p}$-dependency. We conclude that $\F_{\vec{k}}$ is of order $ \Omega_{\vec{k}}^{-1}$, and given by a certain linear combination of $p_\mu \G_{\vec{p}}$, $\D^\alpha_{\vec{p}}$ and $\Dd^\alpha_{\vec{p}}$. Note that the last two terms in Eq.~\eqref{eq:adkList} only contribute higher-order corrections $\mathcal{O}(\Omega_{\vec{k}}^{-2})$ to $\F_{\vec{k}}$.

Next, we confirm that newly generated terms in the transformed Hamiltonian are all included in the universal Hamiltonian \eqref{eq:HuniversalAppdx}. Again, we make a list of such terms. $\Omega_{\vec{k}} \Fd_{\vec{k}} \F_{\vec{k}}$ contains the following cross-products:
\begin{multline}
p_\mu p_\nu \G_{\vec{p}} \G_{\vec{p}'}, \qquad \qquad  \D_{\vec{p}}^\alpha \D_{\vec{p}'}^\beta\\
p_\mu \G_{\vec{p}} \l \Dd_{\vec{p}'}^\alpha + \D_{\vec{p}'}^\alpha \r = p_\mu \G_{\vec{p}} W_{p'} \l \ad_{\vec{p}'} + \a_{\vec{p}'} \r.
\label{eq:FkFklist}
\end{multline}
Note that in the second term, expressions like $\Dd_{\vec{p}}^\alpha \D_{\vec{p}'}^\beta$ can also be included. For the term in the middle we employed that final expression needs to be hermitian, allowing only the combination $\Dd_{\vec{p}'}^\alpha + \D_{\vec{p}'}^\alpha$ to appear. 

Now we check that all terms generated in $\Ud_\Lambda \Ht_\sf \U_\Lambda$ are also included in the list \eqref{eq:FkFklist}. From $\G_{\vec{k}} p_\mu \G_{\vec{p}}$ we obtain terms like $\F_{\vec{k}} p_\mu \G_{\vec{p}}$, included in \eqref{eq:FkFklist}. Moreover, from normal-ordering as described above, we get $[p_\mu \G_{\vec{p}},\F_{\vec{k}}]$. This yields only terms which are at maximum of order $\mathcal{O}(\a_{\vec{k}})$, however, and which hence can be eliminated by a MF shift later. In addition there are terms like $\ad_{\vec{k}} \a_{\vec{k}} p_\mu \G_{\vec{p}}$, which renormalize the fast-phonon frequency. We will return to such contributions below. The remaining terms in Eq.~\eqref{eq:Hmix} lead to similar contributions.

Finally, all terms in Eq.~\eqref{eq:FkFklist} need to be brought into a normal-ordered form. One can check that this only produces terms which are at maximum of order $\mathcal{O}(\a_{\vec{k}})$. As mentioned previously, these linear terms can be eliminated by a unitary MF-shift, $\a_{\vec{k}} \to \a_{\vec{k}} + \delta \alpha_{\vec{k}}$. This in turn leaves the universal Hamiltonian \eqref{eq:HuniversalAppdx} invariant, as will be shown in detail below. Hence we conclude that the RG procedure described above leads to a closed set of RG flow equations for the coupling constants introduced in the beginning. We present the detailed results of our calculations in the following subsection.

%%%%%%%%%%%%%%%%%%%%%
\subsection{Derivation of the RG}
\label{sec:RGderivation}
%%%%%%%%%%%%%%%%%%%%%

As described above, in every RG step we apply the unitary transformation $\U_\Lambda$ to decouple fast from slow phonons, see Eq.~\eqref{eq:defUlambda}. Eliminating terms linear in the fast-phonon operators $\a_{\vec{k}}$ yields
\begin{widetext}
\begin{equation}
\F_{\vec{k}} = \frac{1}{\Omega_{\vec{k}}} \int_\s d^dp \left\{ p_\mu \G_{\vec{p}} \left[ W_k b_\mu + \mathcal{M}_{\mu \nu}^{-1} k_\nu \alpha_{\vec{k}} \right] + W_p \l \ad_{\vec{p}} + \a_{\vec{p}} \r  \left[ b_\mu k_\mu \alpha_{\vec{k}} + W_k G_+ \right] - W_p^{-1} \l \ad_{\vec{p}} - \a_{\vec{p}} \r W_k^{-1} G_- \right\}.
\end{equation}

%%%%%%%%%%%%
\subsubsection{Quantum fluctuations}
%%%%%%%%%%%%
We begin by deriving the properties of quantum fluctuations in the transformed Hamiltonian, which has the form
\begin{equation}
\Ud_\Lambda \Ht_\Lambda \U_\Lambda = \Ht_\s + \delta \Ht_\s + \int_\f d^dk~ \ad_{\vec{k}} \a_{\vec{k}} \l \Omega_{\vec{k}} + \hat{\Omega}_\s(\vec{k}) \r + \int_\s d^dp ~ \l \ad_{\vec{p}} + \a_{\vec{p}} \r \delta \nu_{\vec{p}} + \mathcal{O}(\delta \Lambda^2,\Omega_{\vec{k}}^{-2}).
\end{equation}
The direct renormalization of the slow-phonon Hamiltonian is described by
\begin{multline}
\delta \Ht_\s =  \delta E_0 + \int_\s d^dp ~ \ad_{\vec{p}} \a_{\vec{p}} \frac{1}{2} p_\mu \delta \mathcal{M}_{\mu \nu}^{-1} p_\nu +  \int_\s d^dp  d^dp' \Bigl\{ \frac{1}{2} p_\mu \delta \mathcal{M}_{\mu \nu}^{-1} p_\nu'  : \G_{\vec{p}} \G_{\vec{p}'} : + \\ 
+  : \Dd^\alpha_{\vec{p}} \delta G_{\alpha \beta} \D^\beta_{\vec{p}'}: + p_\mu \delta b_\mu W_{p'}  :\G_{\vec{p}} \l \ad_{\vec{p}'} + \a_{\vec{p}'} \r : \Bigr\}.
\end{multline}

The renormalization of the mass is given by
\begin{equation}
\delta \mathcal{M}_{\mu \nu}^{-1} = -2 \int_\f d^dk ~ \frac{1}{\Omega_{\vec{k}}} \left\{ \alpha_{\vec{k}}^2 \mathcal{M}_{\mu \lambda}^{-1} k_\lambda k_\tau \mathcal{M}_{\tau \nu}^{-1} + \frac{1}{2} \alpha_{\vec{k}} W_k k_\lambda \l b_\mu \mathcal{M}_{\lambda \nu}^{-1} + \mathcal{M}_{\mu \lambda}^{-1} b_\nu \r \right\},
\end{equation}
and the renormalization of $b_\mu$ is given by
\begin{equation}
\delta b_\mu = - \int_\f d^d k ~ \frac{1}{\Omega_{\vec{k}}} \left[ 2 \mathcal{M}_{\mu \nu}^{-1} k_\nu \alpha_{\vec{k}} + b_\mu W_k \right] \left[ W_k G_+ + \alpha_{\vec{k}} k_\lambda b_\lambda \right].
\end{equation}
For $\delta G_{\alpha \beta}$ we obtain (note the symmetries $G_{00}=G_{11}$ and $G_{01} = G_{10}$)
\begin{flalign}
\delta G_{00} &= - \int_\f d^d k ~ \frac{1}{\Omega_{\vec{k}}} \left\{ 2 \alpha_{\vec{k}} b_\mu k_\mu W_k G_+ + W_k^2 G_+^2 + W_k^{-2} G_-^2 + \l \alpha_{\vec{k}} b_\mu k_\mu \r^2 \right\}, \\
\delta G_{01} &= - \int_\f d^d k ~ \frac{1}{\Omega_{\vec{k}}} \left\{ 2 \alpha_{\vec{k}} b_\mu k_\mu W_k G_+ + W_k^2 G_+^2 - W_k^{-2} G_-^2 + \l \alpha_{\vec{k}} b_\mu k_\mu \r^2 \right\}.
\end{flalign}
The flow of the groundstate energy is determined by
\begin{multline}
\delta E_0 = \frac{1}{2} \delta \mathcal{M}_{\mu \nu}^{-1} \l \int_\s d^dp ~ \alpha_{\vec{p}}^2 p_\mu p_\nu \r + \delta b_\mu \l \int_\s d^dp ~ \alpha_{\vec{p}} p_\mu W_p \r + 
 2 G_- \l \int_\f d^dk ~ \frac{W_k^{-1}}{\Omega_{\vec{k}}} \left[ \mathcal{M}_{\mu \nu}^{-1} k_\nu \alpha_{\vec{k}} + b_\mu W_k \right] \r \\
 \times \l \int_\s d^dp ~ \alpha_{\vec{p}} p_\mu W_p^{-1} \r 
+ \int_\s d^dp ~ \Biggl\{  \int_\f d^dk ~ \frac{2}{\Omega_{\vec{k}}} \left[ G_{00}^2 - G_{01}^2 + G_- k_\mu b_\mu \alpha_{\vec{k}} W_k^{-1} \right] -\\
- W_p^2 \int_\f d^dk ~ \frac{1}{\Omega_{\vec{k}}} \left[ \l k_\mu b_\mu \alpha_{\vec{k}} \r^2 + W_k^2 G_+^2 \right] - W_p^{-2} \int_\f d^dk ~ \frac{1}{\Omega_{\vec{k}}} W_k^{-2} G_-^2 \Biggr\}.
\end{multline}

Furthermore we introduced the following renormalization of the fast-phonon frequency,
\begin{equation}
\hat{\Omega}_\s(\vec{k}) = k_\mu \int_\s d^d p ~ \left\{ \mathcal{M}_{\mu \nu}^{-1} p_\nu \G_{\vec{p}} + b_\mu W_p \l \ad_{\vec{p}} + \a_{\vec{p}} \r \right\},
\end{equation}
and the MF-terms are characterized by
\begin{equation}
 \delta \nu_{\vec{p}} = \frac{1}{2} \alpha_{\vec{p}} p_\mu \delta \mathcal{M}_{\mu \nu}^{-1} p_\nu + \frac{1}{2} p_\mu W_p \delta b_\mu + p_\mu W_p^{-1} G_- \l \int_\f d^d k~  \frac{W_k^{-1}}{\Omega_{\vec{k}}} \left[ \mathcal{M}_{\mu \nu}^{-1} k_\nu \alpha_{\vec{k}} + b_\mu W_k \right]  \r.
\end{equation}

%%%%%%%%%%%%
\subsubsection{Mean-field shift}
%%%%%%%%%%%%
Next we eliminate the terms linear in the slow phonon operators, $\sim \delta \nu_{\vec{p}} \l \ad_{\vec{p}} + \a_{\vec{p}} \r$, which give rise to an RG flow of the MF amplitude $\alpha_{\vec{k}}(\Lambda)$. To this end we apply a second unitary transformation, of the form
\begin{equation}
\hat{V}_\MF(\Lambda) = \exp \l \int_\s d^dp ~ \delta \alpha_{\vec{p}} \ad_{\vec{p}} - \hc \r, \qquad \qquad \hat{V}^\dagger_\MF(\Lambda) \a_{\vec{p}}  \hat{V}_\MF(\Lambda) = \a_{\vec{p}} + \delta \alpha_{\vec{p}}.
\end{equation}
Note that $\delta \alpha_{\vec{p}} = \mathcal{O}(\delta \Lambda)$, because the new displacement $\delta \nu_{\vec{p}}$ is also $\mathcal{O}(\delta \Lambda)$. 

The saddle-point is determined by eliminating all terms linear in $\a_{\vec{p}}$ in the (normal-ordered) transformed Hamiltonian $\hat{V}_\MF^\dagger \Ud_\Lambda \Ht_\Lambda \U_\Lambda \hat{V}_\MF$. Collecting such terms yields the self-consistency equations for $\delta \alpha_{\vec{p}}$,
\begin{equation}
\delta \alpha_{\vec{p}} = - \Omega^{-1}_{\vec{p}}  \left[ \delta \nu_{\vec{p}} + \delta I_2^\nu  \l 2 \alpha_{\vec{p}} \mathcal{M}_{\nu \mu}^{-1} p_\mu + 2 b_\nu W_p \r  + 2\delta I_1 \l W_p G_+ + \alpha_{\vec{p}} p_\mu b_\mu \r  \right].
\label{eq:selfConsdeltaAlpha}
\end{equation}
Here we defined the following integrals,
\begin{equation}
\delta I_1 = \int_\s d^d p' ~ W_{p'} \delta \alpha_{\vec{p}'} , \qquad \qquad \delta I_2^\nu = \int_\s d^d p' ~ p_\nu' \alpha_{\vec{p}'} \delta \alpha_{\vec{p}'} .
\label{eq:defI1I2}
\end{equation}

Next we calculate the transformed Hamiltonian, making use of the fact that the MF shift commutes with normal ordering, $\hat{V}_\MF^\dagger : \hat{O} : \hat{V}_\MF = : \hat{V}_\MF^\dagger \hat{O} \hat{V}_\MF:$. We introduce new $\G$-operators by
\begin{equation}
\tilde{\Gamma}_{\vec{p}} \equiv \G_{\vec{p}}(\Lambda - \delta \Lambda) = \ad_{\vec{p}} \a_{\vec{p}} + \l \alpha_{\vec{p}} + \delta \alpha_{\vec{p}} \r \l \ad_{\vec{p}} + \a_{\vec{p}} \r, \qquad \hat{V}_\MF^\dagger \G_{\vec{p}} \hat{V}_\MF = \tilde{\Gamma}_{\vec{p}} + 2 \alpha_{\vec{p}} \delta \alpha_{\vec{p}} + \mathcal{O}(\delta \Lambda^2).
\end{equation}

Collecting all terms in the transformed Hamiltonian yields
\begin{multline}
\Ht_{\Lambda-\delta \Lambda} \equiv \hat{V}_\MF^\dagger \l  \Ht_\s + \delta \Ht_\s + \int_\s d^dp ~ \delta \nu_{\vec{p}} \l \ad_{\vec{p}} + \a_{\vec{p}} \r \r \hat{V}_\MF = \Ht_\s(\Lambda) + \delta E_0 + \int_\s d^dp ~ \ad_{\vec{p}} \a_{\vec{p}} \delta \Omega_{\vec{p}} + \\
+ \int_\s d^dp  d^dp' \Bigl\{ \frac{1}{2} p_\mu \delta \mathcal{M}_{\mu \nu}^{-1} p_\nu'  : \G_{\vec{p}} \G_{\vec{p}'} : +  : \Dd^\alpha_{\vec{p}} \delta G_{\alpha \beta} \D^\beta_{\vec{p}'}: + p_\mu \delta b_\mu W_{p'}  :\G_{\vec{p}} \l \ad_{\vec{p}'} + \a_{\vec{p}'} \r : \Bigr\},
\end{multline}
which, finally, is of the universal form \eqref{eq:HuniversalAppdx} again. As a consequence of the MF shift, the phonon frequency acquires additional renormalization,
\begin{equation}
\delta \Omega_{\vec{p}} = \frac{1}{2} p_\mu \delta \mathcal{M}_{\mu \nu}^{-1} p_\nu +  p_\mu  2 \left[ \mathcal{M}_{\mu \lambda}^{-1} \delta I_2^\lambda + b_\mu \delta I_1 \right] = \frac{1}{2} p_\mu \delta \mathcal{M}_{\mu \nu}^{-1} p_\nu - M^{-1} p_\mu  \delta Q_\mu.
\label{eq:deltaOmegap}
\end{equation}
The first term corresponds to mass renormalization, whereas the second term renormalizes $Q_\mu$, see Eq.~\eqref{eq:defGdefOmegaApdx}.

Before solving the self-consistency equations for $\delta \alpha_{\vec{p}}$, we derive the universal form of $\alpha_{\vec{p}}(\Lambda)$ anticipated in Eq.~\eqref{eq:AlphakUniversal}. Comparison of Eqs.\eqref{eq:deltaOmegap}, \eqref{eq:selfConsdeltaAlpha} directly yields
\begin{equation}
\delta \eta_{\vec{p}} = \frac{1}{2} p_\mu \delta b_\mu W_p + p_\mu W_p^{-1} G_- \l \int_\f d^d k~  \frac{W_k^{-1}}{\Omega_{\vec{k}}} \left[ \mathcal{M}_{\mu \nu}^{-1} k_\nu \alpha_{\vec{k}} + b_\mu W_k \right]  \r + 2 W_p \left[ G_+ \delta I_1 + b_\nu \delta I_2^\nu  \right].
\end{equation}
Here we can read off the RG flows $\delta g_1$ and $\delta g_2^\mu$, leading to the RG flow equations presented in the follow subsection. While $\delta g_1$ depends on $\delta I_1$, $\delta I_2^\nu$, the coupling constant $\delta g_2^\mu$ is independent of these integrals,
\begin{equation}
\delta g_2^\mu = G_- \int_\f d^dk~ \frac{W_k^{-1}}{\Omega_{\vec{k}}} \left[ \mathcal{M}_{\mu \nu}^{-1} k_\nu \alpha_{\vec{k}} + b_\mu W_k \right].
\end{equation}

We can formulate a set of linear equations for $\delta I_1$ and $\delta I_2^\mu$ by plugging Eq.~\eqref{eq:selfConsdeltaAlpha} into the definitions of the integrals in Eq.~\eqref{eq:defI1I2}. For simplicity, we will now assume, without loss of generality, that the polaron momentum $\vec{p} = p \vec{e}_x$ points along $x$-direction. Hence, from symmetry, it follows that $b_\mu = \delta_{\mu x} b_x$ and thus $\delta I_2^\nu = \delta_{\nu x} \delta I_2^x$. We arrive at
\begin{equation}
\l  
\begin{array}{cc}
1+2 G_+ J_1  + 2  b_x J_2  &  2 b_x J_1 + 2 \mathcal{M}_{xx}^{-1} J_2 \\
2 G_+ J_2   + 2 b_x J_7  &    1+2 b_x J_2 + 2 \mathcal{M}_{xx}^{-1} J_7
\end{array}
\r
\l 
\begin{array}{c}
\delta I_1 \\
\delta I_2^x
\end{array}
\r = -
\l 
\begin{array}{c}
 J_3 \delta b_x / 2 + J_4 \delta g_2^x + \delta \mathcal{M}_{\mu \mu}^{-1} J_5^\mu/2\\
 J_5^x \delta b_x / 2 + J_6 \delta g_2^x + \delta \mathcal{M}_{\mu \mu}^{-1} J_8^{\mu}/2
\end{array}
\r,
\end{equation}
where the integrals were defined in Eqs.\eqref{eq:IntDef1} - \eqref{eq:IntDef4}. Now it is easy to solve for $\delta I_1$ and $\delta I_2^x$.
\end{widetext}

%%%%%%%%%%%%%%%%%%%%%%%%%%%%%%%%%%%%%%%%%%%%%%%%%%%%%
\section{Dimensional analysis}
\label{apdx:dimAna}
%%%%%%%%%%%%%%%%%%%%%

In this appendix we provide the details of the dimensional analysis which we summarized in Sec.~\ref{subsec:DimAna}. After deriving the results for engineering dimensions, we include the anomalous scaling dimensions of the coupling constants.

\subsection{Engineering dimensions}
\label{sec:engineering}
We start by deriving the engineering dimensions of all operators appearing in the Hamiltonian, i.e. we neglect the cut-off dependencies of coupling constants generated during the RG flow. This gives a good idea about the importance of the various terms in the Hamiltonian. 

%%%%%%%%
\emph{High energies, $k \gg 1/\xi$.--} 
In this high energy regime we have $\Omega_{k} \simeq k^2/2 m_{\rm red}$, such that $\Omega_\Lambda \simeq \Lambda^2$. Therefore we obtain $\a_\Lambda \simeq \Lambda^{-1-d/2}$. Because $V_\Lambda \simeq \Lambda^0$ it follows that $\alpha_\Lambda \simeq \Lambda^{-2}$. The operator $\G_{\vec{k}}$ is a sum of two terms which scale like $\G_{\Lambda} \simeq \Lambda^{-d-2} + \Lambda^{-3-d/2}$. Depending on dimensionality, the first or the second term dominates the behavior of $\G_{\vec{k}}$ at small cut-offs,
\begin{equation}
\G_\Lambda \simeq 
\begin{cases}
\Lambda^{-7/2}, \qquad d=1, \qquad \alpha_{k} \a_{\vec{k}} ~~ \text{dominates},\\
\Lambda^{-4}, \qquad d=2,\\
\Lambda^{-5}, \qquad d=3, \qquad \ad_{\vec{k}} \a_{\vec{k}} ~~ \text{dominates}.
\end{cases}
\end{equation}
Because for $k \gg 1/\xi$ it holds $W_\Lambda, W_\Lambda^{-1} \simeq 1$, the scalings of $\delta \hat{n}_{\Lambda} \simeq \hat{\vartheta}_\Lambda \simeq \Lambda^{-d/2-1}$ coincide.

%%%%%%%%
\emph{Low energies, $k \ll 1/\xi$.--} 
In this low energy regime we have $\Omega_{k} \simeq c k$, such that $\Omega_\Lambda \simeq \Lambda^1$. Therefore we obtain $\a_\Lambda \simeq \Lambda^{-1/2-d/2}$. Because $V_\Lambda \simeq \Lambda^{1/2}$ it follows that $\alpha_\Lambda \simeq \Lambda^{-1/2}$. The operator $\G_{\vec{k}}$ is a sum of two terms of which $\ad_{\vec{k}} \a_{\vec{k}}$ dominates, and it scales as $\G_{\Lambda} \simeq \Lambda^{-d-1}$. In the low energy regime we obtain $W_\Lambda \simeq \Lambda^{1/2}$. The phase-fluctuations have the same scaling as before, $\hat{\vartheta}_\Lambda \simeq \Lambda^{-d/2-1}$, while the particle-number scattering is suppressed, $\delta \hat{n}_{\Lambda} \simeq \Lambda^{-d/2}$.

%%%%%%%%%%%%%%%%%%%%%
\subsection{Analysis in the absence of divergencies}
\label{sec:ScalingWithoutDivergencies}
In the analysis so far we ignored the RG flows of the coupling constants $\mathcal{M}(\Lambda)$ and $G_\pm(\Lambda)$. If they change with sufficiently large powers, however, the relative importance of various terms in the Hamiltonian may change. Below we derive the scaling of the coupling constants from the RG flow equations \eqref{eq:RGflowMSymm} - \eqref{eq:RGflowg1Symm}, i.e. their anomalous dimensions. 

Here we will consider the regime when neither of the two couplings $G_\pm$ diverges during the RG flow. The resulting dimensional analysis is presented in Tab.\ref{tab:dimAnNoDivergenciesIR} for small energies $k \ll 1/\xi$ and in Tab.\ref{tab:dimAnNoDivergenciesUV} for large energies $k \gg 1/\xi$.

%%%%%%%%%%%%%%%%%%%%%
\begin{table}[b]
\begin{tabular}{  p{4cm} | p{1.2cm} | p{1.2cm} | p{1.2cm} }
 scaling for $k \ll 1/\xi$  & $d=1$ & $d=2$  & $d=3$ \\[1.2ex]
 \hline \hline
 $\a_{\vec{k}}$  & $\Lambda^{-1}$   & $\Lambda^{-3/2}$   & $\Lambda^{-2}$  \\ [1.2ex]
 \hline
  $\frac{1}{\mathcal{M}}  \int^\Lambda d^d k  d^d k'  ~ \vec{k} \cdot \vec{k}' :\G_{\vec{k}} \G_{\vec{k}'}: $  & $\bf \Lambda^0$  & $\bf \Lambda^0$ & $\bf \Lambda^0$ \\[1.2ex]
   \hline
 $G_+  \int^\Lambda d^d k  d^d k' ~ : \delta \hat{n}_{\vec{k}} \delta \hat{n}_{\vec{k}'} : $  &$ \Lambda^1$  & $ \Lambda^2$ & $ \Lambda^3$ \\[1.2ex]
  \hline
 $G_-  \int^\Lambda d^d k  d^d k'  ~ : \hat{\vartheta}_{\vec{k}} \hat{\vartheta}_{\vec{k}'} : $  &$\bf \Lambda^{0}$   & $\bf \Lambda^{0}$  & $ \Lambda^{1}$
\end{tabular}
\caption{Dimensional analysis of quantum fluctuations around the MF polaron state at low energies for $k\ll 1/\xi$. The RG flow of the coupling constants is included and the regime where neither of the coupling constants $G_\pm$ diverges is considered. Relevant and marginal terms are highlighted.}
\label{tab:dimAnNoDivergenciesIR}
\end{table}
%%%%%%%%%%%%%%%%%%%%%

%%%%%%%%%%%%%%%%%%%%%
\begin{table}[t]
\begin{tabular}{  p{4cm} | p{1.2cm} | p{1.2cm} | p{1.2cm} }
 scaling for $k \gg 1/\xi$  & $d=1$ & $d=2$  & $d=3$ \\[1.2ex]
 \hline \hline
 $\a_{\vec{k}}$  & $\Lambda^{-3/2}$ & $\Lambda^{-2}$  &  $\Lambda^{-5/2}$ \\ [1.2ex]
 \hline
  $\frac{1}{\mathcal{M}}  \int^\Lambda d^d k  d^d k' ~ \vec{k} \cdot \vec{k}' :\G_{\vec{k}} \G_{\vec{k}'}: $  & $\bf \Lambda^{0}$ & $\bf \Lambda^{0}$ & $\bf \Lambda^{-1}$ \\[1.2ex]
   \hline
 $G_+  \int^\Lambda d^d k  d^d k' ~ : \delta \hat{n}_{\vec{k}} \delta \hat{n}_{\vec{k}'} : $  & $\bf \Lambda^{0}$ &  $\bf \Lambda^{0}$ &  $\Lambda^{1}$  \\[1.2ex]
  \hline
 $G_-  \int^\Lambda d^d k  d^d k' ~ : \hat{\vartheta}_{\vec{k}} \hat{\vartheta}_{\vec{k}'} : $  & $\bf \Lambda^{0}$ &  $\bf \Lambda^{0}$  &  $\Lambda^{1}$
\end{tabular}
\caption{The same as Tab.\ref{tab:dimAnNoDivergenciesIR} but in the high energy regime where $k \gg 1/\xi$.}
\label{tab:dimAnNoDivergenciesUV}
\end{table}
%%%%%%%%%%%%%%%%%%%%%

%%%%%%%%
\emph{Scalings of the RG coupling constants.--} 
We begin with the analysis of the effective polaron mass. Its RG flow in the high energy regime ($k \gg 1/\xi$) yields $\partial_\Lambda \mathcal{M} \simeq \Lambda^{d-5}$. In $d \leq 3$ dimensions this leads to a monotonic growth of $\mathcal{M}(\Lambda)$, and we obtain $\mathcal{M} \simeq \Lambda^{d-4}$ and thus $\mathcal{M}^{-1} \simeq \Lambda^{4-d}$. At small energies, $k \ll 1/\xi$, the situation is different. There we obtain $\partial_\Lambda \mathcal{M} \simeq \Lambda^{d-1}$, meaning that additional renormalization $\Delta \mathcal{M} \simeq \Lambda^{d}$ at small cut-offs becomes irrelevant. We should set $\mathcal{M}(\Lambda) \simeq \Lambda^0$ because $\mathcal{M}(\Lambda) \approx \mathcal{M}(1/\xi)$, and accordingly $\mathcal{M}^{-1}(\Lambda) \simeq \Lambda^0$.

For $G_\pm$ we can perform a similar analysis. For $k \gg 1/\xi$ we obtain that $G_\pm^{-1}(\Lambda) = G_\pm^{-1}(\Lambda_0) + \Delta G_\pm^{-1}(\Lambda)$, where $\Delta G_\pm^{-1}(\Lambda) \simeq \Lambda^{d-2}$. In one dimension $G_\pm^{-1}$ grows indefinitely and the term $G_\pm^{-1}(\Lambda_0)$ can be discarded. Here we obtain $G_\pm(\Lambda) \simeq \Lambda$. In $d=2,3$ dimensions $\Delta G^{-1}_\pm(\Lambda)$ stops to flow (we ignored a slow logarithmic divergence in two dimensions). As long as we are not in the vicinity of a resonance where $G_\pm^{-1}=0$ we can thus assume that $G_\pm^{-1}(\Lambda) \approx G_\pm^{-1}(\Lambda_0) \simeq \Lambda^0$. A similar analysis for $k \ll 1/\xi$ shows that $G_+(\Lambda) \simeq \Lambda^0$ in any dimension, whereas $G_-(\Lambda)$ scales exactly as it does for $k \gg 1/\xi$.

Finally we need to study the RG flow of $\beta(\Lambda)$, which can be done using Eq.~\eqref{eq:g1FormalSolutionSymm}. In $d=2,3$ dimensions the changes of both $G_+^{-1}(\Lambda)$ and $J_1(\Lambda)$ become irrelevant at high and low energies (again discarding a peculiar log-divergence in two dimensions). In $d=1$ this is also true for $k \ll 1/\xi$. In these cases we obtain $\beta(\Lambda) \simeq \Lambda^0$. The same is true even in one dimension for $k \gg 1/\xi$ as long as we can assume that $\Omega_\Lambda \simeq \Lambda^2$, which is the case when $m_{\rm B} \leq \mathcal{M}$. The situation where $\mathcal{M} \ll m_{\rm B}$ requires a separate treatment.

%%%%%%%%%%%%%%%%%%%%%
\subsection{Analysis in the presence of divergencies}
\label{sec:ScalingWithDivergencies}
Finally we briefly discuss the physics in the vicinity of divergencies of $G_\pm$. The behavior of $\mathcal{M}$ is not modified by this. As before we obtain the following scaling for $k \gg 1/\xi$, $\Delta G_\pm^{-1}(\Lambda) \simeq \Lambda^{d-2}$. Because at the resonance (at a cut-off $\Lambda_c$, say) $G_\pm^{-1}(\Lambda_c)=0$, it follows that $G_\pm(\Lambda) \simeq \Lambda^{2-d}$. Accordingly we obtain for $k \ll 1/\xi$ that $G_-(\Lambda) \simeq \Lambda^{2-d}$ and $G_+(\Lambda) \simeq \Lambda^{-d}$.

In the dimensional analysis of Tabs. \ref{tab:dimAnNoDivergenciesIR} and \ref{tab:dimAnNoDivergenciesUV} the scalings of two-phonon terms are modified when the respective couplings $G_\pm$ diverge. For $d=1,2,3$ the terms become marginal both for $k \ll 1/\xi$ and $k \gg 1/\xi$. I.e. when $G_+ \to \infty$ diverges,
\begin{equation}
 G_+  \int^\Lambda d^d k  d^d k' : \delta \hat{n}_{\vec{k}} \delta \hat{n}_{\vec{k}'} :  ~\simeq ~   \Lambda^0,
\end{equation}
and a similar result holds when $G_- \to \infty$ diverges. In this case there can be a competition between two-phonon and Fr\"ohlich terms.

%%%%%%%%%%%%%%%%%%%%%%%%%%%%%%%%%%%%%%%%%%%%%%%%%%%%%
\section{Two-phonon terms and\\ the logarithmic UV divergence}
\label{apdx:2PhUVlogDiv}
In this appendix we show that two-phonon terms do not modify the Fr\"ohlich type UV log-divergence in the order of the RG that we consider. To this end we show that the second term in Eq.~\eqref{eq:RGflowEnergySymm} for the polaron energy is UV convergent. 

We will make use of the asymptotic expression 
\begin{equation}
W_k ~ \stackrel{\rm UV}{\simeq} ~ 1 - \frac{1}{2 k^2 \xi^2} + \mathcal{O}(k^{-4}).
\end{equation}
Furthermore we note that $m_{\rm red}^* = 1 / \l m_{\rm B}^{-1} + \mathcal{M}^{-1} \r = m_{\rm red} \l 1+ \mathcal{O}(\Lambda^{-1},\Lambda_0^{-1}) \r $, see Eq.~\eqref{eq:MUVsol}. Hence we can write to leading order in $\Lambda^{-1}$:
\begin{multline}
 \int_\s d^3p  \int_\f d^{2}k ~ \frac{\left[ W_k W_p G_+ - W_p^{-1} W_k^{-1} G_- \right]^2}{\Omega_{\vec{k}}} =32 \pi^2 m_{\rm red}\\
 \times \int_0^\Lambda dp~p^2  \left[  G_+ - G_-   - \frac{G_+ + G_-}{2 \xi^2} \l \frac{1}{\Lambda^2} + \frac{1}{p^2} \r \right]^2.
 \label{eq:E2evaluate}
\end{multline}

Now we solve for the RG flows of $G_\pm(\Lambda)$. Because the corresponding RG flow equations are separable, we obtain exact expressions in terms of integrals,
\begin{multline}
G_\pm^{-1}(\Lambda) = G_\pm^{-1}(\Lambda_0)  + 16 \pi m_{\rm red} \int_\Lambda^{\Lambda_0} d \Lambda ~ \l 1 \mp \Lambda^{-2} \xi^{-2} \r\\ 
\times \left[ 1 + M^{-2} m_{\rm red}^4  \frac{8}{\pi^2} n_0 \l \beta_\MF g_\IB^{(0)} \r ^2 \l \Lambda^{-1} - \Lambda_0^{-1} \r \right],
\end{multline}
again to leading order in $\Lambda$, $\Lambda_0$. Note that $G_\pm^{-1}(\Lambda_0)$ is UV divergent itself, 
\begin{equation}
G_\pm^{-1}(\Lambda_0) = - \Lambda_0 16 \pi m_{\rm red} + 8 \pi \frac{m_{\rm red}}{a_\IB} + \mathcal{O}(\Lambda_0^{-1}).
\label{eq:GpmInitialCond}
\end{equation}
Using these expressions, we find that
\begin{equation}
G_+ + G_- = - \frac{\Lambda^{-1}}{8 \pi m_{\rm red}}  \left[ 1 + \mathcal{O} \l \log  \l \frac{\Lambda}{\Lambda_0} \r \frac{1}{\Lambda} \r \right].
\end{equation}
Therefore, the second term in the second line of Eq.~\eqref{eq:E2evaluate} gives rise to energy contributions scaling like
\begin{equation}
\partial_\Lambda E_0 \sim \Lambda^3 \Lambda^{-2} \Lambda^{-4} = \Lambda^{-3} \qquad \Rightarrow \qquad E_0 \sim \Lambda_0^{-2},
\end{equation}
which are UV convergent. For the first term in the second line of Eq.~\eqref{eq:E2evaluate} we use that
\begin{equation}
G_+ - G_- = \frac{\Lambda^{-2}}{8 \pi m_{\rm red} \xi^2}  \l \Lambda^{-1} - \Lambda_0^{-1} \r,
\end{equation}
giving rise to contributions
\begin{align}
\partial_\Lambda E_0 &\sim \Lambda^3 \Lambda_0^{-2} \Lambda^{-4} = \Lambda^{-1} \Lambda_0^{-2} \\ 
 &\Rightarrow \qquad E_0 \sim \Lambda_0^{-2} \log (\Lambda_0),
\end{align}
which are UV convergent as $\Lambda_0 \to \infty$.

%%%%%%%%%%%%%%%%%%%%%%%%%%%%%%%%%%%%
\section{Relation to other logarithmic corrections to the ground state energy}
\label{apdx:relationLogDivergence}
The log-divergence discussed in Eq.~\eqref{eq:UVlogDivergence} is closely related to the logarithmic corrections of the ground state energy in an interacting Bose gas \cite{Wu1959,Sawada1959,Hugenholtz1959}. By considering every particle as an impurity interacting with the surrounding bosons we expect a similar correction $\sim - a^4 n_0^2 \log \l \Lambda_0 / \Lambda_1 \r$ to the energy per particle $E_0/N$ in this case, where $a$ is the scattering length of the bosons. When the UV cut-off is provided by $\Lambda_0 \sim 1/a$ as assumed in Ref.~\cite{Hugenholtz1959}, and considering the case when the mass renormalization is large, we obtain a correction to the energy density $\Delta E_0/L^3 \sim a^4 n_0^3 \log \l  n_0 a^3 \r$. Indeed this is the correct scaling with $a$ and $n_0$ of the leading order correction to the ground state energy due to quantum fluctuations \cite{Wu1959,Sawada1959,Hugenholtz1959}. When the mass renormalization is small the same scaling is obtained by taking into account the dependence of the healing length on the scattering length $a_{\rm BB}=a=a_\IB$, see Ref. \cite{Christensen2015}. Because $\xi = 1/\sqrt{2 m_{\rm B} g_{\rm BB} n_0} \sim \sqrt{n_0 a}$ we obtain $\log (\Lambda_0 \xi) \sim \log n_0 a^3$, again for $\Lambda_0 \sim 1/a$.

In Ref.~\cite{Christensen2015} corrections to the Bose polaron energy were calculated exactly up to order $\mathcal{O}(a_\IB^3)$ using diagrammatic perturbation theory. The authors derived a logarithmic UV divergence which scales like $\sim n_0 a_\IB^3  \xi^{-2} \log (\Lambda_0 \xi)$. This term is not included in our RG and we expect that it corresponds to a  contribution of order $\mathcal{O}(\Omega_k^{-2})$  or higher, neglected in our analysis. The authors of \cite{Christensen2015} pointed out that this term also gives rise to the correct scaling $\Delta E_0 / L^3 \sim n_0^3 a^4 \log (n_0 a^3)$ for the interacting Bose gas (where $a_\IB = a_{\rm BB} = a$) \cite{Wu1959,Sawada1959,Hugenholtz1959}. 

Our RG approach suggests a unified picture of all logarithmic corrections discussed above. We showed in \cite{Grusdt2015Varenna} that the polaronic mass renormalization, which is generically given by $\mathcal{M}(k) = M + \mathcal{O}(1/k)$ in the UV limit $k \lesssim \Lambda_0$, gives rise to a correction of the impurity energy that diverges logarithmically with the UV cut-off $\Lambda_0$. On generic grounds we expect that interaction terms in the Hamiltonian lead to some mass renormalization, connected with the appearance of generic log-divergent terms in the energy. While the log-divergence found by Christensen et al. is related to polaron mass generated by two-phonon terms \cite{Christensen2015}, we study mass renormalization due to the Fr\"ohlich terms in the Hamiltonian of the Bose polaron problem. These different mechanisms of mass renormalization explain why the scaling with $a_\IB$ in front of the log-divergencies is different for the two situations.

\newpage

%%%%%%%%%%%%%%%%%%%%%%%%%%%%%%%%%%%%%%%%
%\bibliography{/Users/fgrusdt/Documents/Library/dataBase_JabRef2.bib}
%\bibliographystyle{unsrt}

\end{document}